\documentclass[structabstract]{aa}

\usepackage{epsfig,natbib,url,twoopt}
\usepackage{graphicx} %xetex
\usepackage{txfonts}
\usepackage{wrapfig}
\usepackage{longtable}
\usepackage{supertabular}
\usepackage{lscape}
\usepackage[labelfont=footnotesize,textfont=footnotesize]{caption}  
\newcommand{\lya}{\mbox{Ly$\alpha$}}

\bibpunct{(}{)}{;}{a}{}{,}    

\begin{document}

\title{The Lyman alpha reference sample: IV. Morphology at low and high redshift 
  \thanks{Based on observations made with the NASA/ESA Hubble Space Telescope. These observations are associated with programme 12310.}}
\author{L. Guaita\inst{1}
\and J. Melinder\inst{1}
   \and M. Hayes\inst{1}
    \and G. {\"O}stlin\inst{1}
\and J. Gonzalez\inst{1} 
\and G. Micheva\inst{1,}\inst{2}
\and A. Adamo\inst{1}
\and  J. M. Mas-Hesse\inst{4}
\and A. Sandberg\inst{1}
\and H. Ot\'i-Floranes\inst{5,14}
\and D. Schaerer\inst{3,6} 
\and A. Verhamme\inst{6}
\and E. Freeland\inst{1}
\and I. Orlitov\'a\inst{7}
\and P. Laursen\inst{8}
\and J. M. Cannon\inst{9}
\and F. Duval\inst{1}
\and T. Rivera-Thorsen \inst{1}
\and E. C. Herenz\inst{10}
\and D. Kunth\inst{11}
\and H. Atek\inst{12}
\and J. Puschnig \inst{1}
\and P. Gruyters \inst{13}
\and S. A. Pardy \inst{14}
}
\offprints{Lucia Guaita, \email{lguai@astro.su.se}}
\institute{Department of Astronomy, Oskar Klein Centre for cosmoparticle physics, Stockholm University,  AlbaNova, Stockholm SE-10691, Sweden
 \and Subaru Observatory, Hilo, Hawai, USA
\and CNRS, IRAP, 14, avenue Edouard Belin, F-31400 Toulouse, France
\and Centro de Astrobiolog\'{\i}a (CSIC--INTA), Departamento de Astrof\'{\i}sica, POB 78, Villanueva de la Ca\~nada, Spain
\and Instituto de Astronom\'{\i}a, Universidad Nacional Aut\'onoma de M\'exico, Apdo. Postal 106, Ensenada B. C. 22800, Mexico
\and Geneva Observatory, University of Geneva, 51 Chemin des Maillettes, CH-1290 Versoix, Switzerland
\and Astronomical Institute of the Academy of Sciences of the Czech Republic, Bo\v{c}n{\'\i} II 1401/1a, CZ-141 00 Praha 4, Czech Republic
\and Dark Cosmology Centre, Niels Bohr Institute, University of Copenhagen, Juliane Maries Vej 30, DK-2100 Copenhagen, Denmark
\and Department of Physics and Astronomy, Macalester College, 1600 Grand Avenue, Saint Paul, MN 55105, USA
\and Leibniz-Institut fur Astrophysik (AIP), An der Sternwarte 16, D-14482 Potsdam, Germany
\and Institut d'Astrophysique de Paris, UMR 7095 CNRS \& UPMC, 98 bis Bd Arago, F-75014 Paris, France
\and Laboratoire d'Astrophysique, \'Ecole Polytechnique F\'ed\'erale de Lausanne (EPFL), Observatoire, CH-1290 Sauverny, Switzerland
\and Department of Physics and Astronomy, Uppsala University, Box 515, 751 20, Uppsala, Sweden
\and Centro de Radioastronom\'{\i}a y Astrof\'{\i}sica, UNAM, Campus Morelia, Mexico
\and Astronomy Department, University of Wisconsin - Madison , 475 N. Charter Street, Madison, Wisconsin, 53706, USA
}
\date{Accepted date: Dec 19th, 2014}
\abstract
%Context
%{The transport of Ly$\alpha$ photons in galaxies is a complex process and the conditions under which Ly$\alpha$ photons manage to escape from certain galaxies is still under investigation. The Lyman alpha reference sample (LARS) is a sample of 14 local star-forming galaxies, designed to study Ly$\alpha$ in detail and relate it to rest-frame UV and optical emission.
%}
{}
%Aim
{With the aim of identifying rest-frame UV and optical properties, typical of Ly$\alpha$ emitters (LAEs, galaxies with EW(Ly$\alpha)>20$ {\AA}) at both low and high redshift, we investigated the morphological properties of the LARS galaxies, in particular the ones that exhibit intense Ly$\alpha$ radiation.
}
%Method
{We measured sizes and morphological parameters in the continuum, Ly$\alpha$, and H$\alpha$ images. We studied morphology by using the Gini coefficient vs M20 and asymmetry vs concentration diagrams. We then simulated LARS galaxies at $z\sim2$ and 5.7, performing the same morphological measurements. We also investigated the detectability of LARS galaxies in current deep field observations. The subsample of LAEs within LARS (LARS-LAEs) was stacked to provide a comparison to stacking studies performed at high redshift.
}
%Results
{LARS galaxies have continuum size, stellar mass, and rest-frame absolute magnitude typical of Lyman break analogues in the local Universe and also similar to $2<z<3$ star-forming galaxies and massive LAEs. LARS optical morphology is consistent with the one of merging systems, %(G$^{SB-rP20S}_{rest-optical}>0.5$, M20$_{rest-optical}<-2.0$)
 and irregular or starburst galaxies. For the first time we quantify the morphology in Ly$\alpha$ images: even if a variety of intrinsic conditions of the interstellar medium can favour the escape of Ly$\alpha$ photons, LARS-LAEs appear small in the continuum, and their Ly$\alpha$ is compact. LARS galaxies tend to be more extended in Ly$\alpha$ than in the rest-frame UV. It means that Ly$\alpha$ photons escape by forming haloes around HII regions of LARS galaxies. 
}
%Conclusions
{The stack of LARS-LAE Ly$\alpha$ images is peaked in the centre, indicating that the conditions, which make a galaxy an LAE, tend to produce a concentrated surface brightness profile. On the other hand, the stack of all LARS galaxies is shallower and more extended. This can be caused by the variety of dust and HI amount and distribution, which produces a more complex, patchy, and extended profile, like the one observed for Lyman break galaxies that can contribute to the stack. We cannot identify a single morphological property that controls whether a galaxy emits a net positive Ly$\alpha$ flux. However, the LARS-LAEs have continuum properties consistent with merging systems.
}


\titlerunning{LARS at low and high $z$}
\authorrunning{L. Guaita}
\maketitle 
%

%%%%%%%%%%%%%%%%%%%%%%%%%%%%%%%%%%%%%%%%%%%%%%%%%%%%%%%%%%%%%%%%%%%%%%%%%%%%
\section{Introduction}     \label{sec:introduction}
%%%%%%%%%%%%%%%%%%%%%%%%%%%%%%%%%%%%%%%%%%%%%%%%%%%%%%%%%%%%%%%%%%%%%%%%%%%%

Originating mainly in recombining gas being ionized by hot O and B stars, Lyman alpha (Ly$\alpha$) radiation has proved an excellent probe of star-forming galaxies at both low \citep[e.g.,][]{Cowie2011} and high \citep[e.g.,][]{Ouchi:2010} redshift. 

\citet{Hayes2005,Hayes2009} and \citet{Ostlin2009}  developed a method to separate the rest-frame UV and Ly$\alpha$ emission in Hubble Space Telescope (HST) data.
In these papers it was demonstrated observationally that, in nearby galaxies ($z<0.1$), 
Ly$\alpha$ emission extends away from the star-forming regions where the Ly$\alpha$ photons were originally generated, forming the so-called Ly$\alpha$ haloes.

Young starburst galaxies are expected to be very bright in Ly$\alpha$ \citep{PP1967}. 
For the past 15 years, star-forming galaxies have been successfully detected at $z>2$ by identifying their strong Ly$\alpha$ emission line. The principal method used is the narrow-band technique \citep[e.g.,][]{CHu1998,Rhoads:2000,Ouchi:2008,Gronwall:2007,Nilsson:2009}: Ly$\alpha$ emitters (LAEs) present an excess in a narrow band (covering the redshifted Ly$\alpha$ wavelength) with respect to a broad-band filter (covering the rest-frame UV continuum). Because Ly$\alpha$ photons are easily absorbed by dust grains and are scattered by neutral hydrogen (HI), LAEs were thought to be a $special$ population of galaxies with $special$ dust and HI amounts and distribution. Although extensive studies have been carried out to characterize LAE physical properties and their special conditions \citep[][among the most recent ones]{Nilsson2010,Acquaviva2012,McLinden2014,Vargas2014,Hagen2014}, the results have been inconclusive. The mechanisms (e.g., interstellar medium geometry and kinematics) controlling the escape of Ly$\alpha$ photon are still debated.

The morphology of the rest-frame UV and optical continua provides information about galaxy formation and evolution \citep[e.g., star-forming region distribution, merger events,][]{Conselice2003,Lotz2004}. From the ground LAEs were observed to be compact in the rest-frame UV, but multiple components were identified in deep HST-resolution images \citep{Bond2009, Bond2012}. There have also been a few attempts to quantify the morphology of the Ly$\alpha$ emission itself. \citet{Bond2010} explored a sample of seven LAEs placed at $z\simeq3.1$ %(\@5000 {\AA}) 
(observed-frame $\lambda$(Ly$\alpha) \sim5000$ {\AA}) by using HST WFPC2 (Wide Field Camera2) F502N narrow-band imaging. They found that, for one source, Ly$\alpha$ emission extended till 1.5 kpc ($\leq1$ kpc for the other six), a just-slightly-larger scale than the UV continuum. Also, this source was composed of two main clumps both in the rest-frame UV and in Ly$\alpha$. \citet{Finkelstein2011b} spatially resolved three spectroscopically confirmed LAEs placed at $z\simeq4.5$ %(\@6570 {\AA}) 
(observed-frame $\lambda$(Ly$\alpha) \sim6570$ {\AA}) by using the HST ACS (Advanced Camera for Surveys) F658N narrow band. Two out of the three systems showed Ly$\alpha$ emission significantly more extended than the UV continuum.

Recently, evidence of extended Ly$\alpha$ emission was found in the stack of a large sample of Lyman break galaxies, \citep{Steidel2011}, generally more massive and dustier than LAEs, and of galaxies located in overdense and not-overdense regions \citep{Matsuda2012}. 
By stacking hundreds of $z\simeq2.2$, $z\simeq3.1$, $z\simeq3.7$, and $z\simeq5.7$ LAEs from deep ground-based imaging, \citet{Momose2014} 
discovered extended Ly$\alpha$ emission, with scale lengths of $5-10$ kpc. However, by stacking their sample of LAEs \citet{Feldmeier2013} just found a marginal detection at $z\sim3.1$ and a non-detection ($z\simeq2.07$). It is clear that depth and image resolution were the main factors affecting their results.

Instead, giant Ly$\alpha$ nebulae, powered by active galactic nuclei, have been studied by a few authors to assess the role of HI scattering and Ly$\alpha$ radiative transfer effects \citep[e.g.,][]{Humphrey2013,Prescott2014}

Local starbursts \citep[][]{Overzier2008,Overzier2009,Overzier2010,Hayes2013,Hayes2014,Petty2014,GO2014} are unique laboratories to study the rest-frame UV in detail and optical light distribution, morphology, and to investigate the mechanisms, which allow Ly$\alpha$ photons to escape. 
In \citet{GO2014} (hereafter Paper I) we presented the Lyman alpha reference sample (LARS), which is composed of 14 star-forming galaxies at $z<0.2$. These galaxies were observed during HST cycle 18 (P.I. G. {\"O}stlin) in a set of rest-frame UV (ACS/SBC F125LP, F140LP, F150LP) and optical (WFC3/UVIS F336W/F390W, F438W/F475W, F775W/F850LP, F502N, F656N, and ACS/WFC F502N/F505N/F551N, F656N/F716N/F782N) filters. Ly$\alpha$ maps were generated by estimating the continuum at rest-frame $\lambda$(Ly$\alpha)=1216$ {\AA}, through modelling the galaxy spectrum as a composite population of young stars, old stars, and nebular gas.
LARS images were published in \citet{Hayes2013} (hereafter Paper 0) and further analysed in \citet{Hayes2014} (hereafter Paper II). We found that the Ly$\alpha$ emission profile appeared different from the rest-frame UV and it flattened on scales larger than the rest-frame UV. The majority of the 14 galaxies showed a negative Ly$\alpha$ equivalent width at small radii and then an increase farther out. We concluded that this was due to scattering on neutral hydrogen, which is able to shape the Ly$\alpha$ emission into the form of haloes. Also, by comparing LARS Ly$\alpha$ with global physical properties, it appeared that the Ly$\alpha$ photon escape was favoured in the system with weaker dust reddening and small stellar mass.

The neutral hydrogen content of LARS galaxies was presented in \citet{Pardy2014} (hereafter Paper III). The spectral lines of HI were detected in 11 of the 14 observed LARS galaxies and it was also found that the Ly$\alpha$ escape was favoured in low HI-mass systems.
LARS interstellar medium kinematics will be presented in \citet{Rivera2014} (submitted), in Duval et al. (in prep), and in Orlitov\'a et al. (in prep). 

In this paper, number IV of the series, we address the question whether specific galaxy morphological properties could be related to the escape of Ly$\alpha$ photons and escape in haloes. Note that in Paper I the present paper was termed paper 7, due to a previous numbering. In Sec. \ref{sec:method}, we briefly explain how we measured morphological parameters (details are give in Appendix A) and the process adopted to simulate how some local ($z<0.2$) galaxies would appear at high redshift ($z>2$). In Sec. \ref{sec:Loriginal}, we describe the morphological properties of the sample of local galaxies and compare them with local-Universe and high-$z$ galaxy populations. 
In Sec. \ref{sec:LARSz2}, we study the morphological properties of the high-$z$-simulated galaxies. In Sec. \ref{sec:Stack}, we present the stacking of the high-$z$-simulated sample and compare with high-$z$ stacks in the literature. In Sec. \ref{sec:discussion} and \ref{sec:conclusions}, we discuss and summarize the main results of the paper. 

Throughout we adopt AB magnitudes and assume a $\Lambda$CDM cosmology of ($H_0$, $\Omega_\mathrm{m}$, $\Omega_{\Lambda}$) = (70 km~s$^{-1}$~Mpc$^{-1}$, 0.3, 0.7) as in Hayes et al. 2013, 2014.

%%%%%%%%%%%%%%%%%%%%%%%%%%%%%%%%%%%%%%%%%%%%%%%%%%%%%%%%%%%%%%%%%%%%%%%%%%%%
\section{Method}    
\label{sec:method}
%%%%%%%%%%%%%%%%%%%%%%%%%%%%%%%%%%%%%%%%%%%%%%%%%%%%%%%%%%%%%%%%%%%%%%%%%%%%

We present the morphology of the local ($0.03<z<0.2$) LARS galaxies and investigate how it would change if the same galaxies were observed at high redshift. As explained above, in Paper II we isolated the contributions of the rest-frame UV ($\sim1220$ {\AA}), optical ($\sim6570$ {\AA}), Ly$\alpha$ (\@1216 {\AA}), and H$\alpha$ (\@6563 {\AA}).
In this work, we measure the morphological parameters of these contributions. The LARS galaxies are hereafter referred to as L$n$, where $n$ ranges from 01 to 14 (see Paper 0).

\subsection{Morphological parameter estimation}
\label{sec:morphology}

With the aim of quantifying the morphology of LARS galaxies, we calculated their sizes and performed non-parametric measurements of morphological parameters (see Appendix A and Fig. \ref{eqMORPH} for details).

We calculated sizes, in terms of Petrosian semi-major axis \citep[rP20, e.g.,][]{Lotz2004,Lisker2008}, circular Petrosian radius \citep{Petrosian1976}, and radii containing 20\%, 50\%, 80\% of the flux (r$_{20}$, r$_{50}$, r$_{80}$). A comparison between these radii gives an idea of the distribution of the light in the galaxy.
We also estimated asymmetry (A), concentration (C), clumpiness (S), Gini coefficient (G), and second-order moment of the brightest 20\% of the galaxy's flux \citep[M20, see e.g.,][]{Conselice2003,Lotz2004,Scarlata2007,Micheva2013}. 

The asymmetry quantifies the symmetry of a galaxy with respect to a 180-degree rotation; the concentration describes how much the light is concentrated in the centre of a galaxy; the clumpiness measures the amount of small-scale structures within a galaxy; the Gini coefficient provides the information on how uniform is the light distribution; 
M20 traces the spatial distribution of any bright knots, and also off-centre clumps, its definition is very similar to that of C, but M20 is more sensitive to merger structures, such as off-centre components.

We first ran the Source Extractor (SExtractor) software \citep{bertin1996}. It provided the galaxy centroid and the elliptical aperture, containing the entire galaxy and characterized by semi-major axis (sma) equal to rP20. The photometry was performed within this SExtractor detection aperture.

We adopted configuration parameters like in \citet{Bond2009} (DETECT\_THRESH =1.65, DETECT\_MINAREA=30, DEBLEND\_MINCONT=1). 
They were optimized to provide morphological measurements in deep HST rest-frame UV observations at $z>2$. 
To prevent SExtractor from breaking up the clumpy, resolved $z\sim0$ LARS galaxies into smaller fragments, we assumed a larger value of DETECT\_MINAREA. This parameter sets the number of contiguous pixels required for a detection to be accepted by SExtractor. We measured fluxes at SExtractor centroid within elliptical apertures, by using the ELLIPSE task in $iraf.stsdas.isophote$ and within circular apertures, by using the PHOT task in $iraf.digiphot.apphot$.
ELLIPSE and PHOT outputs served to infer sizes, A, and C at minimum asymmetry (C$^{minA}$), as explained in Appendix A and previously adopted in \citet{Bershady2000, Conselice2003, Micheva2013}.

The non-parametric measurements and signal-to-noise estimations were performed counting the flux of pixels belonging to a segmentation map. We defined the segmentation map in two ways, one is an ellipse with semi-major axis equal to rP20 \citep{Scarlata2007} and orientation given by SExtractor; one contains the pixels with surface brightness larger than the value at the Petrosian radius \citep{Lotz2004} measured in the smoothed image (smoothed by a kernel of width rP20/5). We calculated M20, S and Gini coefficient by considering the pixels within these segmentation maps.
The Gini coefficients measured in these two segmentation maps are denoted by G$^{rP20}$ and G$^{SB-rp20S}$ respectively. 
As described in \citet{Scarlata2007}, G$^{rP20}$ was defined to be consistent for redshift comparisons, thus we prefer it over G$^{SB-rP20S}$ throughout the paper when we compare with high redshift.

To test our code, we applied it to template galaxies with known profiles and compared the output to the results by \citet{Bershady2000} and \citet{Lotz2006}. We recovered the expected values as described in Appendix A. 

\subsection{Combination of morphological parameters}
\label{sec:useofparam}

As shown in \citet{Conselice2003} and \citet{Lotz2004}, combinations of morphological parameters can give information about galaxy history (e.g., star-formation and merging episodes). First of all the rest-frame UV morphology is sensitive to the current star formation; the rest-frame optical traces the structure of the entire galaxy stellar population \citep{Lee2013,Bond2014}. The combinations of parameters (see previous section) we adopted are,

\begin{itemize}
\item  Asymmetry vs concentration, asymmetry vs clumpiness, together with clumpiness vs concentration, as presented by \citet{Conselice2003}
\item  Gini coefficient vs M20 bright-pixel moment, as presented in \citet{Lotz2004}. 
\end{itemize}

The concentration depends on the galaxy star-formation history in the sense that a rapid gravitational collapse can produce high concentration. The presence of disk and intergalactic gas which cools onto the disk tends to produce a lower concentration value.
Disk galaxies are characterized by $3<$ C $<4$, ellipticals by C $>4$ \citep{Bershady2000}. 
The asymmetry is sensitive to any feature that produces asymmetric light distributions (e.g., star-formation knots, interactions, and mergers).
It is commonly assumed also at high $z$ that large asymmetry (A $>0.38$) indicates a major merger \citep{Aguirre2013, Conselice2003}. Spiral galaxies and systems composed of more than one component are characterized by A $>0.1$. 
The clumpiness is sensitive to the presence of star-forming clumps as well, but background noise can make it difficult to detect low surface brightness regions and increase the appearance of the galaxy as a mix of clumps.
The Gini coefficient can be strongly correlated with C. By definition, G $=1$ means that the light is all concentrated in one pixel, G $=0$ that the light is equally distributed across the galactic body. In the case of a shallow light profile, both G and C are low. When more than one clump contains a significant fraction of light, G can be much larger than zero, but C still low. M20 traces the spatial distribution of off-centre bright regions.  

In general, starburst and irregular galaxies are expected to have large A, large S, and intermediate C, merging systems and perturbed disks show large M20 and intermediate G.

\subsection{High-redshift simulation}    
\label{sec:reg}

We simulated the observations of LARS galaxies (all at $z<0.2$) at higher redshift by transforming their original science- and weight-map images (Paper I) according to the following steps \citep[see also][]{Overzier2008,Adamo2013}.  

\begin{enumerate} 
\item The images were resampled preserving the flux ($IDL$ frebin function). The size of the output image was defined by fixing the physical size of the galaxies.
We chose mainly a $z\sim2$ sampling to be able to compare with the interesting results obtained by surveys of Lyman alpha emitters in the last recent 5 years \citep[][Sandberg in prep.]{Nilsson:2009, Guaita2010, Hayes2010, Nakajima2012a}. Also, the size changes a little with redshift.
\item Continuum subtraction \citep{Hayes2009} was applied to the resampled images to generate rest-frame UV continuum and Ly$\alpha$ line, rest-frame optical continuum and H$\alpha$ line images. The line images are in units of flux (erg sec$^{-1}$cm$^{-2}$), while the continuum images are in units of flux densities (erg sec$^{-1}$cm$^{-2}${\AA}$^{-1}$).
\item The image pixel values were scaled based on luminosity distance and surface brightness dimming \citep[i.e.,][]{Hubble1935,Bouwens2004}. 
\item Gaussian noise, corresponding to a certain simulated survey depth, was added to the resampled and rescaled images by running the MKNOISE task in $iraf.artdata$. To calculate uncertainty on galaxy sizes and morphological parameters we performed Monte Carlo simulations by repeating 100 realizations of a noisy image. The noise applied was defined as the 10$\sigma$ detection within a $\sim$50 pixel (equivalent to a square aperture of $\sim$0.2'' on a side for HST ACS optical filters) area, similar to the limits given for the HUDF \citep[Hubble Ultra Deep Field,][]{Beckwith2006}. We do not show simulations in which we only resampled the pixel scale to that of a ground-based telescope and instrument, because the main effect on continuum and line images was produced by survey depth and ground-based point spread function (PSF, see Sec. \ref{sec:Stack}). 
\end{enumerate}

To choose reasonable ranges of detection limits (Table \ref{tab:depths}) to apply, we referred to the MUSYC (MUlti-wavelength Survey by Yale-Chile) NB3727 narrow band \citep{Guaita2010, Bond2012}, to the triple narrow band by \citet{Nakajima2012a}, CANDELS/HUDF \citep{McLure2013} broad band, and to the dual narrow-band survey by \citet{Lee2012}. 

\begin{table*}
\centering
\caption{10$\sigma$ detection limits applied to high-$z$ simulated LARS images}
\label{tab:depths}
\scalebox{0.9}{
\begin{tabular}{|c|c|c|c|}
\hline\hline 
F(Ly$\alpha$) & m(rest-frame UV) &  F(H$\alpha$) & m(rest-frame optical)\\
\hline
erg sec$^{-1}$ cm$^{-2}$ &AB & erg sec$^{-1}$ cm$^{-2}$& AB \\
\hline
5E-19 &30 &2E-19& 29 \\
3E-18 & 29 & 6E-19 & 28\\
8E-18 & 28 &1E-18 & 27\\
2E-17 & 27 & 3E-18 & 26\\
5E-17  &26 & 1E-17 &25\\
\hline
\end{tabular}
}
\tablefoot{
%A 10$\sigma$ (pixel-by-pixel) noise was added to LARS continuum and line images. 
Noise corresponding to the detection limits given in the table was added to LARS continuum and line images. As LARS Ly$\alpha$ and H$\alpha$(rest-UV and optical continua) images are in units of flux(flux density), the image depths are given in units of erg sec$^{-1}$cm$^{-2}$(AB magnitudes). %As LARS continuum images are in units of flux density, the image depths are given in AB magnitudes. These values are chosen based on real surveys. 
The MUSYC \citep{Guaita2010, Bond2012} NB3727 survey implied a 10$\sigma$ detection limit of F(Ly$\alpha$)=5E-17 erg sec$^{-1}$cm$^{-2}$. MUSYC U,B(HUDF V606) 10$\sigma$ detection limits were about 26(29.5). Assuming a NB3727 filter width and transmission profile, a source with m$_{AB}$(rest-frame UV)=30 and EW(Ly$\alpha)=20$ {\AA} is characterized by F(Ly$\alpha$)=5E-19 erg sec$^{-1}$cm$^{-2}$. 
m$_{AB}$(rest-frame UV) $=30,29,28$ are consistent with HUDF09, GOODS, and GEMS survey depths \citep{Bond2009}.
\citet{Lee2012} survey was characterized by a 10$\sigma$ detection limit of 22.9 in NB210 and 23.7 in $K$. %FWHM(NB210, Ha at z=2.2) = 260 A.
%Their sources spanned a range in F(H$\alpha$) between 1.5E-17 and 1.9E-16 erg sec$^{-1}$cm$^{-2}$.
CANDELS wide(deep) F160W 10$\sigma$ detection limit was 25.8(26.5).
Assuming a NB210 filter width, a source with m$_{AB}$(rest-frame optical) $=27$ and EW(H$\alpha)=20$ {\AA} is characterized by F(H$\alpha$)=2E-19 erg sec$^{-1}$cm$^{-2}$.  }
\end{table*}

%%%%%%%%%%%%%%%%%%%%%%%%%%%%%%%%%%%%%%%%%%%%%%%%%%%%%%%%%%%%%%%%%%%%%%%%%%%%
\section{LARS galaxies at $z\sim0$}
\label{sec:Loriginal}
%%%%%%%%%%%%%%%%%%%%%%%%%%%%%%%%%%%%%%%%%%%%%%%%%%%%%%%%%%%%%%%%%%%%%%%%%%%%

To be able to compare the Ly$\alpha$, H$\alpha$, and continuum properties of LARS galaxies with those of high-$z$ Ly$\alpha$ emitters (LAEs), we focused on the twelve LARS galaxies with EW(Ly$\alpha)>1$ {\AA} as measured in Paper II. Thus, we excluded from this analysis the two galaxies of the sample (L04 and L06) characterized by null Ly$\alpha$ maps. LARS galaxies with integrated EW(Ly$\alpha)>20$ {\AA} composed the subsample of LARS-LAEs (consisting of six galaxies). 
Various physical characteristics of the LARS galaxies (including their coordinates) are discussed in Paper II.

\begin{figure*}
\includegraphics[width=5.5cm]{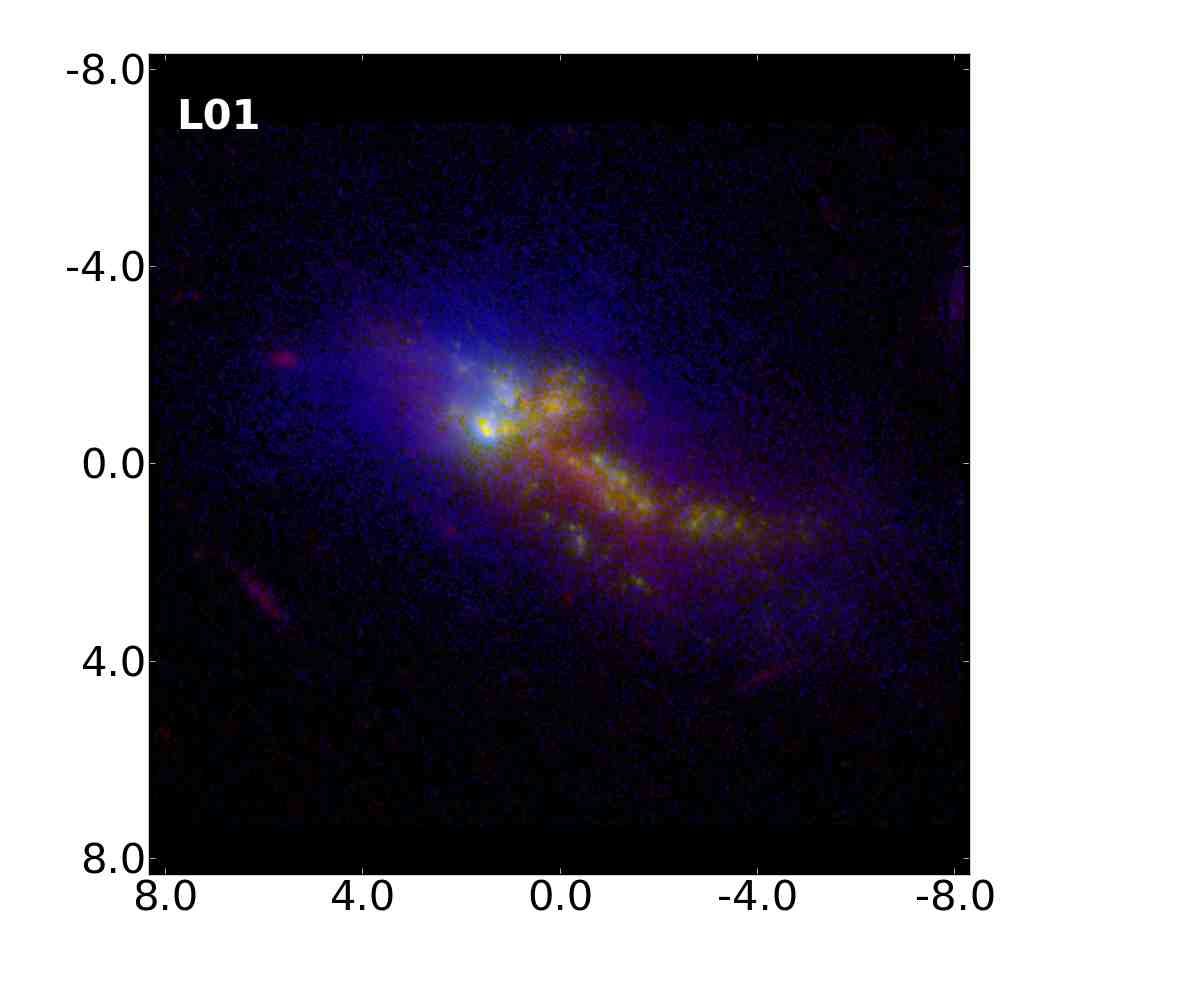}\hspace{-2.5cm}
\includegraphics[width=5.5cm]{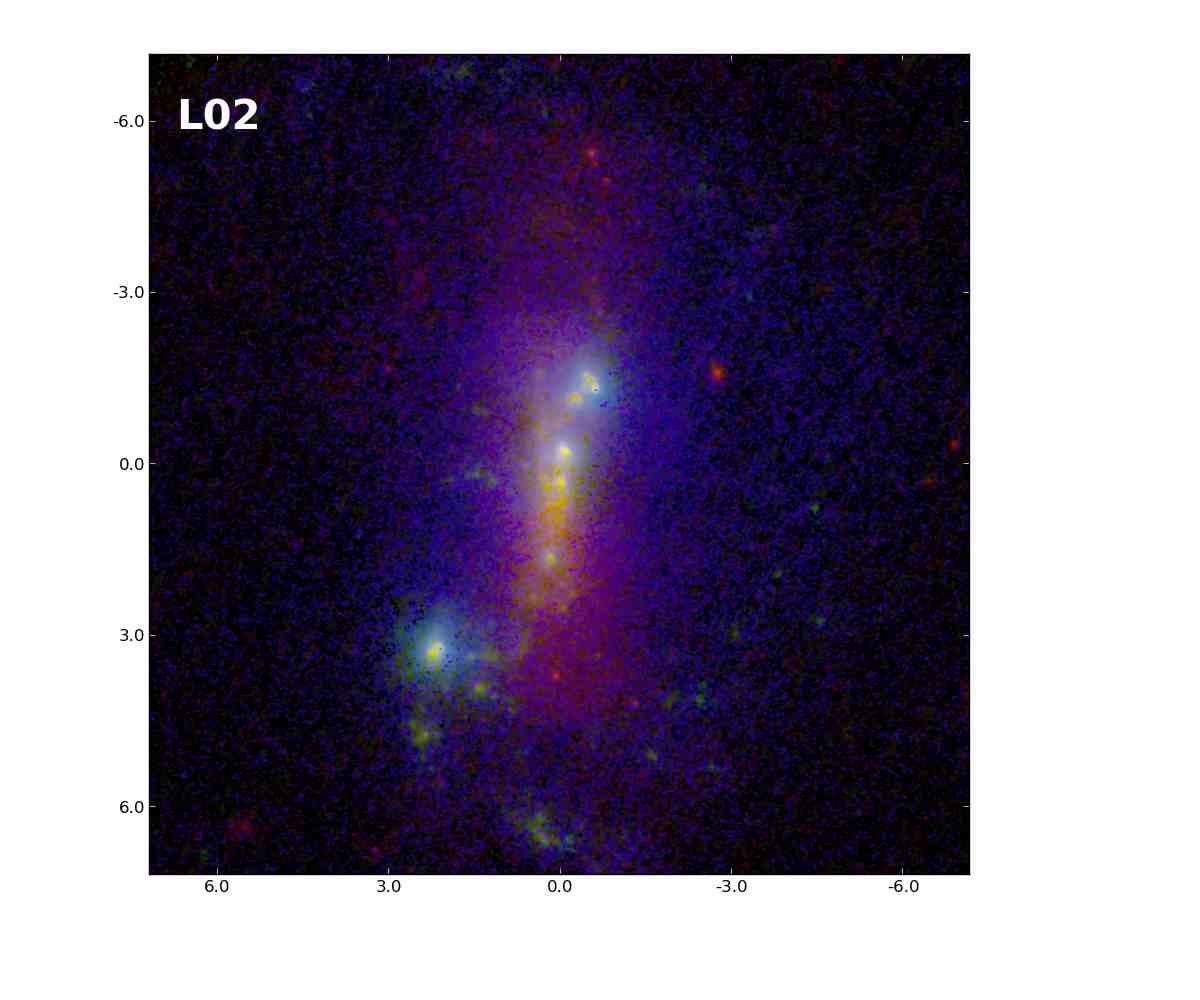}\hspace{-2.5cm}
\includegraphics[width=5.5cm]{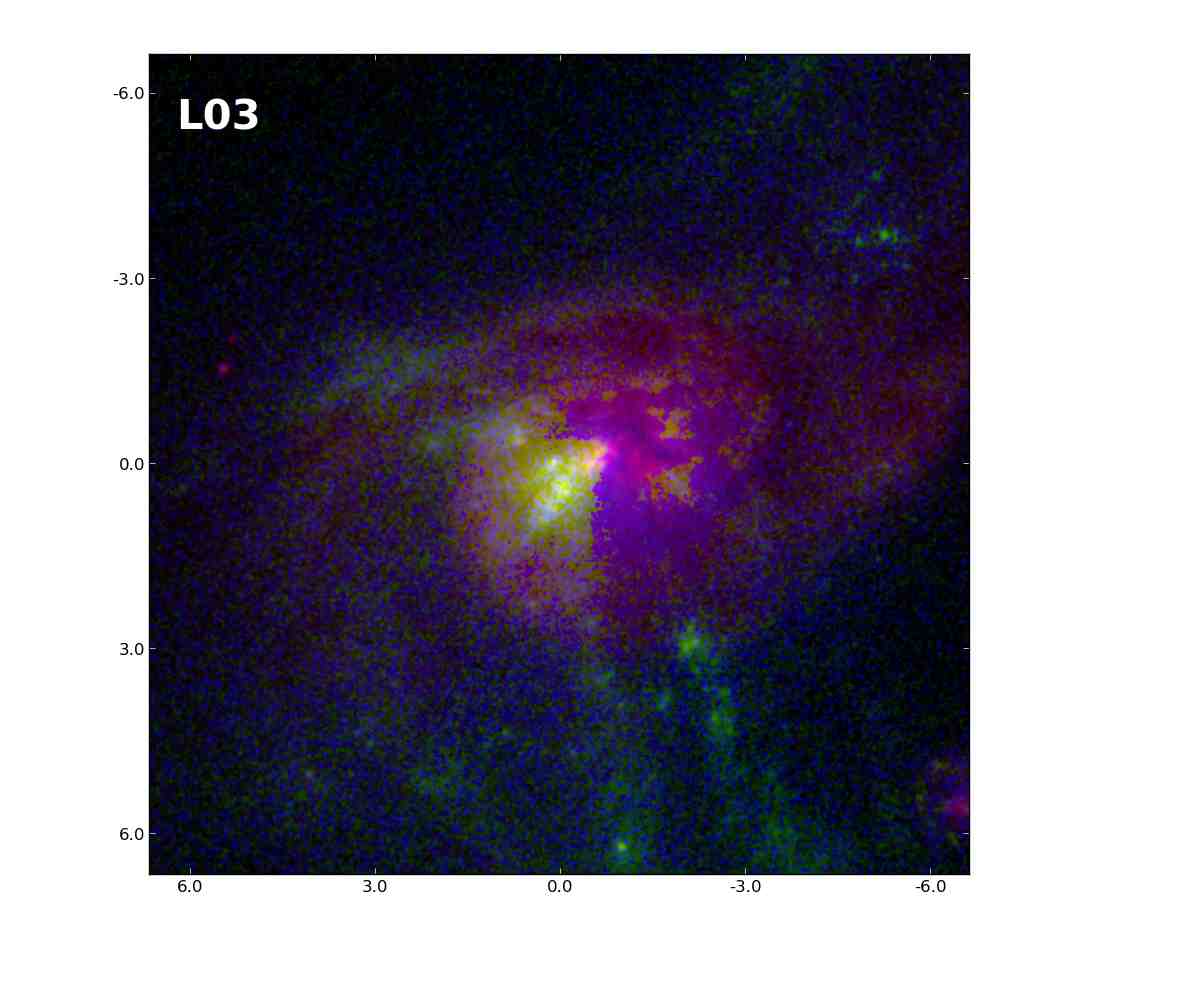}\hspace{-2.5cm}
\includegraphics[width=5.5cm]{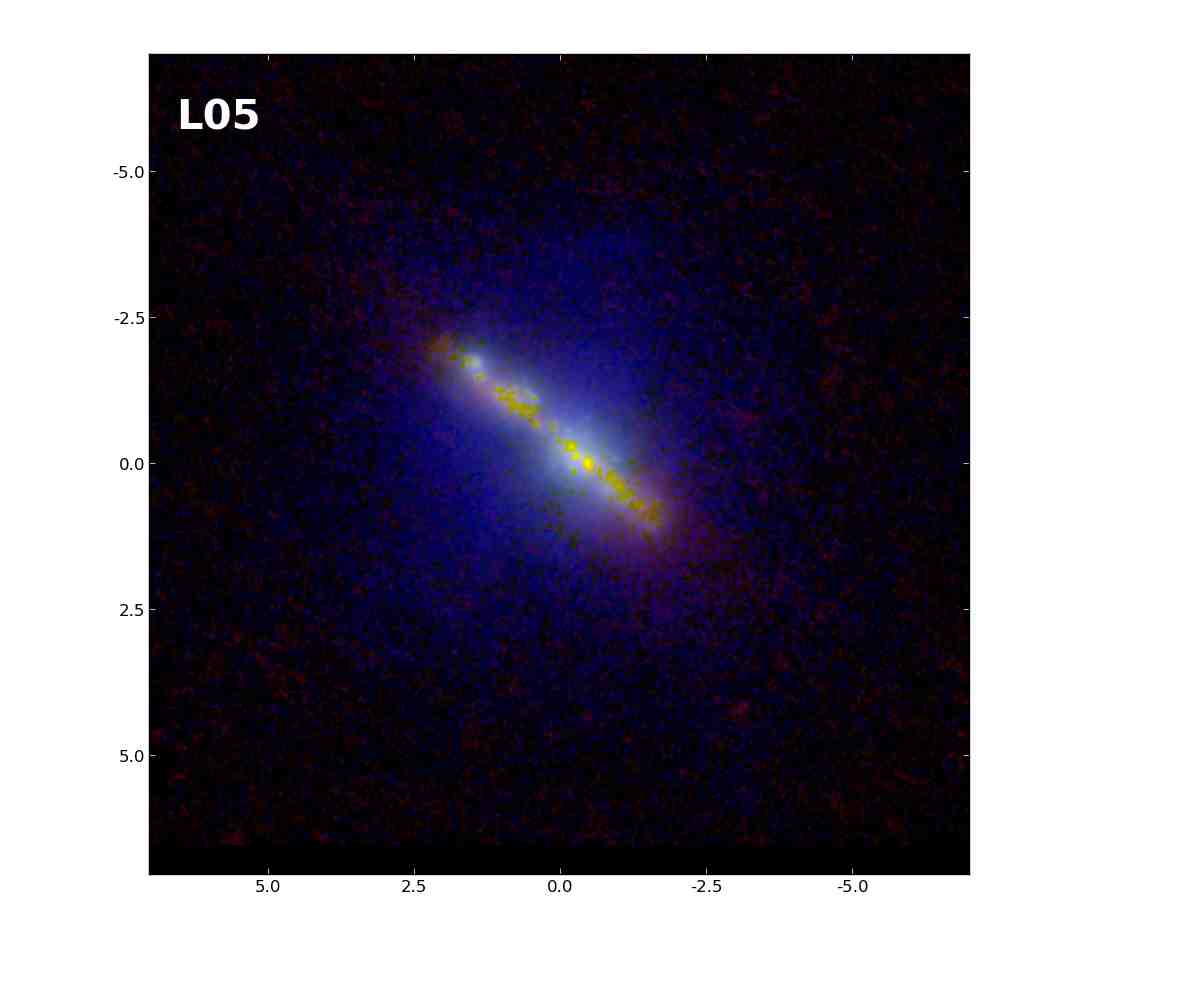}
\includegraphics[width=5.5cm]{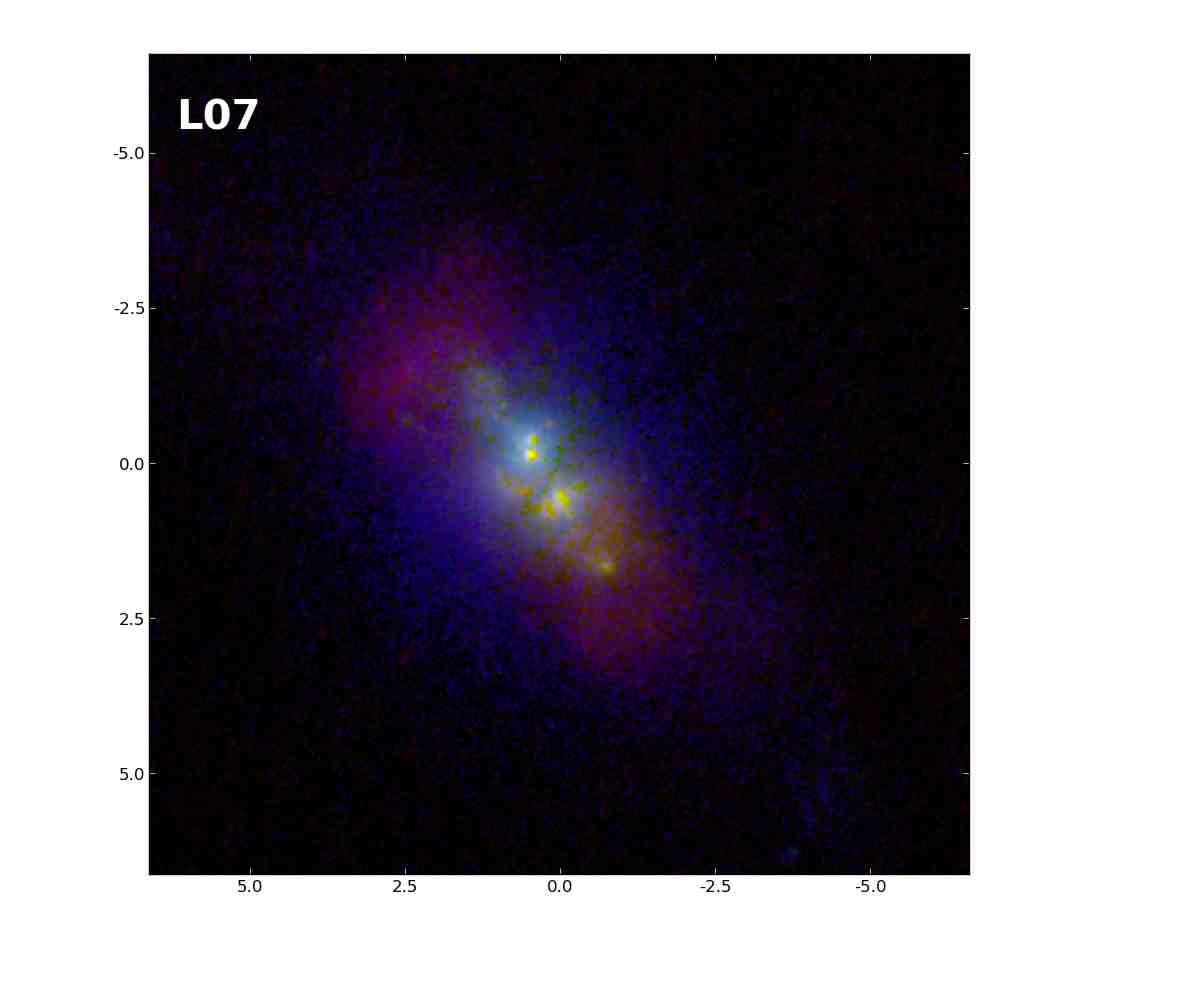}\hspace{-2.5cm}
\includegraphics[width=5.5cm]{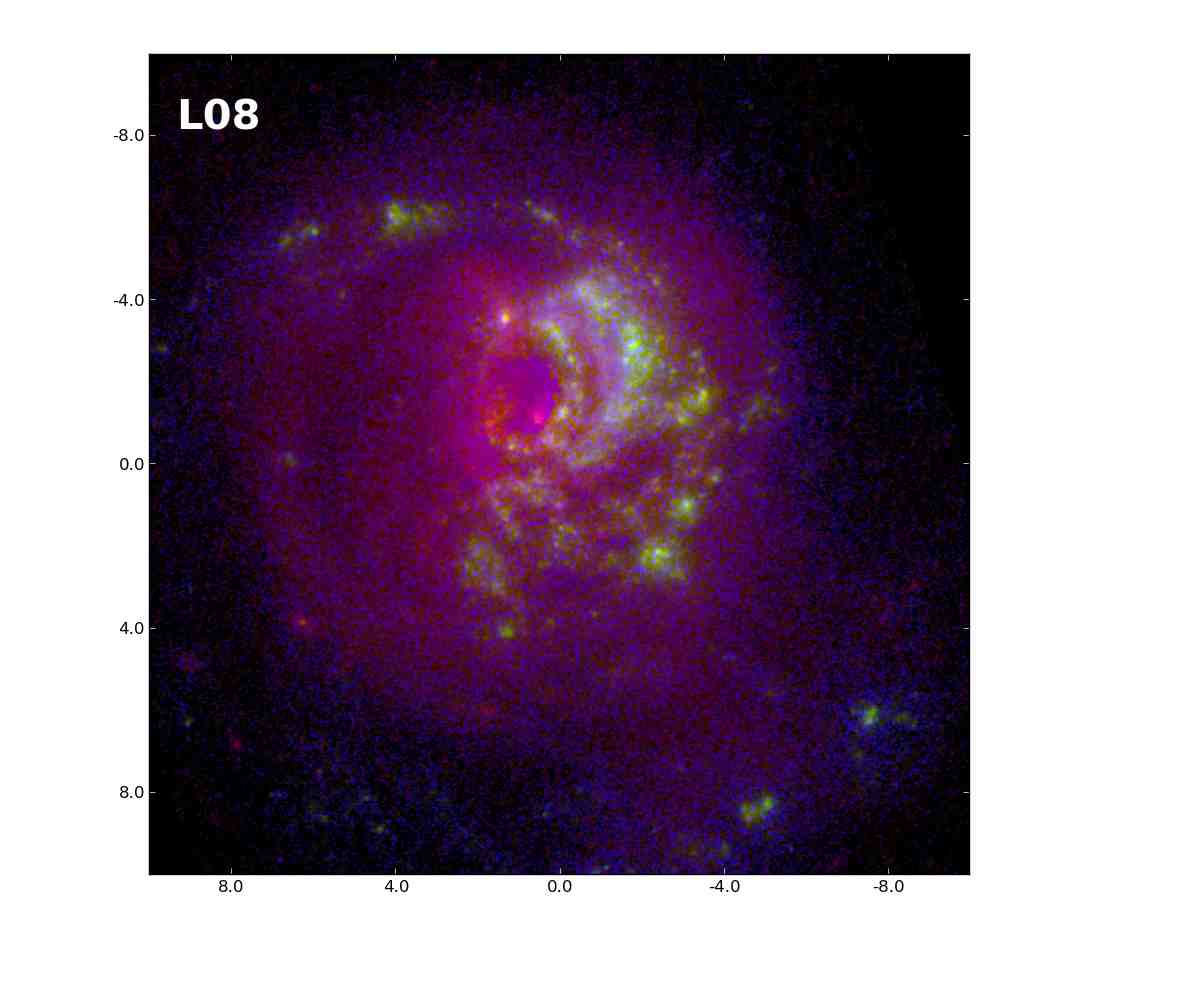}\hspace{-2.5cm}
\includegraphics[width=5.5cm]{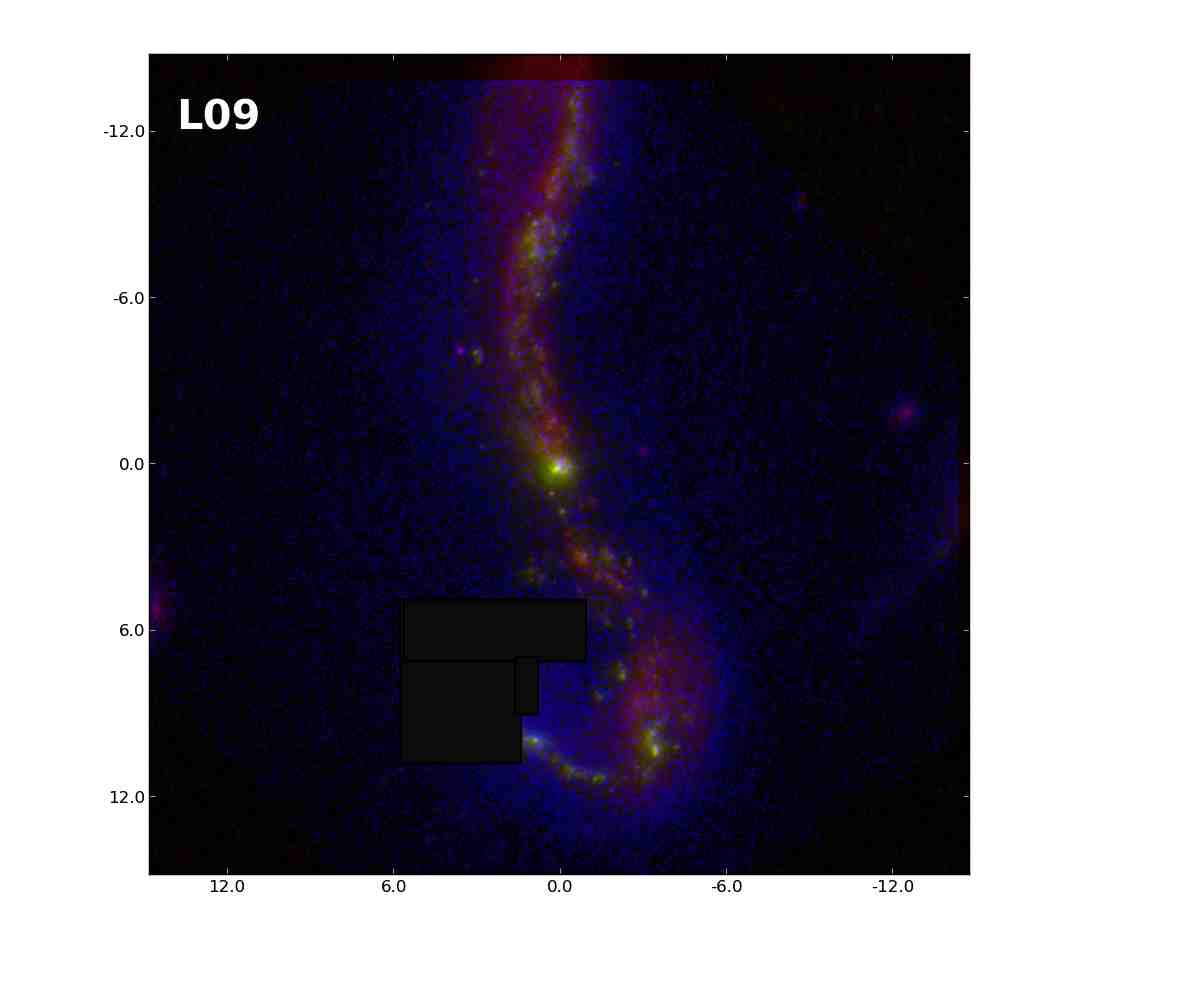}\hspace{-2.5cm}
\includegraphics[width=5.5cm]{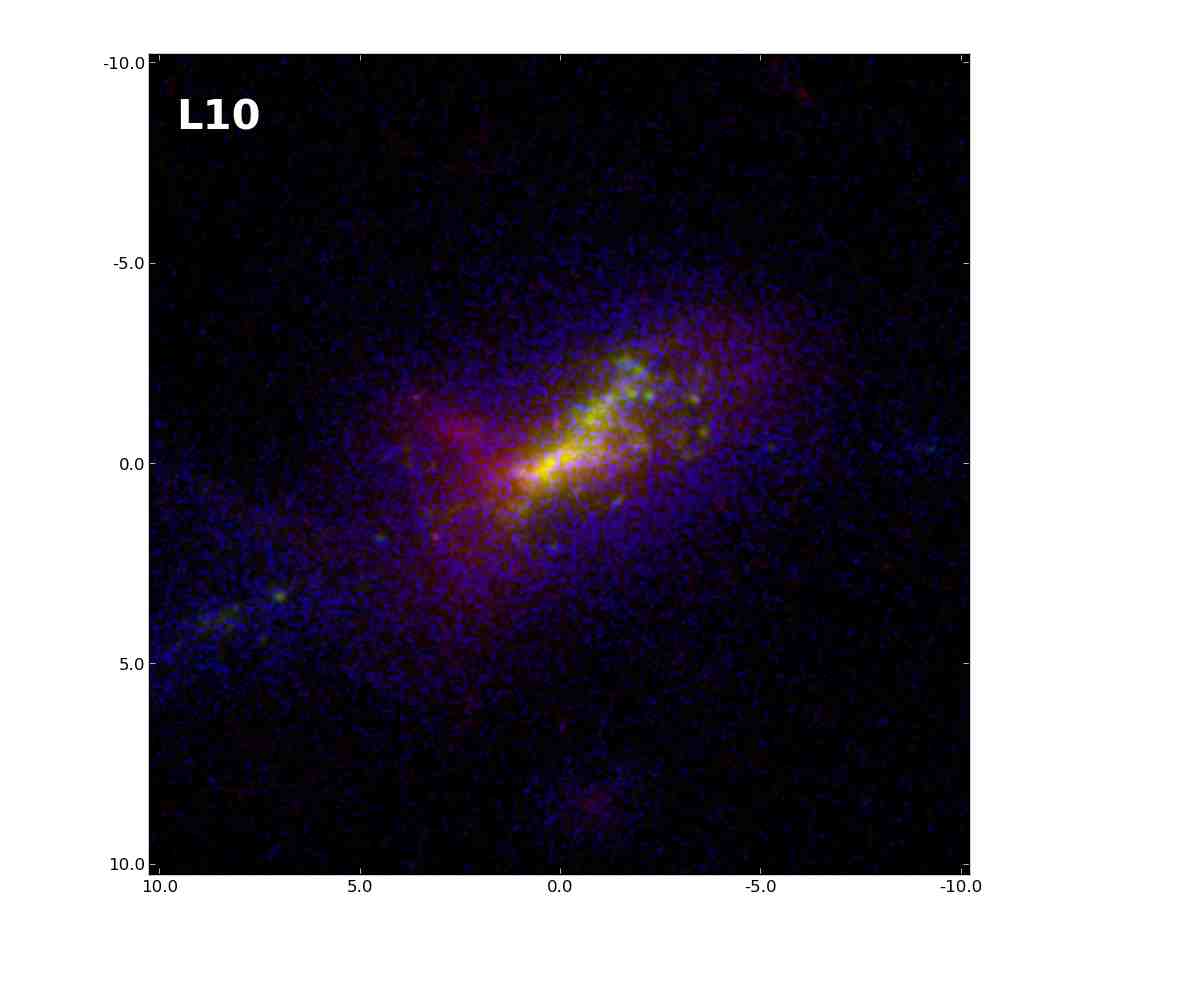} 
\includegraphics[width=5.5cm]{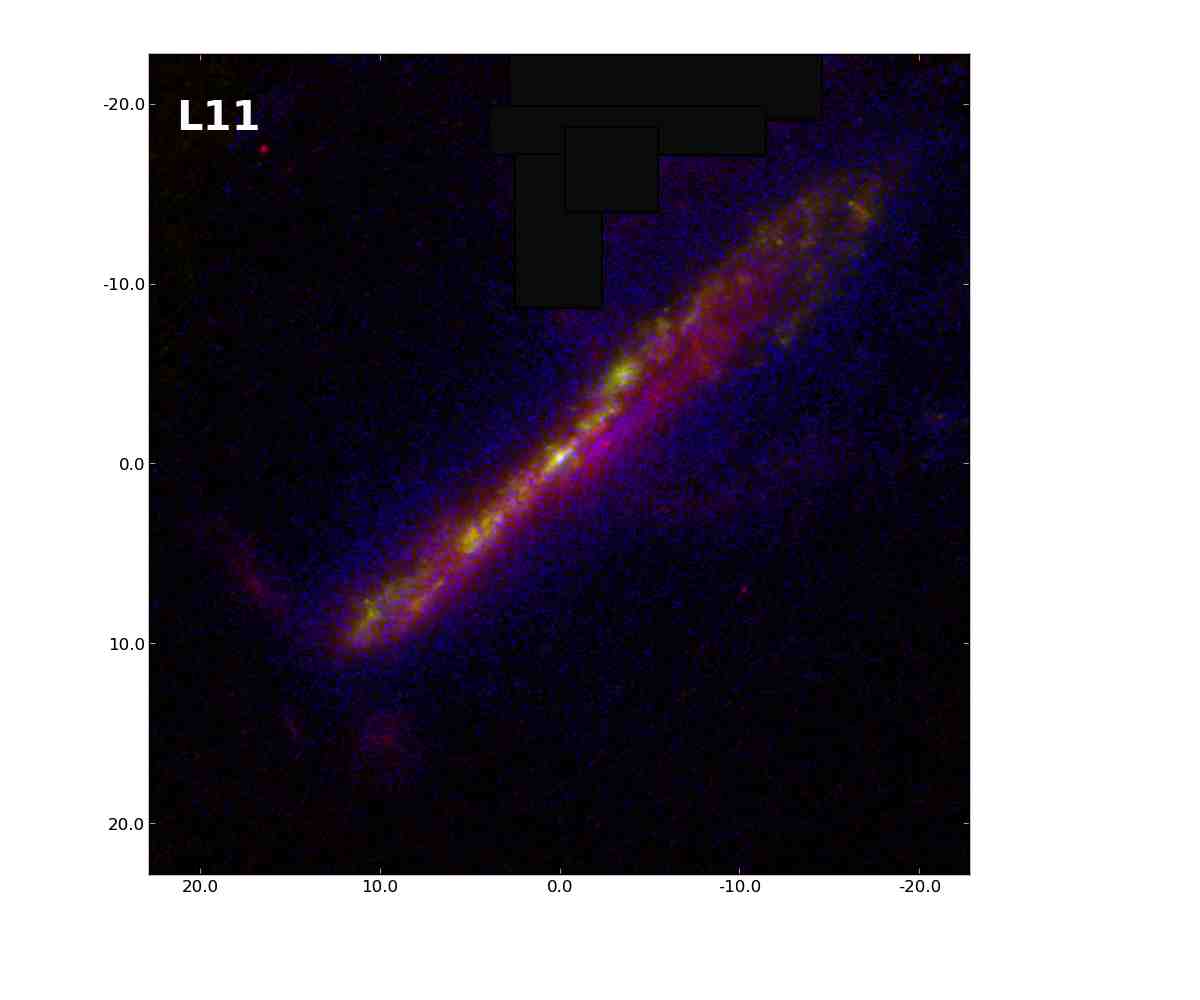}\hspace{-1.1cm}
\includegraphics[width=5.3cm]{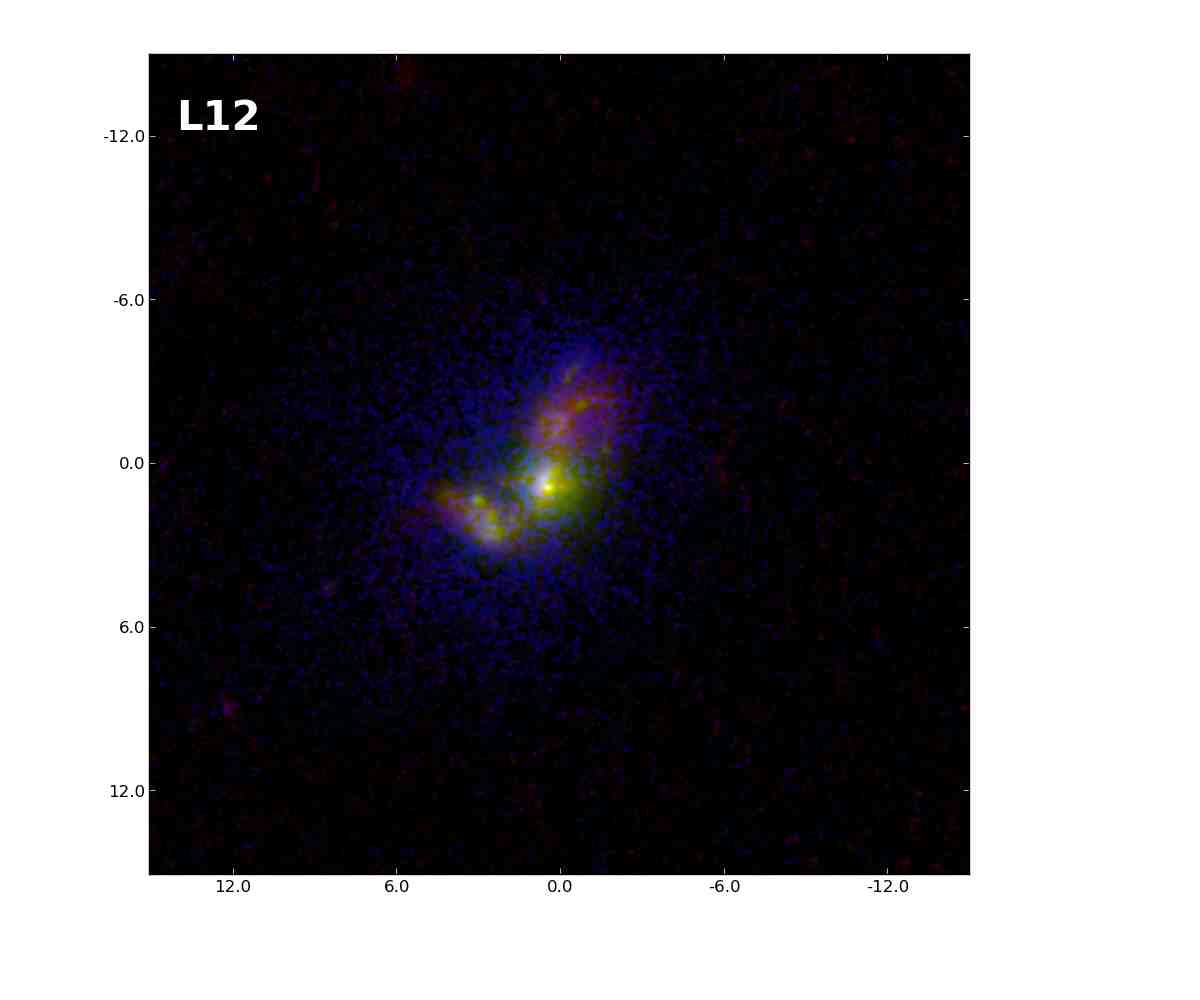}\hspace{-1.05cm}
\includegraphics[width=5.3cm]{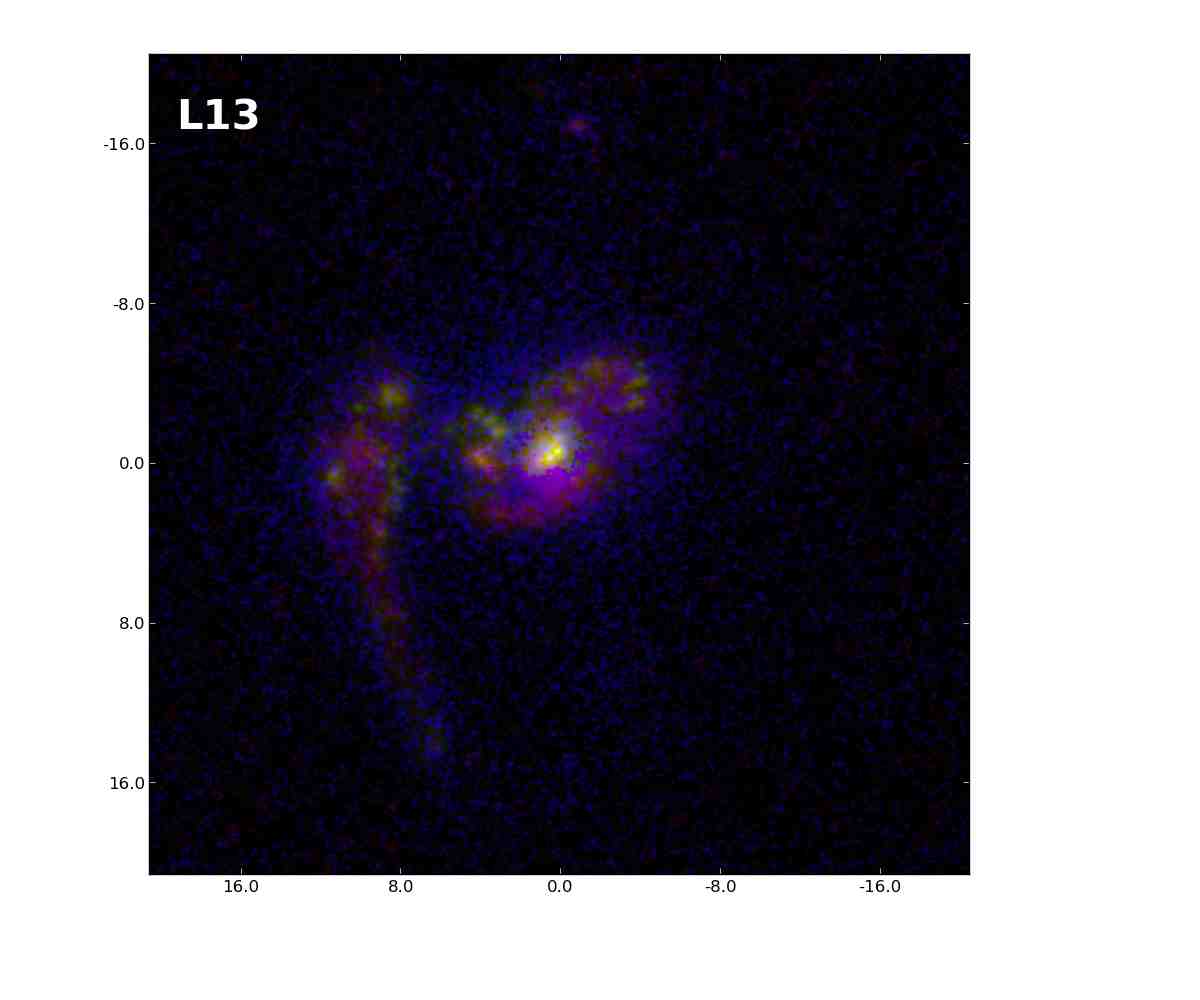}\hspace{-1.1cm}
\includegraphics[width=5.3cm]{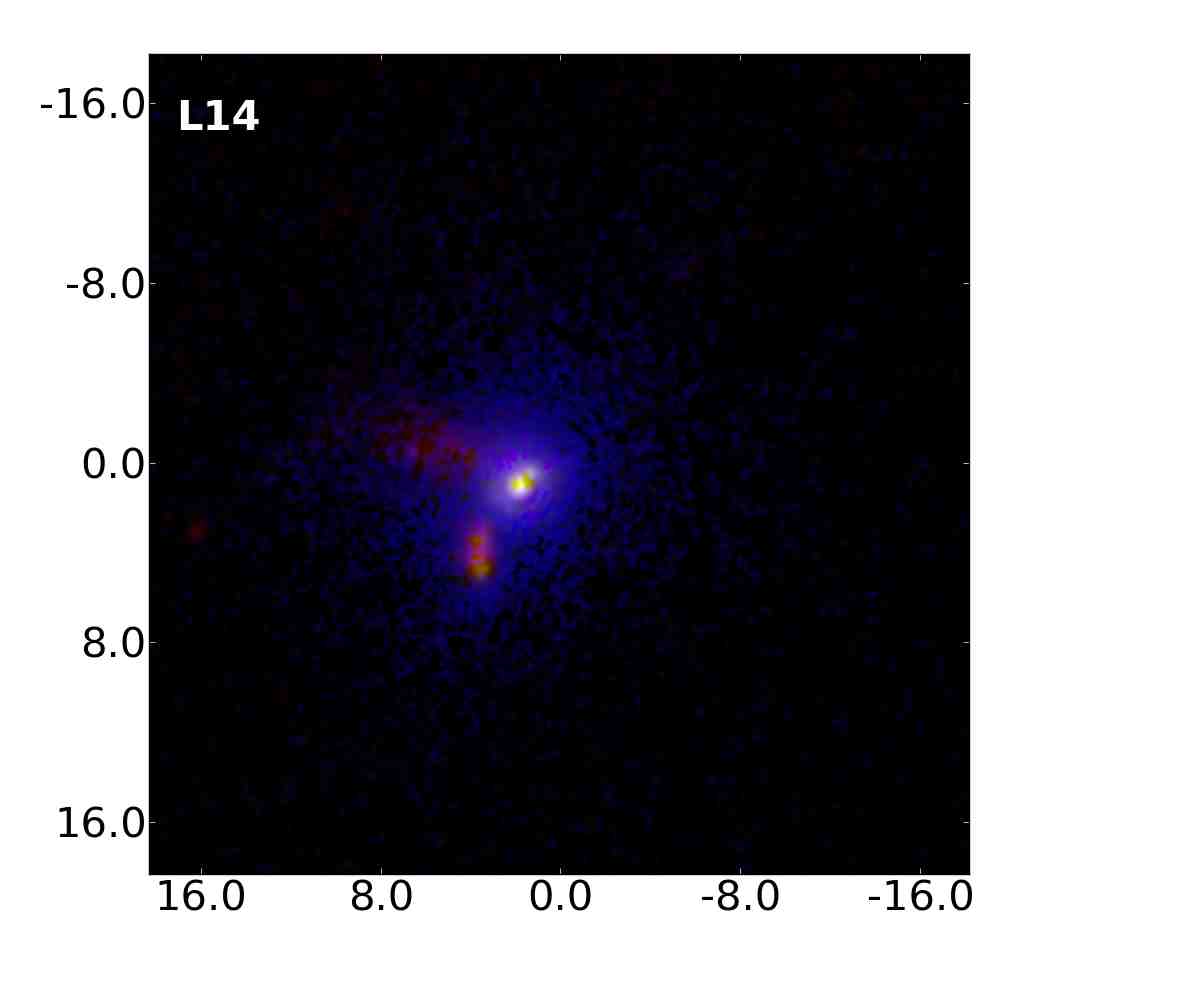}
  \caption{False-colour images of the LARS galaxies analysed in this paper. Red encodes rest-frame optical continuum, green rest-frame UV continuum, and blue shows continuum-subtracted Ly$\alpha$. Scales in kpc are given on the side. Intensity cut levels are set to show details.
}
  \label{RGB}
\end{figure*}

In Fig. \ref{RGB}\footnote{We took advantage of this codification of the \citet{Lupton2004} prescription to produce RGB images:   http://dept.astro.lsa.umich.edu/$\sim$msshin/science/code/\\Python\_fits\_image/} 
we present the RGB images of the twelve LARS galaxies. Most of the galaxies show localized knots of star formation superposed on extended rest-frame optical emission; the Ly$\alpha$ emission is extended on larger angular scales (the Ly$\alpha$ haloes).  

\begin{figure*}
\centering
\includegraphics[width=17cm]{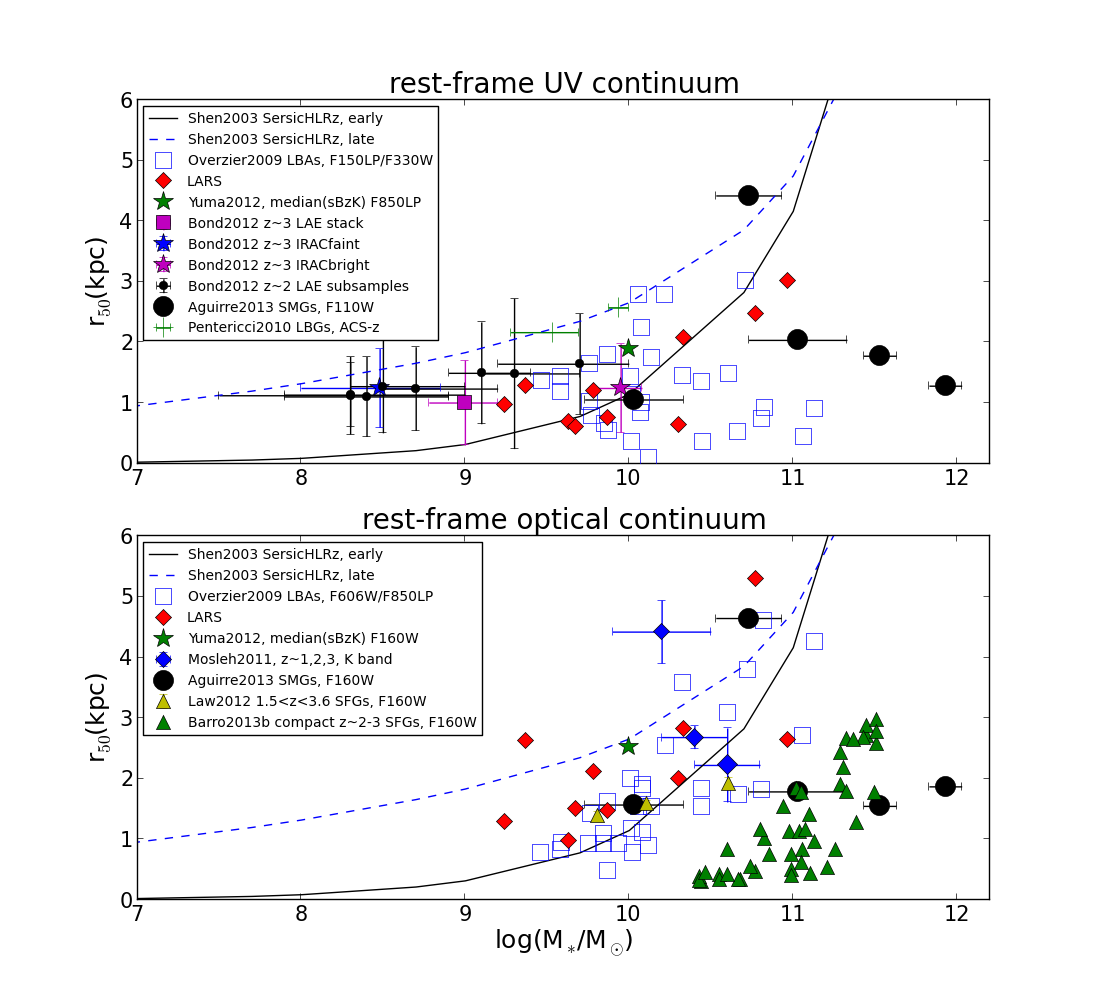} 
\caption{Half-light radius measured using elliptical apertures in the rest-frame UV ($upper$ panel) and optical ($lower$ panel) as a function of stellar mass. LARS values at $z\sim0$ from this work (red diamonds) are shown together with previous rest-frame UV measurements: \citet{Overzier2009} individual LBA values estimated in the HST ACS/SBC F150LP and ACS/HRC F330W filter (open blue squares), \citet{Yuma2012} median estimations in the F850W filter for sBzK (green star), \citet{Bond2012} LAE sample stack and subsamples (magenta square, blue and magenta stars, small black circle), \citet{Aguirre2013} individual SMG values measured in the HST F110W filter (big black dots), and \citet{Pentericci2010} average values of LBGs with and without Ly$\alpha$ in emission (green dots); and rest-frame optical measurements: \citet{Overzier2009} individual LBA values estimated in the HST (Wide Field and Planetary Camera2) WFPC2/F606W and ACS/WFC F850LP filter (open blue squares), \citet{Yuma2012} median estimations in the F160W filter for sBzK (green star), \citet{Mosleh2011} median values of UV-bright sources (GALEX-LBGs at $z\sim0.6-1.5$, LBG at $z\sim2.5-3.5$, and continuum-selected star-forming galaxies at $z\sim1.5-2.5$, blue diamonds), \citet{Aguirre2013} individual SMG values measured in F160W filter, \citet{Law2012} mean value of all the sample of star-forming galaxies at $1.5<z<3.6$ estimated in the F160W filter (yellow triangles), and \citet{Barro2013b} values for compact star-forming galaxies at $2<z<3$ also calculated in the F160W filter (green triangles). We also show the curve derived by \citet{Shen2003} for local SDSS early- and late-type galaxies. As these curves were obtained in z-bands for local galaxies, it is more meaningful to compare them to the radii in the rest-frame optical. However, for reference, we show them in the $upper$ panel as well. The stellar masses are all corrected to Salpeter-IMF values and the size measurements are all scaled to be comparable to half-light radii. LARS stellar masses were calculated in Paper II. L09 and L11 are outside the graph, due to their half-light radius larger than 6 kpc.}
  \label{type1}
\end{figure*}

\begin{figure*}
\centering
\includegraphics[width=14cm]{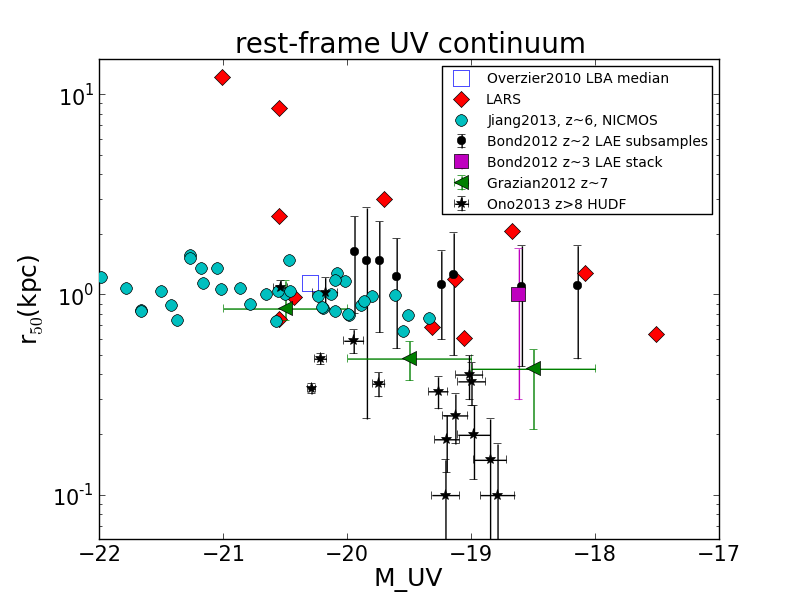}
  \caption{Half-light radius measured using elliptical apertures in the rest-frame UV image as a function of the absolute rest-frame UV magnitude. LARS measurements from this work (red diamonds) are shown together with the literature estimations by \citet{Overzier2010} corresponding to the median value of their $z<0.3$ LBAs, observed in the near-infrared bands, by \citet{Jiang2013} for a sample of $z\sim5.7, 6.5, 7.0$ LAEs (cyan dots), by \citet{Bond2012} for the stack and subsamples of LAEs at $z\simeq2-3$ (black circles, magenta squares), by \citet{Grazian2012} which measured SExtractor half-light radii for a sample of z-drop outs (green triangles), and by \citet{Ono2013} for a sample of high signal-to-noise z- and Y-drop outs detected in Hubble Ultra Deep Field, UDF12 (black stars).}
  \label{type2}
\end{figure*}

We investigated the properties of LARS galaxies in the context of other galaxy populations, to assess the fairness of our comparison. In Figs. \ref{type1} and \ref{type2} we show the location of the LARS galaxies in the half-light radius vs stellar mass (r$_{50}$ vs log(M$_*$/M$_{\odot}$)) and the half-light radius vs UV absolute magnitude (r$_{50}$ vs M$_{UV}$) diagrams, to understand if the LARS galaxies harbour stellar masses and UV magnitudes comparable to values in the literature. These diagrams have been designed for local galaxies, for which sizes could be easily measured in the rest-frame optical bands \citep{Shen2003}. However, measurements in the rest-frame UV could also be performed at high redshift. Following the method described in Sec. \ref{sec:method}, we estimated the half-light radius as r$_{50}$ in the rest-frame optical and also in the rest-frame UV images. 

The high-$z$ studies we adopted for comparison all performed size and morphological measurements by using HST images. These include,

\begin{itemize}
\item Continuum-selected Lyman break galaxies (LBGs) at $z\sim3$, with and without Ly$\alpha$ in emission \citep{Pentericci2010}; at $z\sim1$, 2, and 3 \citep{Mosleh2011}, at $1.5<z<3.6$ \citep{Law2012}; at $z\sim1.8$ \citep{Lotz2004}; z-drop outs at $z\sim7$ \citep{Grazian2012}; high signal-to-noise z- and Y-drop outs detected in the Hubble Ultra Deep Field, UDF12 \citep{Ono2013}

\item  Compact star-forming galaxies (cSFGs) at $2<z<3$ \citep{Barro2013b}. These authors have pointed out that, based on their number densities, masses, sizes, and star formation rates, $z\sim2-3$ compact, star-forming galaxies were likely progenitors of compact, quiescent, massive galaxies at $z<2$

\item  Star-forming galaxies selected based on their $B-z$ and $z-K$ colour \citep[sBzK,][]{Yuma2012,Lee2013}; passive and star-forming galaxies selected based on their $B-z$ and $z-K$ colour \citep[pBzK and sBzK][]{Lee2013}

\item  Star-forming galaxies at $z\sim2-3$ by \citet{Law2012}. These authors found a typical value of the Gini coefficient (G$^{SB-rP20S}=0.4$) for the sources with the strongest Ly$\alpha$ emission, characterized by M$_{*} \sim1.5 \times 10^{10}$ M$\odot$

\item  GOODS (Great Observatories Origins Deep Survey) and UDF (Ultra Deep Survey) $z\sim4$ and GOODS $z\sim1.5$ sources from the study of \citet{Lotz2006}

\item  Sub-millimeter galaxies \citep[SMGs,][]{Aguirre2013}

\item  Narrow-band selected Lyman alpha emitters at $z\simeq2.07$ and $z\simeq3.1$ \citep{Bond2009,Bond2012} belonging to the MUSYC survey. We considered the stack of the $z\sim2.07$ entire sample and of subsamples separated by photometric properties, UV-faint(UV-bright) with $R>25.5(<25.5)$, IRAC-faint(IRAC-bright) with f$_{3.6 \mu m}<0.57 (>0.57) ~\mu$J, low-(high-)EW with EW(Ly$\alpha)<66(>66)$ {\AA}, red-(blue-)LAE with $B-R>0.5(<0.5)$ \citep{Guaita2011}.

\item  Narrow-band selected Lyman alpha emitters at $z\sim5.7, 6.5$, and 7.0 \citep{Jiang2013}, the first very-high-redshift sample where non-parametric morphological measurements were performed.
\end{itemize}

The local-Universe studies, we adopted for comparison, include,

\begin{itemize}
\item Sloan Digital Sky Survey (SDSS) early- and late-type galaxy relations obtained from the analysis of images in the z band \citep{Shen2003}

\item Lyman break analogues (LBAs) at $z\simeq0.2$. These are local starbursts that share typical characteristics of high-$z$ LBGs, such as stellar mass, metallicity, dust extinction, star-formation rate, and physical size. We considered a sample of 30 LBAs from \citet{Overzier2009, Overzier2010}. They were characterized by a median absolute UV magnitude of -20.3, almost one magnitude fainter than typical LBGs. 
\end{itemize}

We found that the LARS galaxies occupy a quite wide range of r$_{50}$. 
Their rest-frame UV and optical sizes (Table \ref{tab:Loriginalsize}) are broadly consistent with LBAs, LBGs, and SMGs. Their stellar mass tend to be larger than LAEs, consistent with LBAs and LBGs. However, there is an overlap in stellar mass between LARS-LAEs (M$_{*}<10^{10}$ M$\odot$) and the most massive LAEs in the sample of \citet{Bond2012}. Also, LARS galaxies are less massive than cSFGs. 
The largest half-light radii characterize the LARS galaxies with the most distorted morphology (see also Fig. \ref{stampsDEEP2}).
LARS M$_{UV}$ magnitudes (and so star-formation rate, SFR$_{UV}$) are comparable with those of $z\simeq2.07$ LAEs and $z\geq7$ LBGs. There is an overlap with $z>5$ LAEs. However, the measurements of LARS sizes in the rest-frame UV are larger than those of $z\geq7$ LBGs.

\begin{table*}
\caption{Size of the original LARS galaxies at $z\sim0$}
%\tablewidth{0pt}
\label{tab:Loriginalsize}
\centering
\scalebox{0.9}{
\begin{tabular}{|c|c|c|c|c|c|c|}
\hline\hline 
(1) & (2) & (3) & (4) & (5) &(6) & (7)\\ %& (9) &(10) & (11) & (12) & (13) & (14) & (15) &(16) & (17)& (18)\\
\hline
LARS &rP20$^{ell}$ & rP20$^{circ}$ & rP20$^{minA}$ & r$_{20}$ & r$_{50}$ & r$_{80}$ \\
\hline
& kpc & kpc & kpc & kpc & kpc &kpc \\
\hline
rest-UV &  &  &  &  &  & \\
\hline
%L01 & 2.486 & 1.165 & 0.784 & 0.728 & 1.075 & 1.971 \\
%L02 & 2.442 & 2.349 & 1.744 & 0.779 & 1.500 & 2.709 \\
%L03 & 1.116 & 0.893 & 0.769 & 0.273 & 0.508 & 0.955 \\
%L05 & 1.748 & 1.436 & 0.840 & 0.366 & 0.705 & 1.504 \\
%L07 &1.583 & 0.934 & 0.814 & 0.362 & 0.739 & 1.326 \\
%L08 & 4.024 & 3.135 & 3.165 & 1.161 & 1.500 & 3.888 \\
%L09 & 0.664 & 0.590 & 0.443 & 0.111 & 0.258 & 0.516 \\
%L10 & 5.592 & 3.669 & 3.094 & 0.973 & 2.453 & 4.620 \\
%L11 & 24.144 & 18.810 & 19.757 & 6.312 & 12.498 & 19.726 \\
%L12 & 1.201 & 1.126 & 0.751 & 0.338 & 0.563 & 1.089 \\
%L13 & 3.342 & 2.982 & 1.028 & 1.388 & 1.500 & 2.982 \\
%L14 & 1.341 & 1.219 & 0.732 & 0.488 & 0.732 & 1.219 \\
L01 & 2.63 & 1.34 & 0.81 & 0.86 & 1.20 & 2.11 \\
L02 & 2.42 & 2.28 & 1.49 & 0.69 & 1.28 & 2.17 \\
L03 & 1.40 & 0.97 & 0.77 & 0.33 & 0.63 & 1.17 \\
L05 & 1.71 & 1.46 & 0.84 & 0.35 & 0.69 & 1.45 \\
L07 & 1.13 & 0.90 & 0.81 & 0.33 & 0.60 & 0.95 \\
L08 & 5.98 & 3.35 & 2.08 & 1.67 & 3.01 & 4.96 \\
L09 & 18.72 & 1.11 & 0.59 & 2.73 & 8.52 & 13.81 \\
L10 & 4.73 & 3.71 & 3.09 & 0.77 & 2.08 & 3.93 \\
L11 & 23.77 & 18.56 & 18.56 & 6.12 & 12.21 & 19.41 \\
L12 & 1.84 & 1.20 & 1.05 & 0.41 & 0.75 & 1.76 \\
L13 & 4.42 & 3.70 & 1.23 & 1.95 & 2.47 & 3.75 \\
L14 & 1.83 & 1.71 & 0.98 & 0.67 & 0.98 & 1.52 \\
\hline
\hline
rest-optical &  &  &  &  &  &   \\
\hline
L01 & 4.87 & 2.15 & 2.44 & 1.28 & 2.11 & 4.01 \\
L02 & 7.12 & 2.23 & 2.51 & 1.08 & 2.63 & 5.62 \\
L03 & 5.73 & 1.98 & 1.83 & 0.72 & 2.00 & 4.59 \\
L05 & 2.32 & 1.79 & 1.84 & 0.41 & 0.98 & 1.79 \\
L07 & 3.95 & 2.89 & 1.96 & 0.62 & 1.51 & 3.18 \\
L08 & 5.28 & 3.17 & 3.29 & 1.48 & 2.63 & 4.22 \\
L09 & 20.54 & 3.06 & 2.07 & 5.98 & 10.27 & 14.74 \\
L10 &  7.14 & 3.93 & 3.93 & 1.08 & 2.83 & 5.57 \\
L11 &  19.13 & 14.52 & 14.52 & 4.83 & 9.82 & 15.43 \\
L12 &  3.68 & 3.15 & 1.20 & 0.64 & 1.46 & 2.97 \\
L13 &  11.41 & 5.04 & 5.35 & 2.31 & 5.30 & 9.15 \\
L14 &  3.72 & 3.78 & 1.22 & 0.73 & 1.28 & 2.80 \\
\hline
\end{tabular}
}
\tablefoot{Size measurements from elliptical and circular aperture photometry of the original LARS images. (1) LARS id, (2) Petrosian semi-major axis, (3) circular Petrosian radius, (4) Petrosian radius at minimum of asymmetry, (5) radius containing 20\%, (6) 50\%, and (7) 80\% of the total flux. The measurements were performed in the bands corresponding to the rest-frame UV (either F140 or F150) and to the rest-frame optical (either F775, F814, or F850). The step in semi-major axis is 1 pixel ($\sim$0.02 kpc at $z\sim0.03$) and in circular radius is 2 pixels ($\sim$0.05 kpc at $z\sim0.03$).}
\end{table*}

\begin{table*}
\caption{Morphological parameters of the original LARS galaxies at $z\sim0$}
%\tablewidth{0pt}
\label{tab:Loriginal}
\centering
\scalebox{0.9}{
\begin{tabular}{|c|c|c|c|c|c|c|c|c|c|c|}
\hline\hline 
(1) & (2) & (3) & (4) & (5) &(6) & (7)& (8)& (9) &(10) & (11) \\
\hline

LARS & G$^{rP20}$ & G$^{SB-rP20S}$ & M20 & ell & C$^{circ}$ & C$^{ell}$ & C$^{minA}$ & SN$_{pixel}$ & A & S \\
\hline
rest-UV & &  &  &  & & & &  & & \\
\hline
L01 & 0.70 & 0.68 & -0.87 & 0.61 & 1.34 & 1.94 & 1.73 & 251.40 & 0.42 & 0.16 \\
L02 & 0.73 & 0.72 & -0.79 & 0.53 & 3.14 & 2.50 & 3.49 & 52.26 & 0.50 & 0.26 \\
L03 & 0.63 & 0.64 & -1.17 & 0.30 & 2.20 & 2.71 & 2.76 & 83.18 & 0.21 & 0.11 \\
L05 & 0.72 & 0.68 & -1.26 & 0.34 & 3.11 & 3.07 & 3.27 & 408.69 & 0.33 & 0.15 \\
L07 & 0.62 & 0.62 & -1.02 & 0.40 & 1.99 & 2.28 & 1.24 & 696.00 & 0.26 & 0.09 \\
L08 & 0.62 & 0.65 & -0.89 & 0.33 & 1.95 & 2.36 & 1.43 & 21.98 & 0.37 & 0.29 \\
L09 & 0.71 & 0.75 & -2.33 & 0.80 & 1.76 & 3.52 & 3.35 & 23.86 & 0.21 & 0.24 \\
L10 & 0.67 & 0.68 & -1.52 & 0.52 & 3.53 & 3.53 & 2.76 & 32.83 & 0.33 & 0.24 \\
L11 & 0.58 & 0.79 & -1.22 & 0.87 & 3.40 & 2.51 & 3.40 & 44.05 & 0.30 & 0.31 \\
L12 & 0.67 & 0.68 & -1.50 & 0.34 & 2.56 & 3.15 & 2.82 & 583.81 & 0.28 & 0.04 \\
L13 & 0.72 & 0.67 & -0.77 & 0.21 & 1.23 & 1.42 & 2.61 & 89.17 & 0.33 & 0.23 \\
L14 & 0.74 & 0.68 & -0.71 & 0.27 & 1.51 & 1.78 & 2.72 & 622.68 & 0.21 & 0.02 \\
\hline
\hline
rest-optical & &  &  &  & & & &  & & \\
\hline
L01 & 0.59 & 0.61 & -1.09 & 0.64 & 1.72 & 2.48 & 2.53 & 21.91 & 0.29 & 0.09 \\
L02 & 0.59 & 0.64 & -1.36 & 0.69 & 2.96 & 3.58 & 2.89 & 4.25 & 0.14 & 0.31 \\
L03 & 0.61 & 0.62 & -2.16 & 0.45 & 3.32 & 4.02 & 3.42 & 50.72 & 0.09 & 0.05 \\
L05 & 0.64 & 0.67 & -1.57 & 0.73 & 3.27 & 3.22 & 4.77 & 64.96 & 0.33 & 0.15 \\
L07 & 0.60 & 0.61 & -1.68 & 0.60 & 3.28 & 3.56 & 2.97 & 14.60 & 0.24 & 0.10 \\
L08 & 0.54 & 0.55 & -1.05 & 0.23 & 2.10 & 2.28 & 1.78 & 36.82 & 0.31 & 0.16 \\
L09 & 0.62 & 0.74 & -1.13 & 0.83 & 2.03 & 1.96 & 4.35 & 11.38 & 0.45 & 0.34 \\
L10 & 0.57 & 0.58 & -1.87 & 0.54 & 3.08 & 3.56 & 3.31 & 9.95 & 0.13 & 0.15 \\
L11 & 0.50 & 0.60 & -1.25 & 0.85 & 3.46 & 2.52 & 3.46 & 18.70 & 0.20 & 0.23 \\
L12 & 0.64 & 0.63 & -1.90 & 0.46 & 3.70 & 3.34 & 3.35 & 54.74 & 0.33 & 0.13 \\
L13 & 0.58 & 0.62 & -1.52 & 0.55 & 2.43 & 2.99 & 3.67 & 3.98 & 0.18 & 0.30 \\
L14 & 0.76 & 0.72 & -1.32 & 0.10 & 2.92 & 2.92 & 3.01 & 35.24 & 0.32 & 0.13 \\
\hline

\end{tabular}
}
\tablefoot{Morphological parameters estimated for the original LARS galaxies, following the equations in Appendix A. (1) LARS id, (2) Gini coefficient estimated within the fixed-size segmentation map, (3) Gini coefficient estimated within the segmentation map built from the pixels with surface brightness larger than the value corresponding to that at the Petrosian radius, (4) M20, (5) SExtractor ellipticity 1-B/A, (6) concentration from circular apertures, (7) concentration from elliptical apertures, (8) concentration corresponding to the minimum of asymmetry, (9) signal-to-noise per pixel, (10) asymmetry, (11) clumpiness. As discussed in Appendix A, we could expect an uncertainty of $\sim$ 10\% in G, of $<5$\% in M20, and of $\sim$ 20\% in C. We could expect an even larger uncertainty in S. The difference in G$^{rP20}$ and G$^{SB-rP20S}$ is generally marginal.}
\end{table*}

Therefore, LARS galaxies could be considered as LBAs, with size, stellar mass, and star-formation rate similar to $2<z<3$ star-forming galaxies.

\subsection{Continuum morphology of LARS galaxies at $z\sim0$}
\label{sec:LARSmorphology}

Following the method described in Sec. \ref{sec:method}, we estimated the non-parametric measurements for the LARS galaxies (Table \ref{tab:Loriginal}). 

Combinations of morphological parameters (see Sec. \ref{sec:useofparam}) can give information about a galaxy's star-formation history. 
\citet{Lotz2004} proposed a criterion for separating perturbed disks or merging systems from normal galaxies, by studying local ultra luminous infrared galaxies (ULIRGs). The criterion identifies a region in the G vs M20 diagram, which is G$^{SB-rP20S}>-0.115 ~\times$ M20 $+ ~0.384$. 
For $z<1.2$ galaxies observed in a rest-frame optical band (4000 {\AA}) at HST resolution, \citet{Lotz2008a} proposed a slightly different relation to identify merging systems, G$^{SB-rP20S}>-0.14 ~\times$ M20 $+ ~0.33$. 
Also, \citet{Conselice2003} distinguished the region where irregular or starburst galaxies were located in the A-C$^{minA}$ and A-S planes.
\begin{figure*}
\centering
\includegraphics[width=20cm]{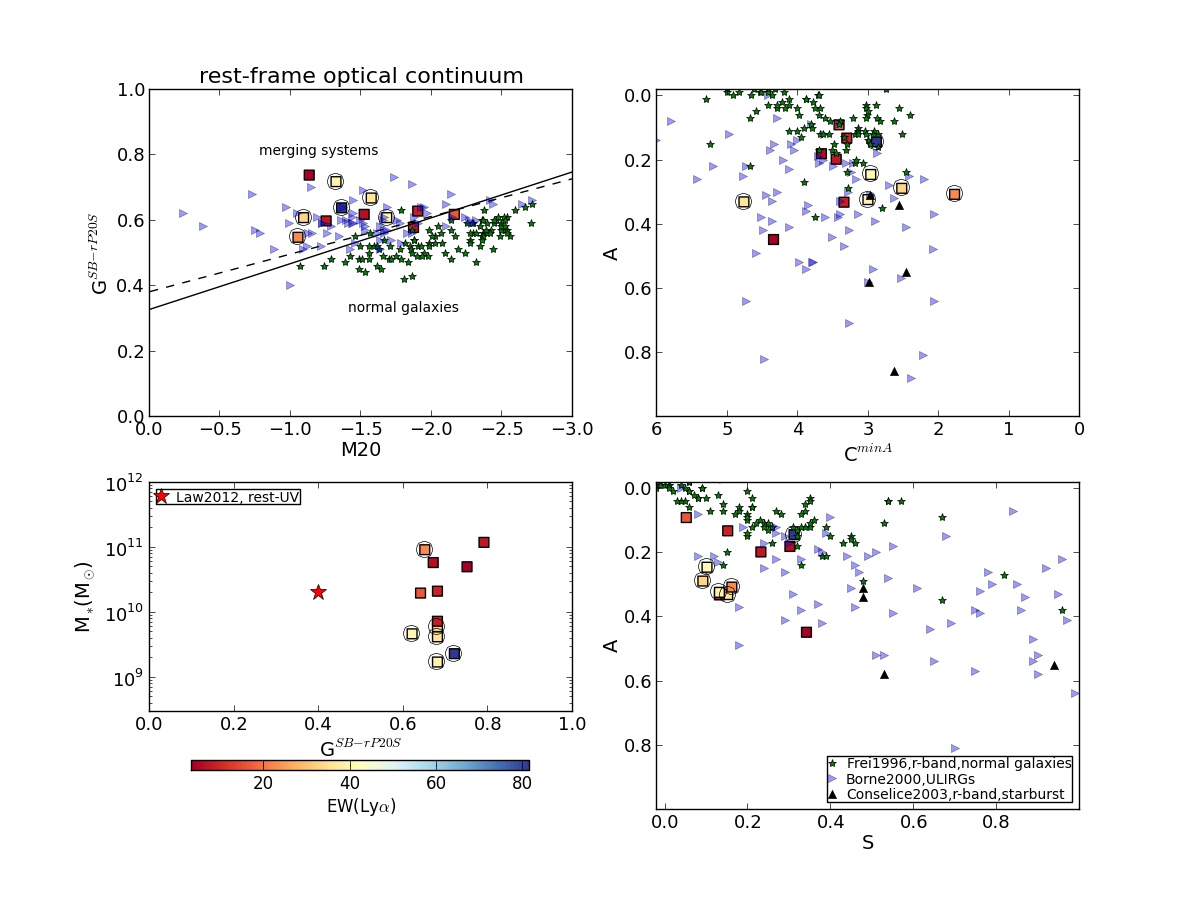}
  \caption{Combinations of rest-frame optical morphological measurements, used in the literature as diagnostics of galaxy past and current history. G$^{SB-rP20S}$ vs M20 ($upper ~left$),  A vs C$^{minA}$ ($upper ~right$), and A vs S ($lower ~right$). The $lower~left$ panel shows stellar mass vs the rest-frame UV G$^{SB-rP20S}$. The typical value of G$^{SB-rP20S}=0.4$ for the strongest Ly$\alpha$ emitters of the sample by \citet{Law2012} is reported as a red star.
The twelve LARS galaxies analysed here are presented as squares, LARS-LAEs are rounded by open circles. The colour scale corresponds to EW(Ly$\alpha$). 
For comparison, green stars correspond to the \citet{Frei1996} sample of normal galaxies and light blue triangles correspond to the ULIRG sample of \citet{Borne2000} as processed by \citet{Lotz2004}. Black triangles correspond to a sample of starburst galaxies presented in \citet{Conselice2003}. Dashed and solid lines correspond to the separation between ULIRGs and normal galaxies, proposed by \citet{Lotz2004} and \citet{Lotz2008a} respectively (see text).
}
  \label{LoriginalCASGM20_coHa}
\end{figure*}

As seen in Fig. \ref{LoriginalCASGM20_coHa} and Table \ref{tab:Loriginal}, LARS galaxies, in particular the LARS-LAEs, tend to avoid the location of normal galaxies and to occupy the region of perturbed disks or merging systems %(G$^{SB-rP20S}>0.5$, M20 $<-2.0$)
 and of irregular or starburst galaxies. % (A $>0.2$, C$^{minA}>2$, S $>0.2$). 
The values of G, M20, C, and A, we calculated for LARS galaxies, are consistent with the ones measured by \citet{Overzier2010} for LBAs. 
Even if our sample is just composed of twelve sources, we do not see any significant dependency between G and EW(Ly$\alpha$).
L08 is the most massive of the LARS-LAEs, but equally concentrated within the segmentation map. L02 is the largest-EW(Ly$\alpha$) emitter, characterized by one of the smallest stellar masses and the largest G$^{SB-rP20S}$ among the LARS-LAEs.

\subsection{Ly$\alpha$ morphology of LARS galaxies at $z\sim0$}
\label{sec:LyaLARSmorphology}

\begin{figure*}
\centering
\includegraphics[width=19cm]{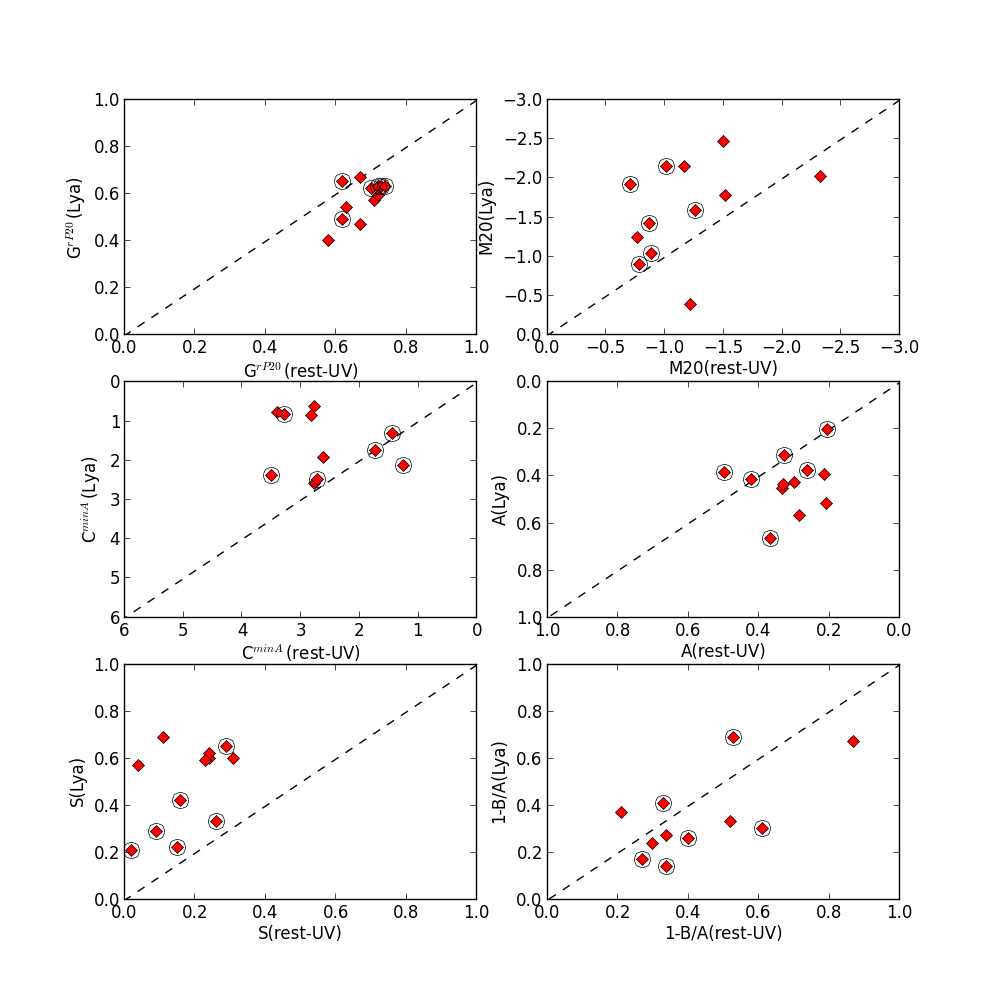}
  \caption{Non-parametric measurements performed in the Ly$\alpha$ images versus the ones performed in the rest-frame UV. From the $upper~left$ to the $lower~right$: G$^{rP20}$, M20, C$^{minA}$, A, S, and SExtractor ellipticity (1-B/A, where A and B are the semi-major and semi-minor axes of the detection ellipse). The dashed line indicates the 1:1 relation. The numbers reported in each panel correspond to the Spearman test coefficient, r, and probability, p, of uncorrelated datasets. r $=0$ indicates no correlation, r $=1$(-1) indicates direct(indirect) proportionality.}
  \label{compcoLaliLa}
\end{figure*}

\begin{figure*}
\centering
\includegraphics[width=19cm]{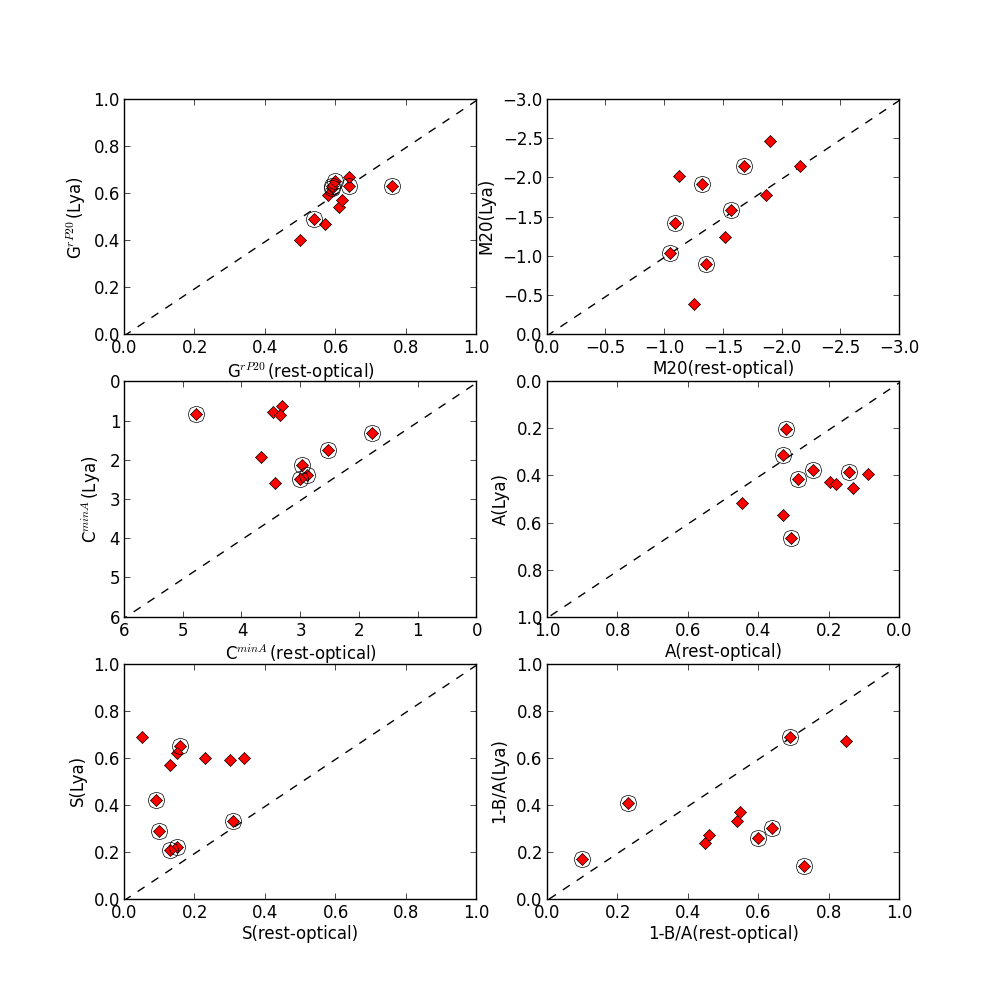}
  \caption{Non-parametric measurements performed in the Ly$\alpha$ images versus the ones performed in the rest-frame optical. From the $upper~left$ to the $lower~right$ we show the same parameters as in Fig. \ref{compcoLaliLa}.}
  \label{compcoHaliLa}
\end{figure*}

One of the goals of our work was to quantify and compare the morphologies of LARS galaxies in Ly$\alpha$ and in the continuum. 
We present morphological parameters measured in the continua and in Ly$\alpha$ of LARS images in Figs. \ref{compcoLaliLa} and \ref{compcoHaliLa}.
G$^{rP20}$, M20, concentration, and ellipticity are smaller; while clumpiness and asymmetry are generally larger in Ly$\alpha$ than in the rest-frame UV continuum. The LARS-LAEs tend to be characterized by the highest concentration, lowest asymmetry, and lowest clumpiness in Ly$\alpha$. G$^{rP20}$ and M20 measured in the rest-frame optical are consistent with the values measured in Ly$\alpha$.

%%%%%%%%%%%%%%%%%%%%%%%%%%%%%%%%%%%%%%%%%%%%%%%%%%%%%%%%%%%%%%%%%%%%%%%%%%%%
\section{LARS galaxies as seen at $z\sim2$}    
\label{sec:LARSz2}
%%%%%%%%%%%%%%%%%%%%%%%%%%%%%%%%%%%%%%%%%%%%%%%%%%%%%%%%%%%%%%%%%%%%%%%%%%%%

We applied the procedure described in Sec. \ref{sec:reg} to simulate LARS galaxies at $z\sim2$. We named the high-$z$ simulated galaxies as z2LARS and the subsample of Ly$\alpha$ emitters as z2LARS-LAEs. We estimated sizes and calculated morphological parameters (Tables \ref{tab:REGsize} - \ref{tab:REGsize4} and \ref{tab:REG}-\ref{tab:REG4}) in the same way we did for the original images in Sec. \ref{sec:morphology}. 

The purpose of this test was to understand whether we could expect to detect LARS-type galaxies and LARS-type Ly$\alpha$ haloes in current high-$z$ surveys.  In particular, we wanted to understand how galaxy size and morphological parameters changed when varying the survey depth (Table \ref{tab:depths}). The results show that, in a sufficiently deep survey, faint galaxy structures in between bright knots remain connected together and SExtractor is able to detect just one source (the entire galaxy) in the image. In a shallower survey, the faint connecting structures tend to be lost in the noise and a galaxy appears to be composed of separated clumps. In that case SExtractor identifies more than one source and photometry is performed by locating the photometric aperture around the brightest clump. In Figs. \ref{stampsDEEP1}, \ref{stampsDEEP2}, and \ref{stampsDEEP3}, we show how LARS galaxies would appear if detected in the deepest continuum and line surveys simulated here, while Figs. \ref{stampsSHALLOW1}, \ref{stampsSHALLOW2}, and \ref{stampsSHALLOW3} show the results for shallower surveys. In Appendix \ref{sec:appendix3}, we present the corresponding surface brightness profiles.

In the following sub-sections, we describe the detection of LARS galaxies in the simulated surveys with 10$\sigma$ detection limits presented in Table \ref{tab:depths}. 
In the first sub-section, we give details on the detection of L01 as an example. We proceed to describe the cases of the LARS-LAEs and of the galaxies with the faintest Ly$\alpha$ emission. Then, we explain the variations in size and ellipticity versus clumpiness owing pixel resampling and survey depth. In Sec. \ref{sec:morphoparC}, we quantify the morphology of z2LARS and compare with high-$z$ observations from the literature.

\subsection{Detection of L01 in high-redshift surveys}
\label{z2LARSdetection}

\begin{figure*}
\centering
\includegraphics[width=20cm]{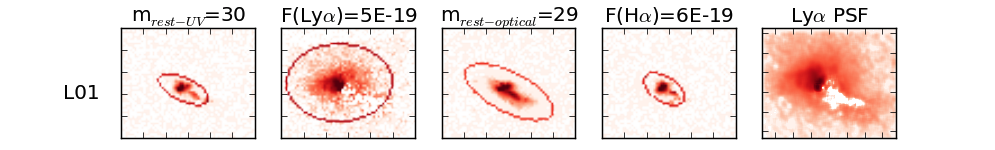}
  \caption{Simulated $z\sim2$ rest-frame UV, Ly$\alpha$, rest-frame optical, and H$\alpha$ emission for L01 as it would be observed in the deepest surveys probed here (first four panels starting from the $left$). The last panel shows the original-pixel-scale Ly$\alpha$ image convolved with a Gaussian kernel, resembling a ground-based seeing Point Spread Function, PSF, of 1.32''. Every panel is 20x17 kpc wide. The reddish ellipses indicate SEx apertures, corresponding to the assumed detection parameters: DETECT\_THRESH =1.65, DETECT\_MINAREA=30, and DEBLEND\_MINCONT=1 from Bond et al. (2009). The log colour scaling is chosen to show a visually consistent background noise.}
  \label{L01imagePSF}
\end{figure*}

In Fig. \ref{L01imagePSF}, 
we show the rest-frame UV, Ly$\alpha$, optical, and H$\alpha$ images of L01 simulated to be at $z\sim2$ (z2L01) as they would be observed in the deepest surveys probed here; SExtractor detection apertures are over-plotted. The detection parameters we adopted are sensitive enough that (at the deepest simulated surveys) this galaxy is detected as a single source. 
As described in detail in Paper I and II, L01 consists of a bright UV star-forming centre with an extended tail, also seen in H$\alpha$ and in the rest-frame optical. 
The Ly$\alpha$ emission is coincident with the bright UV knot and extends in a fan-like structure possibly indicating the presence of an expanding bubble. The main features of emission (dark red pixels in Fig. \ref{L01imagePSF}) and absorption (white pixels), observed in Ly$\alpha$ thanks to the HST resolution and the careful continuum subtraction presented in Paper II, are clearly visible in the $z\sim2$ simulation as well. 
However, the extremely detailed Ly$\alpha$ structures close to the centre of the galaxy (see Paper I Fig. 1) are not visible. 
The last panel of Fig. \ref{L01imagePSF} shows L01 Ly$\alpha$ image, convolved with a ground-based seeing. 
From the ground L01 Ly$\alpha$ morphology would appear smoothed.

\begin{figure*} 
\centering
\includegraphics[width=15cm]{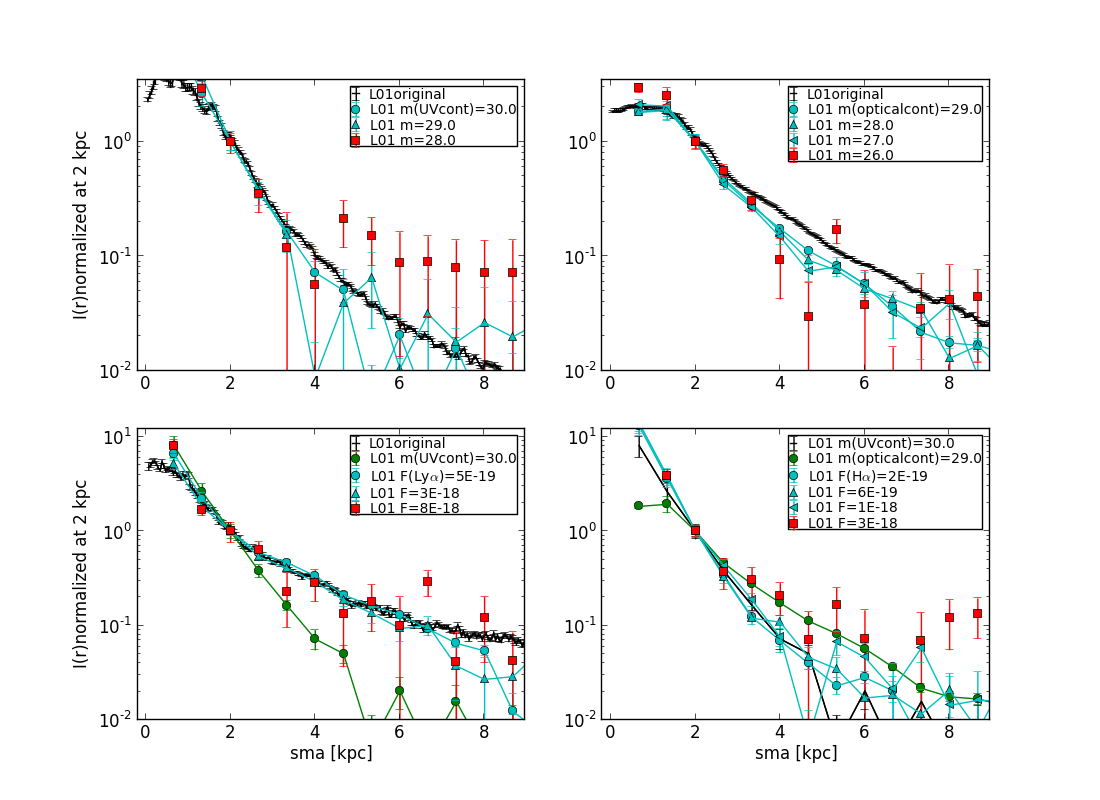}
\caption{Normalized surface brightness profiles of z2L01. The semi-major axis of an elliptical aperture is indicated as sma. Black points with error bars correspond to the surface brightness profile of the original LARS images in the rest-frame UV, optical, and Ly$\alpha$ line. The red squares represent the profiles at the limits of detection. The profiles indicated with circles(triangles) correspond to the deepest(intermediate) depth surveys.
In the $upper ~left$ panel we show the original rest-frame UV profile and the one at a depth of $m_{rest-UV}=30$, 29(cyan) and 28(red). In the $upper~right$ panel the original rest-frame optical profile, the one at a depth of $m_{rest-optical}=29, 28, 27$(cyan), and 26(red). In the $lower ~left$ the original Ly$\alpha$ profile, the one at a depth of $F(Ly\alpha)=5$E-19, 3E-18(cyan), and 8E-18(red), together with the deepest-survey rest-frame UV profile (green circle and line). In the $lower~right$ the H$\alpha$ profile at a depth of $F(H\alpha)=2$E-19, 6E-19(cyan), and 1E-18(red), together with the deepest-survey rest-frame optical profile (green circle and line) and the deepest-survey rest-frame UV profile (black line with error bars).}
\label{SBL01}
\end{figure*}

We show the surface brightness profiles of z2L01 in Fig. \ref{SBL01}. The rest-frame UV and Ly$\alpha$ profiles (left column panels) are preserved when observed in a survey with sensitivity deeper than m$_{rest-UV}^{lim}=28$ and F(Ly$\alpha)^{lim}=8$E-18 erg sec$^{-1}$ cm$^{-2}$. However, on scales larger than 4 kpc, the profiles are indistinguishable from the background noise. In shallower surveys, the profiles start to be affected by the simulated-survey noise on smaller scales and z2L01 could not be detected by adopting a SExtractor detection threshold, DETECT\_THRESH $=1.65$. Therefore, size and morphological parameter measurements could not be performed either.
We define m$_{rest-UV}^{lim}$ and F(Ly$\alpha)^{lim}$ as the limits for detection and morphological parameter measurement.
These limits are m$^{lim}_{rest-optical}=26$ and F(H$\alpha)^{lim}=3$E-18 erg sec$^{-1}$ cm$^{-2}$ for L01 rest-frame optical and H$\alpha$.

The lower left panel of Fig. \ref{SBL01} shows that the rest-frame UV continuum profile is steeper than the Ly$\alpha$ profile. The lower right panel shows that the rest-frame optical continuum tends to be shallower than the rest-frame UV and the H$\alpha$, and more similar to the Ly$\alpha$ profiles. 

\subsection{Detection of the LARS-LAEs in high-redshift surveys}

The LARS-LAE galaxies show more than one bright knot, connected by filaments, in the continua. They also show an intense Ly$\alpha$ emission close to their centres and Ly$\alpha$ structures in their outskirts. The Ly$\alpha$ emission is accompanied by regions of absorption. As a typical trend, in increasingly shallower surveys the filaments, seen in the continua, show lower surface brightness, while their Ly$\alpha$ emission become increasingly localized in the galaxy centre. Only L14, the galaxy brightest in Ly$\alpha$, could be detected in the shallowest Ly$\alpha$ survey probed here. The magnitude and flux limits for z2LARS-LAE detection are shown in Table \ref{tab:limits}.

\subsection{Detection of the Ly$\alpha$-faint z2LARS galaxies}

In general these LARS galaxies are bright enough in UV, optical, and H$\alpha$ (see Paper II) to be detected as a single source in our SExtractor run. The exception is L13 (Fig. \ref{stampsDEEP3}): this galaxy is detected as two possibly blended sources (dashed-line aperture in Fig. \ref{stampsDEEP3}) in UV and H$\alpha$, and photometry was performed locating the aperture on the (brightest) right clump.
In the shallow simulated surveys (Figs. \ref{stampsSHALLOW1}, \ref{stampsSHALLOW2}, and \ref{stampsSHALLOW3}) low surface brightness filaments connecting the main continuum knots disappear into the background noise making the sources appear to be composed of several clumps.
In increasingly shallow surveys, fewer clumps could be detected. For example, only the brightest clump is detected for L13 at m$_{rest-UV}=29$ and m$_{rest-optical}=26$.
Only L09, the brightest galaxy in H$\alpha$, could be detected in the shallowest H$\alpha$ survey probed here.

In general these galaxies could not be detected in Ly$\alpha$ surveys shallower than F(Ly$\alpha)=3$E-18 erg sec$^{-1}$ cm$^{-2}$.  
The galaxies presenting the strongest Ly$\alpha$ absorption (white pixels in Fig. \ref{stampsDEEP2}) of the sample could be detected only in the deepest Ly$\alpha$ survey probed here. 

\begin{table*}

%\tabletypesize{\scriptsize}
%\rotate
\centering
\caption{Limits of detection for LARS galaxies simulated to $z\sim2$}
%\tablewidth{0pt}
\label{tab:limits}

\scalebox{0.9}{
\begin{tabular}{|c|c|c|c|c|}
\hline\hline 
ID & m$_{rest-UV}$  & m$_{rest-optical}$ & F(Ly$\alpha)$ & F(H$\alpha)$ \\
\hline
& AB & AB & erg sec$^{-1}$ cm$^{-2}$ & erg sec$^{-1}$ cm$^{-2}$ \\
\hline
\hline
L01lae & 28 & 26 & 8E-18 & 3E-18 \\
L02lae & 29 & 27 & 8E-18 & 1E-18\\
L03 & 29 & 25 & 3E-18 & 3E-18 \\
L05lae & 27 & 26 & 8E-18 & 3E-18\\
L07lae & 28 & 26 & 2E-17 & 3E-18\\
L08lae & 28 & 25 & 8E-18 & 3E-18\\
L09 & 28 & 27 & 5E-19 & 1E-17\\
L10 & 28 & 25 & 5E-19 & 1E-18\\
L11 & 28 & 26 & 3E-18 & 3E-18 \\
L12 & 26 & 26 & 3E-18 & 3E-18\\ 
L13 & 28 & 27 & 8E-18 & 3E-18\\
L14lae & 27 & 27 & 5E-17 & 1E-17\\
\hline
\end{tabular}
}
\tablefoot{Almost all LARS galaxies would be detected in Ly$\alpha$ at a 10$\sigma$ detection limit depth of 3E-18 erg sec$^{-1}$ cm$^{-2}$. The exceptions are L09 and L10, which show strong absorption in their Ly$\alpha$ images. % (Fig \ref{stampsDEEP2} and \ref{stampsSHALLOW2}) and in surface brightness profile (Fig. \ref{SBL09L10}. 
LARS-LAEs would be detected in shallower surveys. Also, the majority of LARS galaxies would be detected in H$\alpha$ at a depth of 3E-18 erg sec$^{-1}$ cm$^{-2}$.
}
\end{table*}

%%%%%%%%%%%%%%%%%%%
\subsection{Size}
\label{sec:size}
%%%%%%%%%%%%%%%%%%%

\begin{figure*}
\centering
\includegraphics[width=17cm]{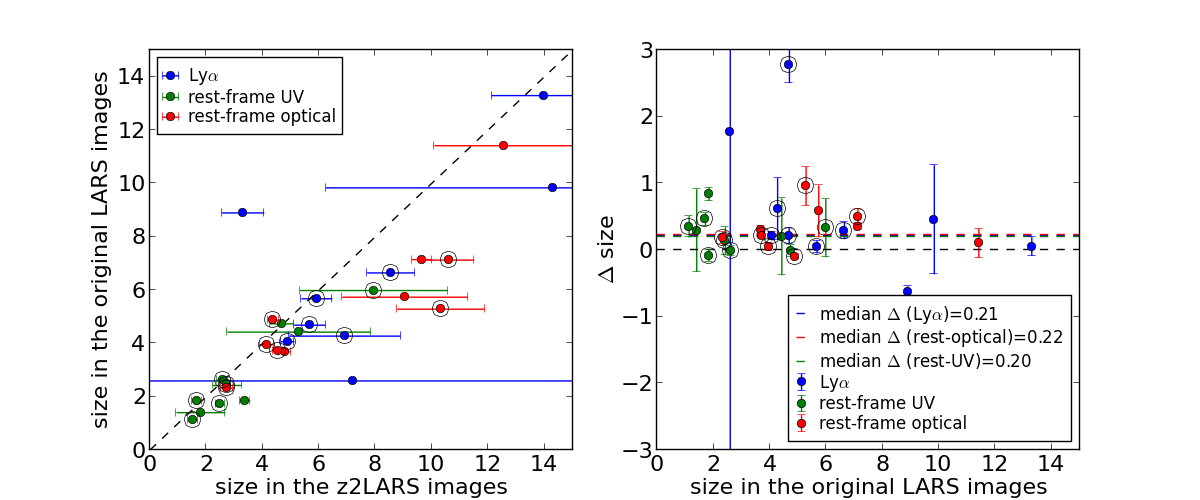} 
  \caption{$Left$: rP20$^{ell}$ (kpc) measured in the original LARS images as a function of the value from the high-$z$ simulated ones. The 1:1 relation is shown as a dashed line. $Right$: $\Delta$size = $\frac{rP20^{ell}(z2LARS) - rP20^{ell}(LARS)}{rP20^{ell}(LARS)}$ as a function of rP20$^{ell}$ measured in the original LARS images. The black dashed line indicates $\Delta=0$; the dashed blue, green, and red lines the median $\Delta$ values for Ly$\alpha$, rest-frame UV, and optical.
Open circles indicate LARS-LAEs; blue, green, and red dots correspond to the measurements in Ly$\alpha$, rest-frame UV, and optical images.}
  \label{size_afterregriding}
\end{figure*}

The Petrosian radius measured within elliptical apertures (rP20$^{ell}$) is the quantity adopted for the comparison of Ly$\alpha$ versus continuum size estimations (Appendix A and Paper 0). The rP20$^{ell}$ estimated in the z2LARS continuum and Ly$\alpha$ images could vary by up to 20\% in median (Fig. \ref{size_afterregriding}). The variation depends on the specific morphology.

\begin{figure*}
\centering
\includegraphics[width=17cm]{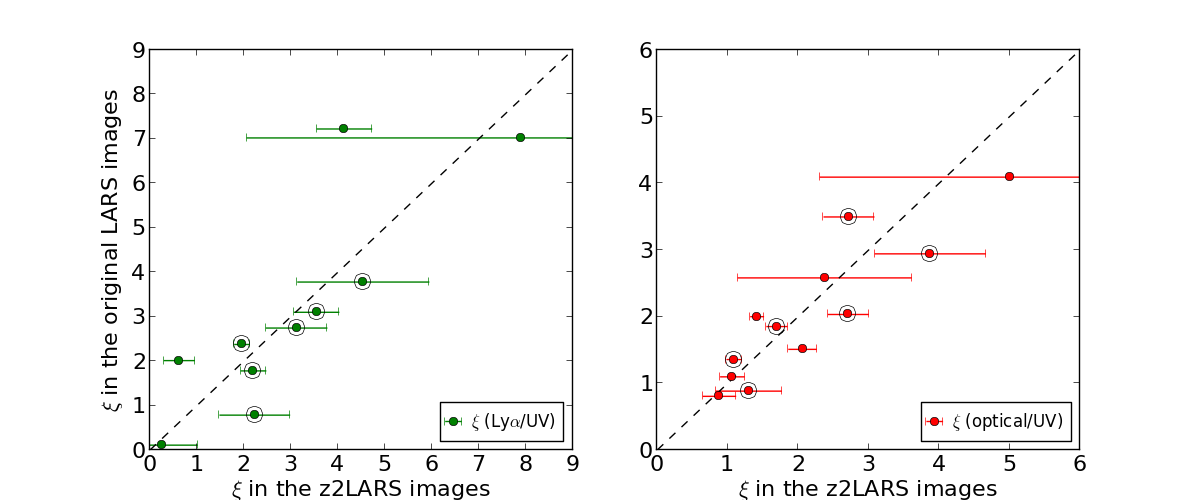}
  \caption{$\xi$(Ly$\alpha$/rest-UV) ($left$) and $\xi$(rest-optical/rest-UV) ($right$) measured in the original as a function of the regridded LARS images. Open circles indicate LARS-LAEs. The dashed line shows a 1:1 relation.}
  \label{eta_afterregriding}
\end{figure*}

As proposed in Paper 0, we define the quantity $\xi$ to estimate the size of Ly$\alpha$ with respect to the size of H$\alpha$ (eq. \ref{eqxiHa}) and continuum (eq. \ref{eqxiUV}) emission,
\begin{equation}
\xi(Ly\alpha/H\alpha) = \frac{rP20^{ell}(Ly\alpha)}{rP20^{ell}(H\alpha)}\\
\label{eqxiHa}
\end{equation}
\begin{equation}
\xi(Ly\alpha/UV) = \frac{rP20^{ell}(Ly\alpha)}{rP20^{ell}(rest-UV)}\\
\label{eqxiUV}
\end{equation}
A $\xi$ larger than 1 indicates that Ly$\alpha$ photons extend to larger scales than UV and H$\alpha$ due to neutral hydrogen scattering; the galaxy presents a Ly$\alpha$ halo. The left panel of Fig. \ref{eta_afterregriding} compares $\xi$ measured in the high-$z$ simulated image with that of the original LARS image.
Even if the rP20$^{ell}$ estimated in the high-$z$ simulation is larger than in the original image, the $\xi$ values remain consistent within the errors. The ratio between $\xi$(Ly$\alpha$/UV) measured in the z2LARS and original LARS images is $1.1\pm0.8$ on average.
The ratios $\xi$ (Ly$\alpha$/UV) or $\xi$(Ly$\alpha$/H$\alpha$) vary between 1 and 5 among the LARS galaxies, implying that Ly$\alpha$ haloes are common in the LARS sample. However, the total extension of the haloes depends on the depth of the simulated survey and the limits of detection in Ly$\alpha$, UV, and H$\alpha$ (Sec. \ref{z2LARSdetection}, 4.2, and 4.3).
The sizes of z2LARS for each simulated survey probed here are presented in Tables \ref{tab:REGsize}, \ref{tab:REGsize2}, \ref{tab:REGsize3}, and \ref{tab:REGsize4}. For comparison \citet{Fin2011} reported size measurements of three LAEs at $z\sim4.5$ of 1-2 kpc in the rest-frame UV and of 1-3 kpc in Ly$\alpha$. Similarly \citet{Bond2010} estimated 2 kpc both in UV and in Ly$\alpha$. 

In Fig. \ref{eta_afterregriding}, we also show 
\begin{equation}
\xi(optical/UV) = \frac{rP20^{ell}(rest-optical)}{rP20^{ell}(rest-UV)}\\
\label{eqoptUV}
\end{equation}
From the right panel of the figure, we see that the ratio between $\xi$(optical/UV) measured in the z2LARS and in the original LARS images is $1.1\pm0.3$ on average.

%%%%%%%%%%%%%%%%%%%
\subsection{Ellipticity and clumpiness}
\label{sec:ELLclump}
%%%%%%%%%%%%%%%%%%% 

As noticed in \citet{Bond2009}, it is not possible to easily understand if clumps in high-$z$ broad-band images are merging components rather than star-forming regions connected by low surface brightness structures. \citet{Gronwall2011,Bond2012} observed that the rest-frame UV emission from a typical LAE was neither smooth nor spheroidal; its ellipticity was about 0.6, suggesting the presence of elongated structures due to merging activity or clumps of star formation. However, \citet{Shibuya2013} found that the LAEs in their sample with EW(Ly$\alpha)>100$ {\AA} tended to be characterized by a small ellipticity both in the rest-frame UV and optical. This is supported by the theoretical results of \citet{Verhamme2012} and \citet{L2009}, which showed that Ly$\alpha$ photons could more easily escape from face-on disks (ellipticity $\sim0$), generally characterized by low column density of HI along the line of sight.

LARS galaxies do not show any clear trend between continuum ellipticity and Ly$\alpha$ equivalent width (Table \ref{tab:Loriginal}). 
After pixel resampling the rest-frame UV, ellipticity varied by up to 60\% for some of the LARS galaxies and no trend remains between continuum ellipticity and Ly$\alpha$ equivalent width for the z2LARS galaxies. 

\begin{figure*}
\centering
\includegraphics[width=8.cm]{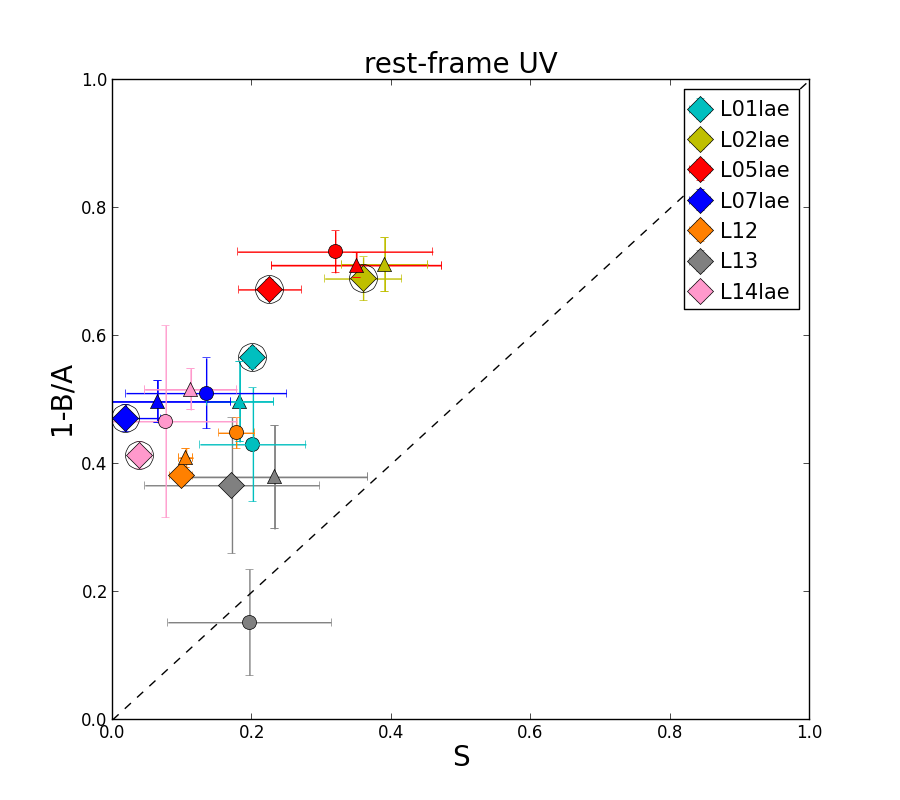}
\includegraphics[width=8.cm]{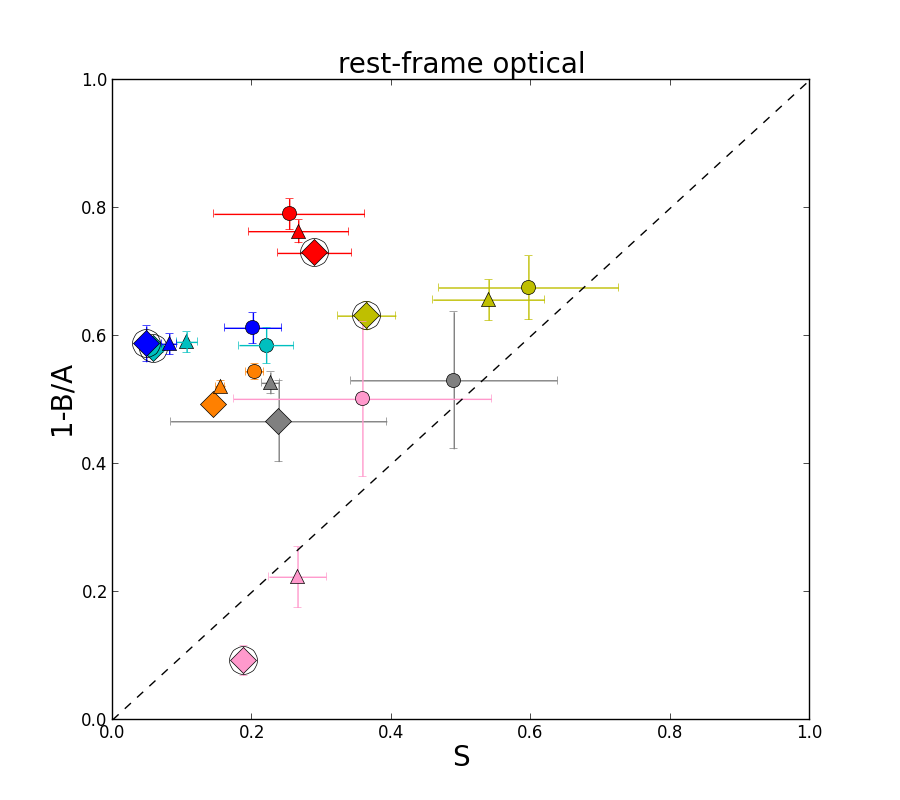}
  \caption{Ellipticity as defined by SEx (1-B/A, where A and B are the semi-major and minor axes of the detection ellipse) as a function of clumpiness for the rest-frame UV and optical continuum. Diamonds represent z2LARS measurements in the deepest survey, triangle and circles in increasingly shallower surveys. Dashed line is a 1:1 relation drawn to aid the eye.}
  \label{ell_Scont}
\end{figure*}

\begin{figure*}
\centering
\includegraphics[width=8.cm]{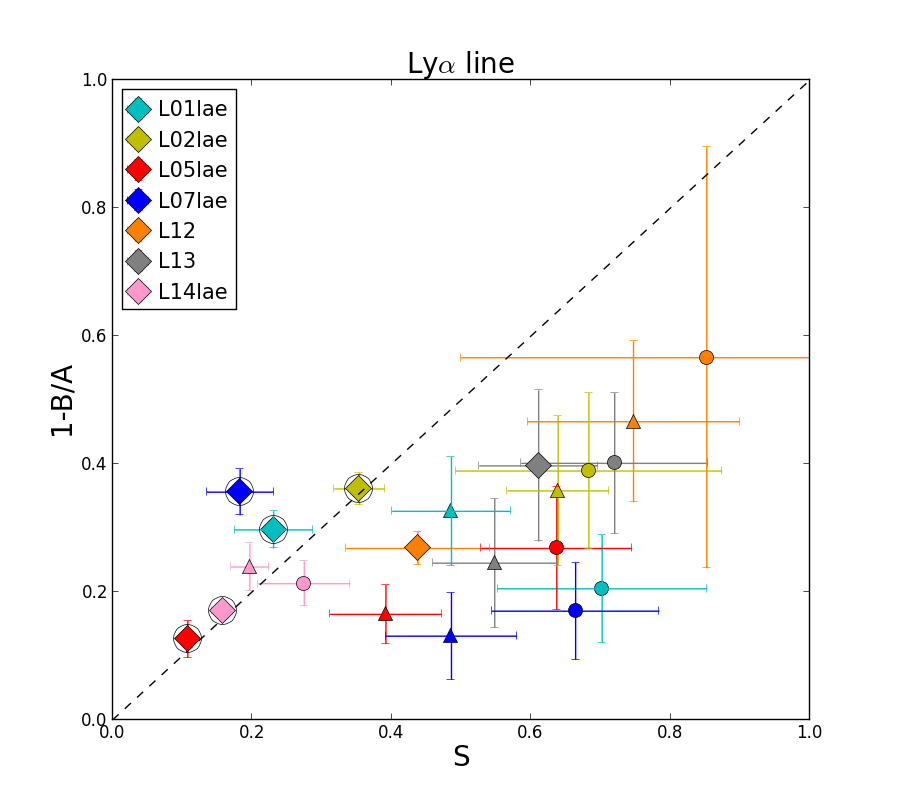}
\includegraphics[width=8.cm]{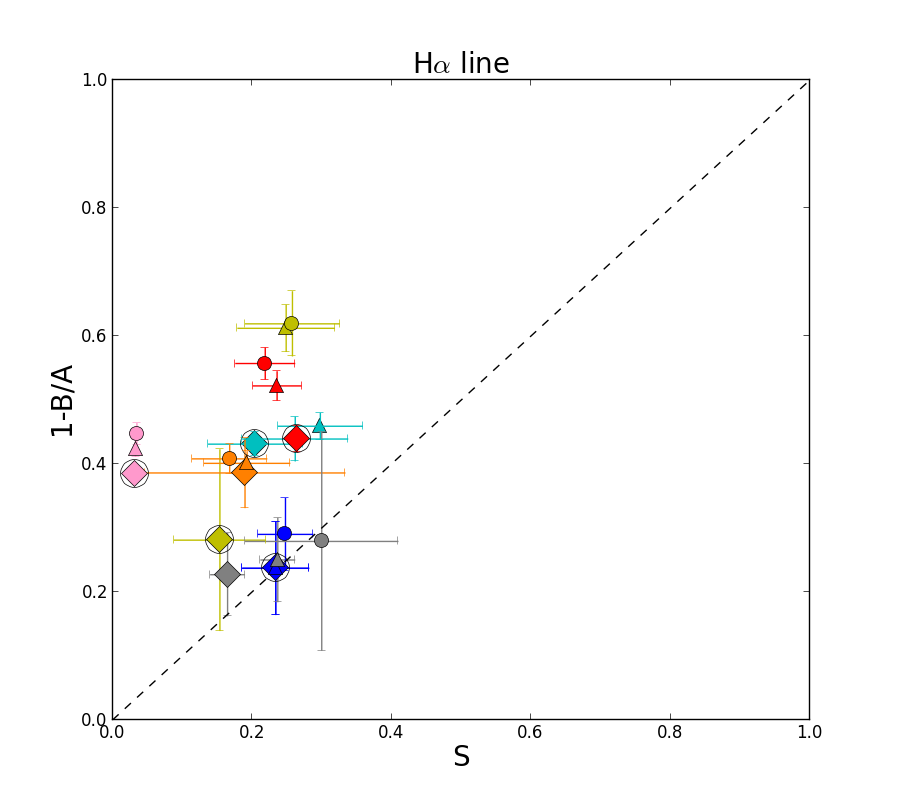}
  \caption{Ellipticity versus clumpiness as in Fig \ref{ell_Scont} for measurements performed in Ly$\alpha$ and H$\alpha$ images.}
  \label{ell_Sline}
\end{figure*}

In shallow surveys, bright knots of star formation, seen without low surface brightness connectors, could appear aligned and lead to the galaxy appearing more elongated. 
Figs. \ref{ell_Scont} and \ref{ell_Sline} show ellipticity versus clumpiness for a few z2LARS for which these measurements are significant. They also show how those parameters change when measured in the range of surveys probed here. 
Quantitatively z2LARS galaxies tend to have lower ellipticity %($0.1<1-B/A<0.5$) 
 in Ly$\alpha$ than in the rest-frame UV, optical continuum, and H$\alpha$. % ($0.3<1-B/A<0.7$). 
Also, z2LARS-LAEs tend to show lower S than the other LARS galaxies in Ly$\alpha$. 

On average the ellipticity values increase in a shallower and shallower survey when measured in the rest-frame optical. We evaluate the correlation between depth and ellipticity or clumpiness by calculating the difference between the morphological parameter measured in shallow surveys and in the deepest one. In the sample of LARS galaxies, 70\% show a Spearman probability p $\sim0.0$ and coefficient r $\sim-1$ for the correlation of ellipticity.  Also, 85\% of the LARS galaxies show a correlation between depth and S estimated in the rest-optical. For the z2LARS-LAEs the ellipticity increases when measured in the rest-frame UV as well.
Therefore, some of the high-$z$ observations of large ellipticity and clumpy systems could be explained in terms of survey depth.
Only a small change in S and ellipticity is seen in the H$\alpha$ images. In Ly$\alpha$ we generally measure a significant increase of S in shallow surveys. Of the LARS galaxies, 85\% are characterized by a correlation between depth and S when measured in Ly$\alpha$. For a few galaxies the Ly$\alpha$ ellipticity also increases with the shallowness of the survey.

%%%%%%%%%%%%%%%%%%%
\subsection{Morphological parameters of LARS at $z\sim2$}
\label{sec:morphoparC}
%%%%%%%%%%%%%%%%%%%

\begin{figure*}
\centering
 \includegraphics[width=9.cm]{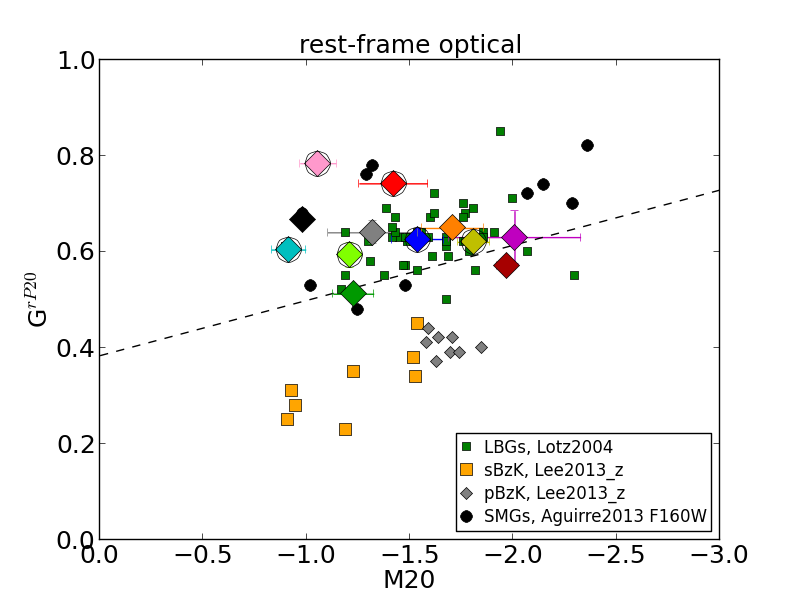}
\includegraphics[width=9.cm]{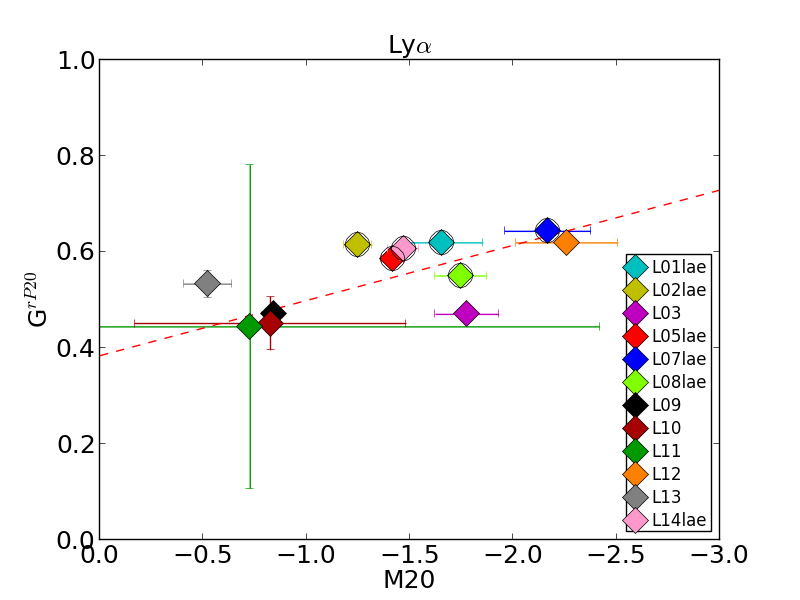}
  \caption{$Left$: G$^{rP20}$ vs M20 measured in the rest-frame optical images of z2LARS (big diamonds). Open circles indicate z2LARS-LAEs. Data from the literature are presented as small symbols: 
LBGs detected in the Hubble Deep Field-North and measured by \citet{Lotz2004} (green squares), passive and star-forming BzK \citep[grey diamonds and yellow squares]{Lee2013}, and $z\sim2.5$ SMGs \citep[black circles]{Aguirre2013}, where the measurements are all performed on HST images. The dashed line indicates the separation between normal galaxies and merging systems by \citet{Lotz2008a}. 
$Right:$ G$^{rP20}$ vs M20 measured in the Ly$\alpha$ images of z2LARS (big diamonds). The colour coding for LARS galaxies is as used throughout this paper. The dashed red line represents the rest-frame optical ($left$ panel) separating region, drawn to guide the eye.}
  \label{LARSreg_location_coHa}
\end{figure*}

As in the case of the original LARS images (Sec. \ref{sec:LARSmorphology}), we performed non-parametric measurements of morphological parameters for z2LARS. To compare with high-$z$ literature observations, we made used of the G vs M20 diagram. We chose G$^{rP20}$ which shows a better consistency for high-$z$ comparisons \citep{Scarlata2007}. As shown in Appendix A, the pixel resampling has little effect on the measurements of G$^{rP20}$ and M20 for LARS galaxies. The change of parameters owing resampling has been studied in the literature by some groups \citep{Lotz2006,Overzier2010,HuertasCompany2014,Petty2014}. 
The changes for irregular galaxies are in agreement with our findings for LARS galaxies.
The non-parametric measurements for all the simulated surveys probed here are presented in Tables \ref{tab:REG}, \ref{tab:REG2}, \ref{tab:REG3}, and \ref{tab:REG4}.
The left panel of Fig. \ref{LARSreg_location_coHa} shows that z2LARS have morphology consistent with that of merging system, like $z\sim2.5$ SMGs and some of the LBGs studied by \citet{Lotz2004}.
They tend to be separated from the location of $z\sim2$ sBzK and pBzK galaxies \citep{Lee2013}. Some z2LARS-LAEs show a larger G$^{rP20}$ per M20 value than the other z2LARS. 

\begin{figure*}
\centering
\includegraphics[width=9.cm]{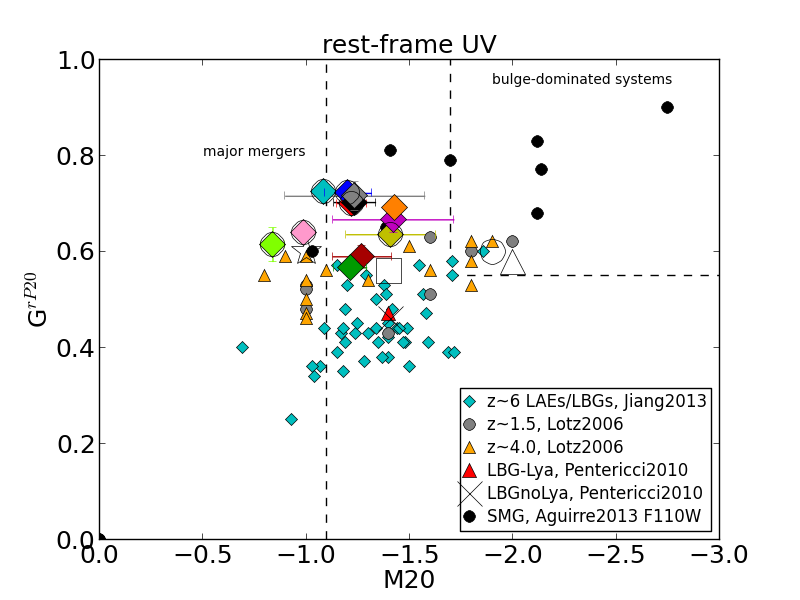}
\includegraphics[width=9.cm]{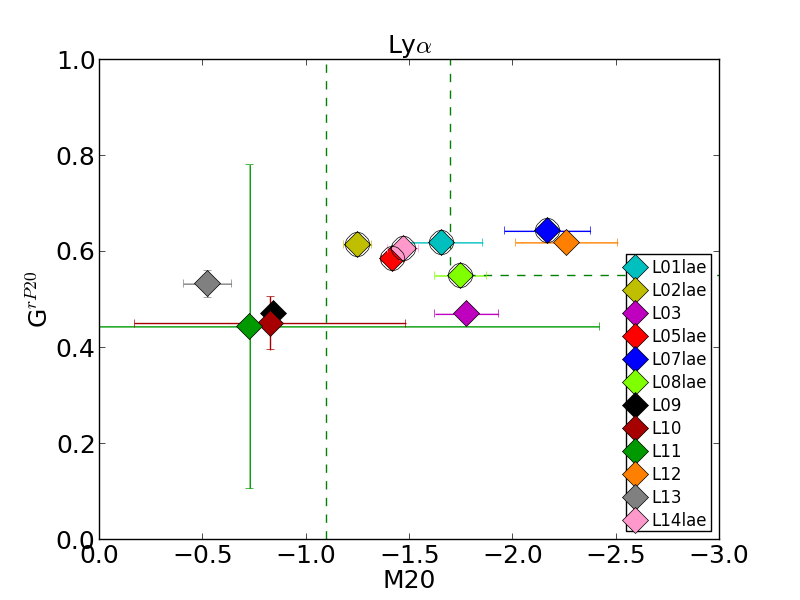}
  \caption{ $Left$: G$^{rP20}$ vs M20 measured in the rest-frame UV images of z2LARS (big diamonds). Open circles indicate z2LARS-LAEs. 
Small symbols are data from literature: LAEs and LBGs at $z>6$ \citep[cyan diamonds]{Jiang2013}, GOODS $z\sim1.5$ and GOODS/UDF $z\sim4$ galaxies \citep[grey circles and yellow triangles]{Lotz2006}, mean values for $z\sim3$ LBGs with and without Ly$\alpha$ emission \citep[red triangle and black cross]{Pentericci2010}, and $z\sim2.5$ SMGs \citep[black circles]{Aguirre2013}, where the measurements are all performed in HST images. For comparison we also show the local galaxies regridded to be at $z\sim1.5$ by \citet{Lotz2006}: an elliptical (open circle), an Sb (open triangle), an Scb (open square), and a merging galaxy (open star). 
The dashed lines indicate the criteria introduced by \citet{Lotz2006} to identify major mergers and bulge-dominated systems. 
$Right$: G$^{rP20}$ vs M20 measured the Ly$\alpha$ images of z2LARS (big diamonds). The dashed green lines represent the rest-frame UV ($left$ panel) separating regions, drawn to aid the eye. The colour coding for LARS galaxies is the same as in Fig. \ref{LARSreg_location_coHa}.
}
  \label{LARSreg_location_coLa}
\end{figure*}

\citet{Lotz2006} built a criterion to identify merging systems based on their rest-frame UV. 
Since the rest-frame UV tends to be more disturbed than the optical, they defined two main regions in the G-M20 diagram. The condition M20 $\geq-1.1$, typical of well-separated double or multiple bright nuclei, was used to identify major mergers, while the condition M20 $<-1.6$, G $>0.6$ to identify bulge-dominated systems. The left panel of Fig. \ref{LARSreg_location_coLa} shows that z2LARS galaxies have morphologies similar to that of high-$z$ LAEs and LBGs in the rest-frame UV, and tend to be less compact than SMGs. In this diagram L01, L08, and L14 have morphologies consistent with that of major-merger systems. The non-LAE z2LARS galaxies, similar to the majority of the star-forming galaxies at high redshift, are characterized by intermediate M20 values. In the right panel of Figs. \ref{LARSreg_location_coHa} and \ref{LARSreg_location_coLa}, we present G$^{rP20}$ vs M20 measured in the Ly$\alpha$ images. The subsample of z2LARS-LAEs tend to have lower G$^{rP20}$ and lower M20 in Ly$\alpha$ than in the rest-frame optical and UV. Some of the z2LARS non-LAEs tend to have a more distorted (composed of more than one clump) morphology in Ly$\alpha$ than in the continua.
It can be seen in Tables \ref{tab:REG}-\ref{tab:REG4} that the morphology of LARS galaxies does not change significantly in the rest-frame optical when non-parametric measurements were performed in simulated surveys with depths comparable with the limits of detection (Table \ref{tab:limits}). For a few z2LARS, G$^{rP20}$ and M20 vary in a way that they approach the dashed line drawn in Fig. \ref{LARSreg_location_coHa}. In the rest-frame UV, G$^{rP20}$ and M20 become characterized by larger uncertainties, while in Ly$\alpha$ z2LARS become increasingly compact and less composed of multiple structures.

\addtocounter{table}{8}

%%%%%%%%%%%%%%%%%%%%%%%%%%%%%%%%%%%%%%%%%%%%%%%%%%%%%%%%%%%%%%%%%%%%%%%%%%%%
\section{Stacking of regridded LARS images}  
%%%%%%%%%%%%%%%%%%%%%%%%%%%%%%%%%%%%%%%%%%%%%%%%%%%%%%%%%%%%%%%%%%%%%%%%%%%%
\label{sec:Stack}

As extensively described in Paper 0 and II, a significant fraction of the
Ly$\alpha$ photons in LARS galaxies is emitted from haloes. These haloes begin in
the inner few kpc and extend outward to scales larger than those characterizing
localized star-forming regions. Ly$\alpha$ maps were shown in Paper II as well
as maps of Ly$\alpha$/H$\alpha$ ratio and dust-reddening maps. Since H$\alpha$
photons are emitted directly from the HII regions (i.e., they do not scatter), a
value of the Ly$\alpha$/H$\alpha$ ratio that exceeds that of case B
recombination is most probably related to HI scattering. Dust reddening and HI scattering together can contribute to the situation in
which the number of Ly$\alpha$ photons observed on a particular sight-line is
reduced. The re-processing of ionizing photons \citep[e.g.,][]{Humphrey2008} in
a region of the galaxy different from that of the star formation could
contribute to an additional production of Ly$\alpha$ photons and a possible
extended Ly$\alpha$ emission. In this case, however, H$\alpha$ radiation will also be produced at large radii by the same recombinations that make Ly$\alpha$.
The availability of H$\alpha$ images, together with those of Ly$\alpha$, has
favoured an interpretation in which HI scattering plays a key role in \lya\
emission, but it has proven difficult to disentangle the contribution of other
factors. However, HI scattering would
also produce a certain degree of polarization in the Ly$\alpha$ emission
\citep[e.g.,][]{Humphrey2013b,Hayes2011}. At high redshift, Ly$\alpha$ haloes have
been detected only in a minority of cases, indicating that their detection is
challenging.

\subsection{Literature results}
\label{sec:literature}

\citet{Rauch2008} stacked \lya\ spectra of a sample of faint ($V>25.5$ for 80\%
of the sample) spectroscopically detected galaxies at $2.67<z<3.75$, reaching a
Ly$\alpha$ surface brightness limit of $\sim2$E-19 erg sec$^{-1}$ cm$^{-2}$
arcsec$^{-2}$ at 1$\sigma$. The resulting Ly$\alpha$ profile extended up to
$\sim$30 kpc from the centre.  \citet{Matsuda2012} stacked narrow-band images of
large subsamples of LBGs, separated based on their environment. They found that
galaxies in overdense regions tend to show Ly$\alpha$ emission on scales up to
40-60 kpc, and are larger than those of isolated galaxies. This could imply that
Ly$\alpha$ haloes follow the dark matter distribution.  \citet{Steidel2011}
(hereafter S11) also stacked \lya\ images of subsamples of LBGs at $z\sim2.65$,
segregated by Ly$\alpha$ equivalent width.  The galaxies came from surveys of
three different fields, which were also characterized by over-densities.
The stack of the entire sample reached a Ly$\alpha$ surface brightness limit of
$\sim1$E-19 erg sec$^{-1}$ cm$^{-2}$ arcsec$^{-2}$  (1$\sigma$), while the LAE
(EW(Ly$\alpha)>20$ \AA) subsample reached $\sim2.4$E-19 erg sec$^{-1}$
cm$^{-2}$ arcsec$^{-2}$. They found that the stacked Ly$\alpha$ profiles had a
characteristic size up to 9 times larger than the stellar continua. Their
detected Ly$\alpha$ emission extended up to 80 kpc from the centre of the
stacked source.

By stacking images of 187 narrow-band selected Ly$\alpha$ emitters at
$z\simeq2.07$, \citet{Feldmeier2013} found no evidence of extended Ly$\alpha$
emission, but did observe a 5-8 kpc halo in a stack of about 200 LAEs at
$z\simeq3.1$. These field-galaxy stacks reached a Ly$\alpha$ surface brightness
1$\sigma$ limit of just 9.9E-19 erg sec$^{-1}$ cm$^{-2}$ arcsec$^{-2}$ at
$z\simeq2.07$ and 6.2E-19 erg sec$^{-1}$ cm$^{-2}$ arcsec$^{-2}$ at
$z\simeq3.1$.  Recently, \citet{Momose2014} (hereafter M14) succeeded in
detecting extended haloes, by stacking over 3500, 300, and 350 narrow-band
detected LAEs at $z\sim2.2$, $\sim3.1$, and $z\sim5.7$ from Subaru surveys (not
necessarily in overdense regions). They reached a Ly$\alpha$ surface brightness
1$\sigma$ limit of 1.6E-20, 1.7E-19, and 5.5E-20 erg sec$^{-1}$ cm$^{-2}$
arcsec$^{-2}$, respectively. Both these last two studies performed LAE selection
in deep \citep[5$\sigma$ detection limit of 25 AB for Feldmeier;][]{Guaita2010,
Gronwall:2007} and very deep \citep[5$\sigma$ detection limit of 25.1--25.7 AB for
Momose;][]{Nakajima2012a} narrow-band images.  Also, they treated the sources of
uncertainty in a very careful way.  However, the detected haloes (r$_e\sim8$ kpc
at $z=2.2$) were not as extended as the ones claimed by S11 (r$_e\sim25$ kpc at
$z=2.65$).  A few attempts to detect Ly$\alpha$ haloes from individual high-$z$
LAEs were performed by \citet{Bond2010} and \citet{Finkelstein2011b} in
HST-filter images.  \citet{Bond2010} presented a morphological study of LAE
Ly$\alpha$ emission at $z\simeq3.1$ (F502N filter, 1$\sigma$ detection limit of
3E-17 erg sec$^{-1}$cm$^{-2}$ arcsec$^{-2}$ for 1 arcsec source). Their images
were very shallow and did not show significant extended haloes
($\xi(Ly\alpha/UV)$ $\sim1$, Table \ref{tab:REGsize}).  An estimation at
$z\sim4.5$ was presented in \citet{Finkelstein2011b} (F658N filter, 1$\sigma$
detection limit of about 2E-18 erg sec$^{-1}$cm$^{-2}$ for 1 arcsec source).
They studied \lya\ emission from three spectroscopically confirmed LAEs and
found evidence of Ly$\alpha$ haloes in two of them ($\xi(Ly\alpha/UV)$
$\sim1.5$).

\subsection{Stacking procedure and stacked surface brightness of LARS at high redshift}

\begin{figure*} 
\centering
\includegraphics[width=17cm]{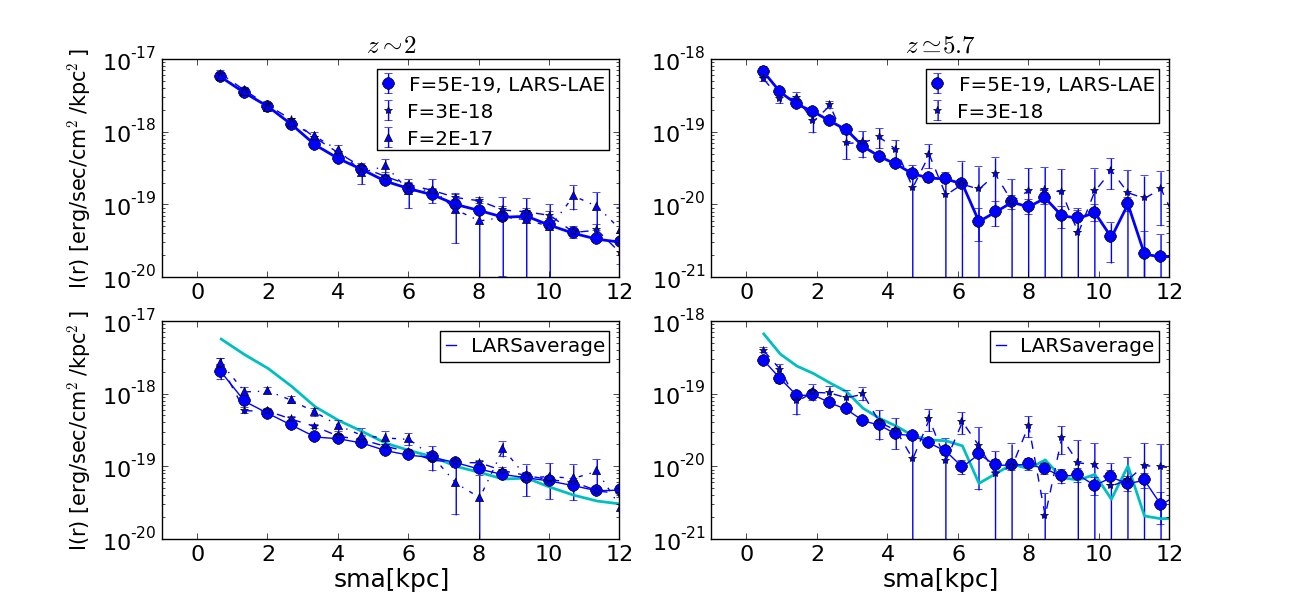}
\caption{\lya\ surface brightness profiles of the two stacks: LARS-LAE
($upper$) and LARSaverage ($lower$; see text for details).  $Left$ and $right$
panels show the profiles at $z\sim2$ and 5.7, respectively, corresponding to three and two
10$\sigma$ detection limits (blue circles, stars, and
triangles). The cyan solid line represents the LARS-LAE profile of the deepest simulated survey and
it is shown in the lower panels for comparison.  The difference in the surface
brightnesses in the $left$ and $right$ panels is the result of surface
brightness dimming.  The error bars correspond to the shallowest
survey in each panel; for the other depths they are usually smaller than the
symbols.  The large error bars at sma $> 5$ kpc are produced by background
noise.}
\label{liLa_LAE_av_med}
\end{figure*}

By following the steps listed in Sec. \ref{sec:method}, we simulated the
appearance of LARS galaxies at $z\sim2$ and $z\simeq5.7$. After adding noise to
the simulated high$-z$ images, we stacked the observations of individual
galaxies the same way it is done at high redshift to increase the signal-to-noise. We used the IMCOMBINE task in $iraf.images.immatch$ to (average) stack every galaxy at the position of SExtractor centroid. We generated an average (LARSaverage), median stack of all the twelve high-$z$ LARS galaxies (like in M11), and also an average stack of the six LARS-LAEs.  
It was only meaningful to look at the stacked profiles up to 12 kpc, which was the physical scale probed by the ACS/SBC detector common to all LARS images. 
However, narrow-band observations at high redshift are able to probe much larger scales.  As the median stack provided very similar results as the average one, 
we only reported results from the LARS-LAE and LARSaverage stacks in the tables and figures.  
We chose to combine individual galaxy frames from simulated survey depth of F(Ly$\alpha$)=5E-19, 3E-18, 2E-17, and 1E-15 erg sec$^{-1}$ cm$^2$. The central value of
F(Ly$\alpha$)=2E-17 is comparable to recent ground-based narrow-band surveys. The shallowest value was chosen to match the depth of \citet{Bond2010}, in which we did
not detect any Ly$\alpha$ emission from the stacks.  The final depths depended on the number of sources in the stack, F(Ly$\alpha$)/$\sqrt{12}$ for LARSaverage and F(Ly$\alpha$)/$\sqrt{6}$ for LARS-LAEs. We applied the same code we ran on the individual z2LARS galaxies to obtain surface brightness
profiles (Fig. \ref{liLa_LAE_av_med}) and sizes (Table \ref{tab:stacksize}) of
the stacks at $z\sim2$ and $z\simeq5.7$. The LARS-LAE Ly$\alpha$ profile was the most
peaked in the centre, and was also the most affected by background noise at large
radii.  The S{\'e}rsic indices of the LARS-LAE stack profiles were found to
be 3.4 and 2.1, while the ones of the LARSaverage stacks are 2.5 and 2.0 at
$z\sim2$ and $z\simeq5.7$, respectively.
\begin{table*}
\caption{Size of stacked Ly$\alpha$ and continuum images}
%\tablewidth{0pt}
\label{tab:stacksize}
\centering
\scalebox{0.9}{
\begin{tabular}{|c|c|c|c|c|c|c|c|}
%\hline\hline 
%(1) & (2) & (3) & (4) & (5) &(6) & (7)& (8)\\ %& (9) &(10) & (11) & (12) & (13) & (14) & (15) &(16) & (17)& (18)\\
\hline\hline
LARS stack & r$_{circ}^{SEx}$ & rP20$^{ell}$ & rP20$^{circ}$ & rP20$^{minA}$ & r$_{20}$ & r$_{50}$ & r$_{80}$ \\
\hline
& kpc &kpc  &kpc  &kpc  &kpc &kpc &kpc  \\
\hline
$z\sim2$ & &  &  &  & & &  \\
F(Ly$\alpha)=5E-19$ & &  &  &  & & & \\
\hline
%OK june2014
%LARS-LAE & 5.26& 9.51 & 6.00 & 3.34 & 3.00 & 5.00 & 7.84  \\ error found June 2014
LARS-LAE & 5.26 & 6.67 & 6.00 & 3.34 & 1.33 & 2.67 & 5.34 \\
LARSaverage & 8.05 & $>12$ & $>12$ & 8.01 & 2.84 & 7.84 & $>12$ \\
%LARSmedian & 8.05 & $>12$ & $>12$ & 8.01 & 2.84 & 7.84 & $>12$ \\
\hline
F(Ly$\alpha)=3E-18$ & &  &  &  & & & \\
\hline
LARS-LAE & 3.55 & 6.84 & 5.00 & 3.67 & 1.50 & 2.67 & 5.67 \\
LARSaverage & 4.84 & $>12$ & $>12$ & 7.00 & 2.84 & 7.34 & $>12$ \\
%LARSmedian & 4.84 & $>12$ & $>12$ & 7.00 & 2.84 & 7.34 & $>12$ \\
\hline
F(Ly$\alpha)=2E-17$ & &  &  &  & & & \\
\hline
LARS-LAE & 2.16 & 4.00 & 5.34 & 3.67 & 1.17 & 2.50 & 4.00 \\
LARSaverage & 2.54 & 7.00 & 9.01 & 4.00 & 1.83 & 3.84 & 6.84 \\
%LARSmedian & 2.54 & 7.00 & 9.01 & 4.00 & 1.83 & 3.84 & 6.84 \\
\hline
$z\simeq5.7$ & &  &  &  & & & \\
F(Ly$\alpha)=5E-19$& &  &  &  & & & \\
\hline
LARS-LAE & 2.40 & 4.70 & 6.34 & 3.52 & 1.17 & 2.35 & 4.35 \\
LARSaverage & 2.72 & 11.98 & 11.75 & 5.87 & 2.11 & 5.17 & 10.34 \\
%LARSmedian & 2.72 & 11.98 & 6.58 & 5.87 & 2.11 & 5.17 & 10.34 \\
\hline
F(Ly$\alpha)=3E-18$ & &  &  &  & & & \\
\hline
LARS-LAE & 0.81 & 4.35 & 3.52 & 2.82 & 1.29 & 2.47 & 4.11 \\
LARSaverage & 0.82 & 5.05 & 3.29 & 4.70 & 1.76 & 3.17 & 5.17 \\
%LARSmedian & 0.82 & 5.05 & 3.29 & 4.70 & 1.76 & 3.17 & 5.17 \\
\hline
\hline
$z\sim2$ & &  &  &  & & & \\
m$_{rest-UV}=30$ & &  &  &  & & & \\
\hline
LARS-LAE & 2.28 & 2.17 & 1.33 & 1.33 & 0.50 & 1.00 & 1.83 \\
LARSaverage & 3.09 & 4.34 & 3.00 & 1.67 & 0.67 & 1.50 & 3.50 \\
%LARSmedian & 3.09 & 4.34 & 3.00 & 1.67 & 0.67 & 1.50 & 3.50 \\
\hline
$z\simeq5.7$ & &  &  &  & & & \\
m$_{rest-UV}=30$ & &  &  &  & & & \\
\hline
LARS-LAE & 1.28 & 1.88 & 1.41 & 0.94 & 0.35 & 0.82 & 1.53 \\
LARSaverage & 1.45 & 3.76 & 1.64 & 1.17 & 0.59 & 1.29 & 3.29 \\
%LARSmedian & 1.46 & 3.88 & 1.64 & 1.17 & 0.59 & 1.29 & 3.41 \\
\hline
\end{tabular}
}
\tablefoot{Size measurements from elliptical and circular aperture photometry. For three(two) simulated survey depths, we reported the Ly$\alpha$ emission sizes of the two stacks (LARS-LAE, LARSaverage) at $z\sim2$($z\simeq5.7$). We also reported the continuum emission sizes of the two stacks in the lower part of the table at a 10$\sigma$ detection limit of m$_{rest-UV}=30.0$. F(Ly$\alpha$) is given in erg sec$^{-1}$ cm$^{-2}$.}
\end{table*}

We find that a depth of F(Ly$\alpha)=5$E-19 erg sec$^{-1}$ cm$^{-2}$
($\sim$1.4E-19 erg sec$^{-1}$ cm$^{-2}$ after stacking twelve sources) enables us to
recover Ly$\alpha$ haloes in the stacks of LARS galaxies at both $z\sim2$ and
$z\simeq5.7$. A depth of F(Ly$\alpha)=2$E-17 erg sec$^{-1}$ cm$^{-2}$
($\sim$5.8E-18 erg sec$^{-1}$ cm$^{-2}$ after stacking twelve sources) is more
realistic in terms of current surveys.  Even the brightest galaxy (L14)
would hardly be detectable individually.  In a simulated $z\simeq5.7$ survey
with depth of F(Ly$\alpha)=3$E-18 erg sec$^{-1}$ cm$^2$ ($\sim$0.9E-18 erg
sec$^{-1}$ cm$^{-2}$ after stacking twelve sources), the Ly$\alpha$ surface
brightness profile reached the level of the
background noise at a radius of 5~kpc.  

\begin{figure*} 
\includegraphics[width=8cm]{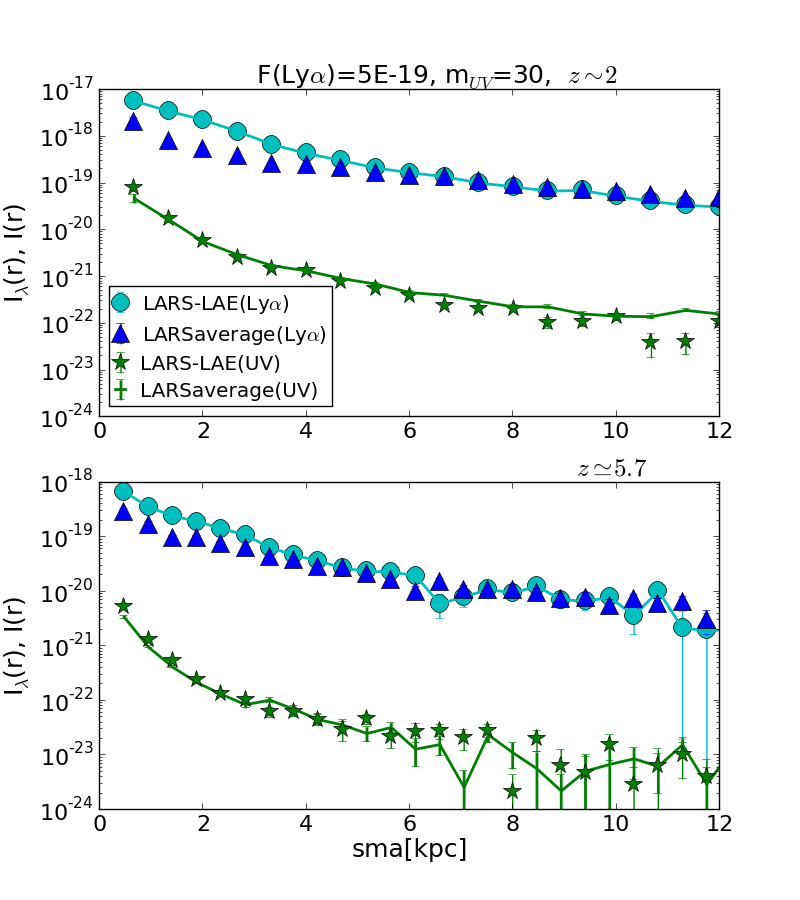} 
\includegraphics[width=9.5cm]{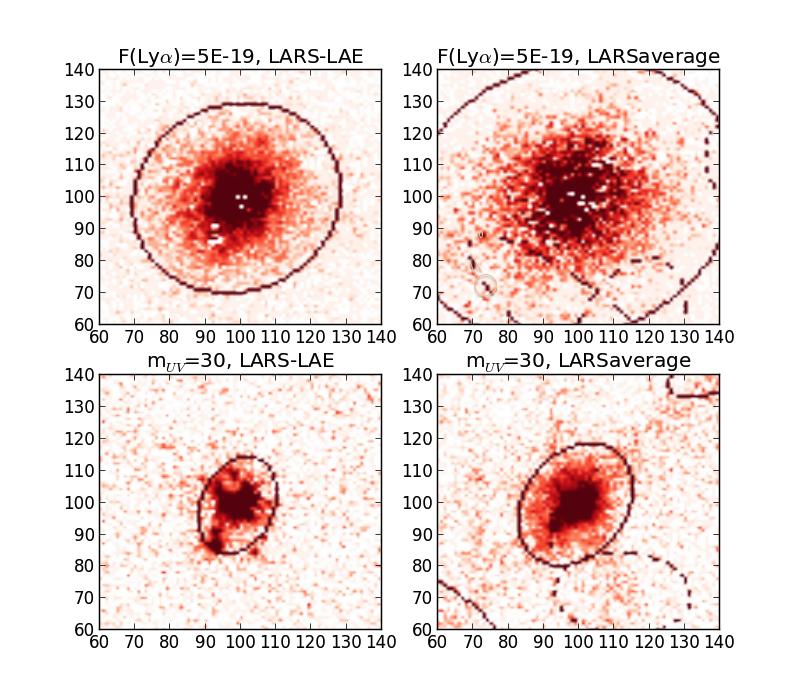}
\caption{$Left$: Surface brightness profiles for Ly$\alpha$ (units of
erg~sec$^{-1}$~cm$^{-2}$~kpc$^{-2}$) at $z\sim2$ ($upper$ panel).  LARS-LAEs are
shown with the cyan circles and line, and LARSaverage with (blue triangles).
Rest UV profiles (units of erg~sec$^{-1}$~cm$^{-2}$~kpc$^{-2}$~{\AA}$^{-1}$) are
shown for the LARS-LAE sample (green stars) and LARSaverage (green line with
errorbars).  The $lower$ panel shows the equivalent profiles at $z=5.7$.
$Right$: From the $upper~left$ to the $lower~right$: Ly$\alpha$ for the LARS-LAE and LARSaverage stacks; rest-frame UV
continuum for LARS-LAEs and LARSaverage. The logarithmic colour scale is adjusted to match the background level. The
elongated structures in the LARSaverage continuum frame come from L09 (vertical)
and L11 (diagonal). The whitish structures in the LARSaverage Ly$\alpha$ come
from the same galaxies, which are weak Ly$\alpha$ emitters but show strong
absorption centrally.  The cut-out size is about 15 kpc. }
\label{liLa_aL_AV_co30}
\end{figure*}

In Fig. \ref{liLa_aL_AV_co30} 
we compare the rest-frame UV and Ly$\alpha$ line surface brightness. The figure shows that the UV profiles for the LARS-LAEs and
LARSaverage stacks are very similar. They are also steeper than any other
Ly$\alpha$-stack at both redshifts, showing S{\'e}rsic indices $n\sim$ 9 and $n\sim 11$ at $z\sim2$ and $z\simeq5.7$, respectively). The consequence of this
is that the sizes of the Ly$\alpha$-stacks are generally larger than the continuum ones (Paper 0, Table \ref{tab:stacksize}).  The profile of the
LARSaverage stack (composed of both EW(Ly$\alpha)>20$ {\AA} and
EW(Ly$\alpha)<20$ {\AA} sources) was also shallower and more extended than that
of the LARS-LAEs; i.e.  the Ly$\alpha$ emission of the LARS-LAE stack was more
compact.  As we described in Section \ref{sec:LyaLARSmorphology}, there are no
unique conditions for Ly$\alpha$ photons to escape or unique morphologies.  The
conditions that make a galaxy a Ly$\alpha$ emitter (mainly Ly$\alpha$ flux
concentrated around rest-frame UV bright star-forming regions) produce a
consistent surface brightness profile, while variations in the dust and HI 
contents and distributions, which made Ly$\alpha$ photons eventually escape along the
line of sight, produce a more complex, patchy, and extended emission.  

\subsection{Ground-based PSF convolution and stacking of LARS at high redshift}

\begin{figure*} 
\includegraphics[width=19cm]{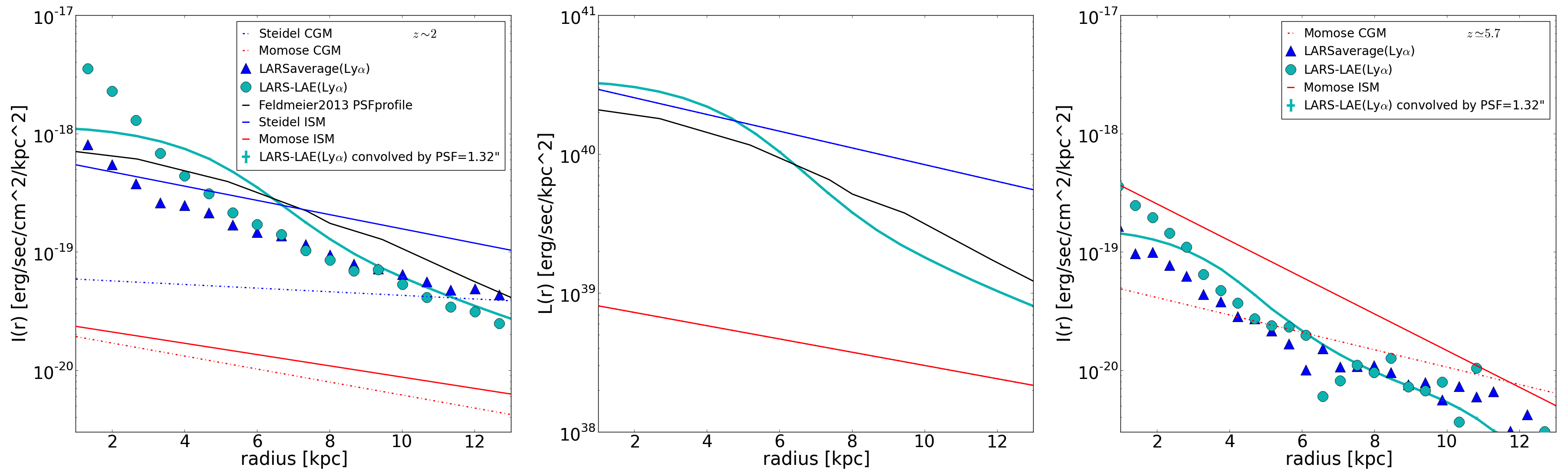}
\caption{$Left$: Ly$\alpha$ surface brightness, I(r), of the $z\sim2$ LARS-LAE
and LARSaverage stacks, before convolution with ground-based PSF (cyan circles
and blue triangles).  The PSF-convolved LARS-LAE stack is shown with the cyan
line.  The ground-based PSF profile from \citet{Feldmeier2013} is shown as a
black line. The blue and red dashed-dotted lines represent the circumgalactic
medium, as fit by \citet{Steidel2011} and \citet{Momose2014} at $z\simeq2.2$, we extrapolated inwards. The blue and red solid lines represent the I(r) fit on scales of 10-20 kpc derived from Fig. 9 of \citet{Steidel2011} and Fig. 3
of \citet{Momose2014}.  $Middle$: Luminosity per kpc$^2$ for our
PSF-convolved LARS-LAE stack in Ly$\alpha$ (cyan line), for the ground-based PSF
profile from \citet{Feldmeier2013} (black line), for the 10-20-kpc-scale fit from
\citet{Steidel2011} (blue line) and \citet{Momose2014} (red line).  $Right$:
Ly$\alpha$ surface brightness of the $z\simeq5.7$ LARS-LAE and
LARSaverage stacks before convolving with ground-based PSF (cyan circles and
blue triangles) together with that of the PSF-convolved LARS-LAE stack (cyan
line). The red dashed-dotted line represents the circumgalactic medium fit by
\citet{Momose2014} at $z\simeq5.7$, we extended inwards.  The red line
shows the fit on scales of 5-15 kpc derived from Fig. 3 of \citet{Momose2014}.
}
\label{convolved}
\end{figure*}

To be able to properly compare with current observations at $z\sim2$ and $z\simeq5.7$, we convolved LARS images with ground-based PSF  
(Fig. \ref{convolved}). As the ground-based PSF is much larger than the HST one, we applied a Gaussian kernel with $\sigma=$ PSF(pixel)/2.3548, using the GAUSS task in $iraf.images.imfilter$. \citet{Feldmeier2013} data at
$z\simeq2.07$ exhibited a $1.4''$ PSF in narrow-band observation and
$1''$ in the broad band. S11 observations at $z\simeq2.64$ were performed in similar
weather conditions (PSF=$0.8''$-$1.2''$). The measurements reported in M14 
were performed on frames smoothed to a FWHM $=1.32''$. To be able to compare
with their observations at both $z\simeq2.2$ and $z\simeq5.7$, we convolved line
and continuum images with that same Gaussian kernel, before stacking. 
This convolution produced a profile close to Gaussian in the
stacked images of both continuum and Ly$\alpha$; the central peak was
suppressed and, as a consequence, the profile became shallower. However, the stacked Ly$\alpha$ profile remained more extended than that of the rest-frame UV (Table \ref{tab:stacksizePSF}).

\begin{table*}
\caption{Size of the LARS-LAE stack convolved with ground-based PSF}
%\tablewidth{0pt}
\label{tab:stacksizePSF}
\centering
\scalebox{0.9}{
\begin{tabular}{|c|c|c|c|c|c|c|c|}
\hline\hline 
%(1) & (2) & (3) & (4) & (5) &(6) & (7)& (8)\\ %& (9) &(10) & (11) & (12) & (13) & (14) & (15) &(16) & (17)& (18)\\
%\hline
LARS stack & r$_{circ}^{SEx}$ & rP20$^{ell}$ & rP20$^{circ}$ & rP20$^{minA}$ & r$_{20}$ & r$_{50}$ & r$_{80}$ \\
\hline
 & kpc & kpc &kpc  &kpc  &kpc & kpc& kpc \\
\hline
$z\sim2$ & &  &  &  & & &  \\
F(Ly$\alpha)=5$E-19 & &  &  &  & & & \\
\hline
%%%LARS-LAE & 5.26 & 9.51 & 6.00 & 3.34 & 3.00 & 5.00 & 7.84 \\error found June 2014
LARS-LAE & 5.26 & 6.67 & 6.00 & 3.34 & 1.33 & 2.67 & 5.34 \\
%Feldmeier2013, r$_{n}^{PSFsub}$ & &  &  &  & & 5.4 &  \\
Feldmeier2013, r$_{n}$ & &  &  &  & & 5.4 &  \\
LARS-LAE,PSF,F & 9.40 & 9.51 & 9.34 & 6.00 & 3.00 & 5.00 & 7.84 \\
Momose2014, r$_{n}$ & &  &  &  & & 7.9 &  \\
LARS-LAE,PSF,M & 9.21 & 9.34 & 9.01 & 5.67 & 2.84 & 4.84 & 7.67 \\
Steidel2011, r$_{n}^{LAEonly}$ & &  &  &  & & 28.4 &  \\
\hline
m$_{rest-UV}=30$ & &  &  &  & & & \\
\hline
LARS-LAE & 2.28 & 2.17 & 1.33 & 1.33 & 0.50 & 1.00 & 1.83 \\
LARS-LAE,PSF,F & 5.69 & 5.50 & 5.34 & $>12$ & 2.00 & 3.34 & 4.67 \\
LARS-LAE,PSF,M & 6.40 & 6.67 & 6.67 & 4.34 & 2.50 & 4.00 & 5.67 \\
Steidel2011, r$_{n}^{LAEonly}$ & &  &  &  & & 2.9 &  \\
\hline
\hline
$z\simeq5.7$ & &  &  &  & & & \\
F(Ly$\alpha)=5$E-19 & &  &  &  & & & \\
\hline
LARS-LAE & 2.40 & 4.70 & 6.34 & 3.52 & 1.17 & 2.35 & 4.35 \\
Momose2014, r$_{n}$ & &  &  &  & & 5.9 &  \\
LARS-LAE,PSF,M & 6.60 & 9.28 & 7.99 & 4.70 & 2.35 & 4.11 & 7.28 \\
\hline
m$_{rest-UV}=30$ & &  &  &  & & & \\
\hline
LARS-LAE & 1.28 & 1.88 & 1.41 & 0.94 & 0.35 & 0.82 & 1.53 \\
LARS-LAE,PSF,M & 5.59 & 5.99 & 7.52 & 4.46 & 2.00 & 3.29 & 4.93 \\
\hline
\end{tabular}
}
\tablefoot{Sizes for the LARS-LAE stack before and after convolving with \citet{Feldmeier2013} (PSF,F) and \citet{Momose2014} (PSF,M) ground-based PSF. The values correspond to the Ly$\alpha$ and rest-frame UV continuum images. For reference we report the e-folding parameters of the best-fit exponential profile, I(r), of the observed surface brightness by \citet{Feldmeier2013}, by \citet{Momose2014}, and by \citet{Steidel2011}, where I(r) $\propto$ exp(-r/r$_{n}$) and r $\gtrsim10$ kpc.}
\end{table*}

In the studies mentioned earlier the surface brightness of the circumgalactic
medium (CGM) was estimated by fitting an exponential curve to radial profiles that
excluded the first $2''$ ($\sim17$ kpc at $z\sim2$ and $\sim11$ kpc at
$z\sim5.7$).  Inside the first $2''$, emission from the interstellar medium (ISM) 
dominates over that from the CGM.  Moreover, the ground-based PSF will most
likely dominate on these (and smaller) scales. 

A fair comparison with high-$z$ observations should have been performed to radii larger than those
allowed by our HST observations, in which we mainly detected
Ly$\alpha$ photons coming also from the interstellar media. 
Looking at Fig. 9 of S11 and Fig. 3 of M14, we derived the high-$z$ Ly$\alpha$ profiles on scales of 10-20 ($z\sim2$) and 5-15 ($z\simeq5.7$) kpc, resembling the profiles typical of high-$z$ interstellar media, with some contribution from the PSF. We show them in Fig. \ref{convolved}, labelled as ``Steidel ISM'' and ``Momose ISM''.
The surface brightness of M14 stacked sample at $z\simeq2.2$ was more than one order of magnitude fainter than Steidel's and ours. LARS surface brightness profile was as bright as Steidel's one and steeper than S11 and M14 samples. The characteristic scale of a fitted exponential profile, I(r) $\propto$ exp$({-r/r_{n}})$ was r$_{n}=4$, almost half the value we calculated for S11 and M14 ISM profiles. At $z\sim5.7$ we obtained a similar slope profile as in M14, with r$_{n}\sim2.7$. 

%%%%%%%%%%%%%%%%%%%%%%%%%%%%%%%%%%%%%%%%%%%%%%%%%%%%%%%%%%%%%%%%%%%%%%%%%%%%
\section{Discussion}  
%%%%%%%%%%%%%%%%%%%%%%%%%%%%%%%%%%%%%%%%%%%%%%%%%%%%%%%%%%%%%%%%%%%%%%%%%%%%
\label{sec:discussion}

In Sec. \ref{sec:Loriginal}, we have presented the morphological properties of LARS galaxies; in Sec. \ref{sec:LARSz2}, we have described the results of the test of simulating LARS galaxies at $z\sim2$, following the methods explained in Sec \ref{sec:method}; in Sec \ref{sec:Stack}, we have performed the stacking of individual LARS galaxy frames to simulate their typical extended Ly$\alpha$ emission at $z\sim2$ and $z\simeq5.7$.  Here we discuss the main results.

\subsection{Local Universe LARS}
\label{sec:lowzLARS}

As described in Paper II, LARS galaxies are irregular, star-forming galaxies. Compared to the non-LAE LARS galaxies, the LARS-LAEs are younger and are characterized by lower star-formation rates, by Ly$\alpha$ escape fractions larger than 10\% (except L08), by low dust content in terms of the ratio H$\alpha$/H$\beta$, and by lower masses. 
Figs \ref{type1} and \ref{type2} tell us that the LARS galaxies have sizes, stellar masses, and rest-frame absolute magnitudes similar to those of Lyman break analogues, and that they are also comparable to $2<z<3$ star-forming galaxies. 
The stellar masses of the LARS galaxies tend to be larger than those estimated for LAEs at $z\sim2-3$. Even if most of the LARS-LAEs have Ly$\alpha$ luminosities twice fainter than those of high-$z$ Ly$\alpha$ emitters, they have stellar masses and sizes comparable to those of the subsample of the most massive LAEs from \citet{Guaita2011} and \citet{Bond2012}. Therefore, LARS galaxies are analogues of 10$^9$-10$^{11}$ M$_{\odot}$ high-$z$ star-forming galaxies; they also share some properties with the most massive Ly$\alpha$ emitters at high $z$.

The ratio between half-light radii estimated in the rest-frame optical and rest-frame UV have been adopted in the literature to identify the presence of multiple stellar populations. Within LARS, this ratio varies from 0.8 to 3, with a median value larger than 1 ($\sim1.4$). This shows that the young stellar populations are more localized than the old ones, consistent with the findings for $z\sim2$ SMGs \citep{Swinbank2010}, sBzK \citep{Yuma2012}, and star-forming galaxies at $1.4 < z < 2.9$ \citep{Bond2011}. 

The optical morphologies of LARS are typical of irregular or starburst galaxies and merging systems. They vary from the typical of early-stage mergers to closely-gathered clumps of intense star-formation. 
G and M20 are smaller when measured in Ly$\alpha$ than in the rest-frame UV. However, they are comparable when measured in the rest-frame optical and in Ly$\alpha$ (Fig. \ref{compcoHaliLa}). This indicates quantitatively that the Ly$\alpha$ emission of LARS galaxies is generally characterized by one component in a structure that tends to be more extended in Ly$\alpha$ than in the bright UV continuum and seems to follow the entire galaxy stellar populations. 
Also, the LARS-LAEs are the ones characterized by the highest concentration, lowest asymmetry and lowest clumpiness in Ly$\alpha$. A Ly$\alpha$ emitting galaxy with these properties would easily satisfy the EW(Ly$\alpha$) requirement adopted at high redshift to identify Ly$\alpha$ emitters. 

\begin{figure*} 
\centering
\includegraphics[width=18cm]{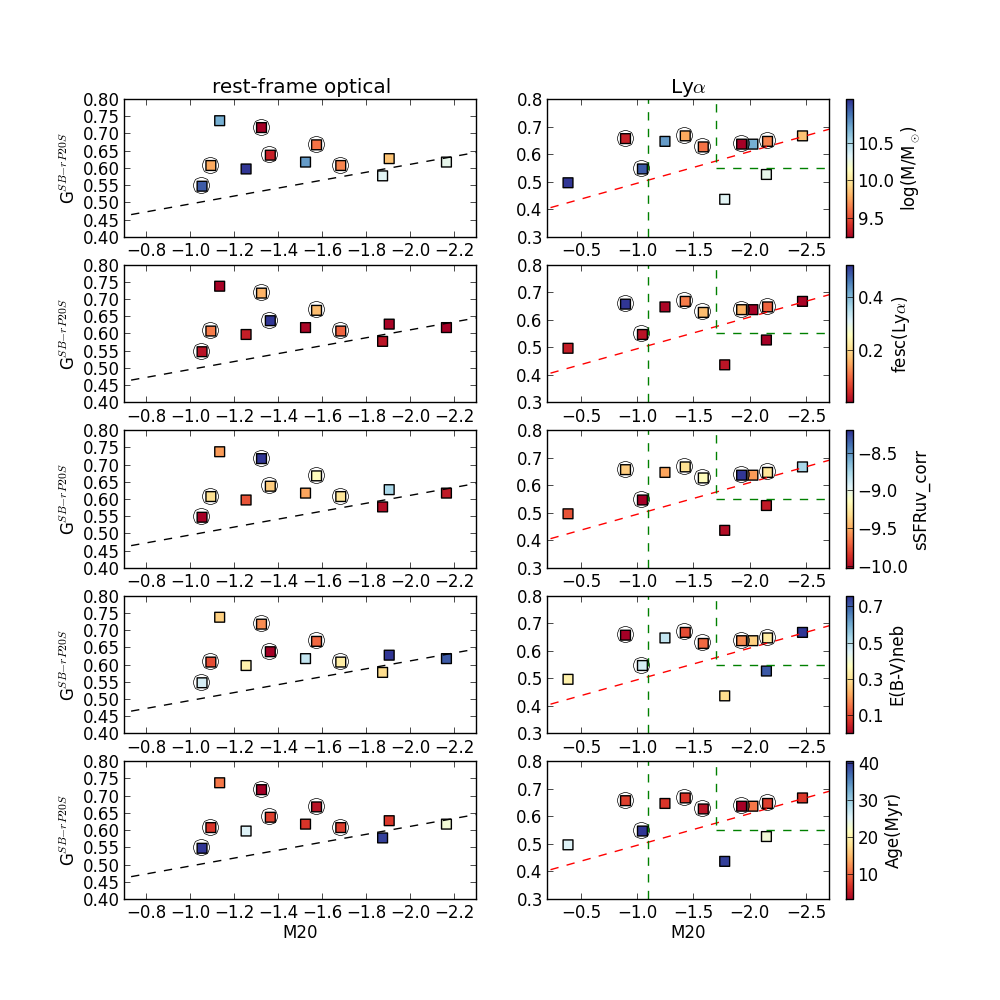}
\caption{G$^{SB-rP20S}$ vs M20 measured in the rest-frame optical ($left$ panels) and Ly$\alpha$ ($right$ panel) of the original LARS galaxies (squares). LARS-LAEs are indicated with open circles. In each raw the colour scale corresponds to an integrated physical property derived in Paper II (vertical colour bar). From the top raw the integrated physical properties are stellar mass, Ly$\alpha$ escape fraction, dust-corrected specific star formation rate, nebular reddening, and age. The dashed lines indicate the regions of separation between merging system, normal galaxies, and bulge-dominated systems as presented in Figs. \ref{LARSreg_location_coHa} and \ref{LARSreg_location_coLa}.}
\label{ConclusivePlot}
\end{figure*}

With the aim of identifying any continuum morphological property that characterizes Ly$\alpha$ emitters and star-forming galaxies, we study the correlations between morphological and physical parameters in Table \ref{tabST}.
The LARS galaxies characterized by larger EW(Ly$\alpha$) tend to be smaller in the rest-frame UV and optical than the other galaxies. They also present a more symmetric Ly$\alpha$ emission. The galaxies which are younger (Age $\leq10$ Myr), less massive (M$_{*}<10^{10}$ M$\odot$), and characterized by larger specific star-formation rate (dust-corrected sSFRuv $>-9.3$ yr$^{-1}$) tend to present larger Gini coefficient for a fixed M20 value. These are mainly the LARS-LAEs where the merging components appear separated, like in early-stage mergers. 
In Fig. \ref{ConclusivePlot} we show the G vs M20 diagram, in which the colour scale corresponds to the integrated physical parameters (Paper II). 

In Ly$\alpha$, LARS-LAEs tend to be composed of one bulge-like component and to harbour lower S and ellipticity. The other LARS galaxies are composed of patchy Ly$\alpha$ emission (see also Appendix B). Some LARS galaxies with large E(B-V)neb tend to be characterized by the lowest M20 in the rest-frame optical, typical of a bulge-like morphology. 
According to these results, it seems that early-stage mergers could be characterized by younger stellar populations and symmetrical, somewhat homogeneous Ly$\alpha$ emissions, which could satisfy high-$z$ LAE selections. 
Late-stage mergers could instead be characterized by turbulent star-formation episodes and, as a consequence, patchy Ly$\alpha$ emissions, just like the ones observed within one galaxy with numerous star-forming regions. 
This may indicate that the clumpy Ly$\alpha$ emitting galaxies observed at high redshift \citep{Bond2012,Gronwall2011,Shibuya2013} could also be experiencing early-stage merging events \citep{Cooke2010}. Of the twelve LARS galaxies, whose continuum morphologies are consistent with them being merging systems, six are LAEs. This does not mean that every observed LAE must necessarily be a merging system \citep{Shibuya2013,Law2012}. However, we can not verify the claim by \citet{Shibuya2013} that mergers are rare in LAEs with EW(Ly$\alpha)>100$ {\AA}, because LARS galaxies all present lower Ly$\alpha$ equivalent widths. 

\begin{table*}
\centering
\caption{Spearman coefficient and probability of the correlations between morphological and physical parameters}
\label{tabST}
\scalebox{0.9}{
\begin{tabular}{|c|c|c|c|c|c|c|c|c|}
\hline\hline 
parameter & rP20$^{ell}$  & G$^{SB-rP20S}$ & M20& C$^{minA}$ & A & S & ell & $\xi$(Ly$\alpha$/rest-UV) \\
\hline
\hline
rest-UV & & & & & &&& \\
\hline
EW(Ly$\alpha$) & {\bf{(-0.64,0.03)}} & (0.25,0.44)& (0.49,0.11) & (-0.20,0.53) & (0.22,0.50) & (-0.29,0.37) & (-0.11,0.74) & \\
fesc(Ly$\alpha$) & (-0.22,0.48)  &   (0.24,0.45) & (0.43,0.16) &(0.09,0.77) &  (0.38,0.22)  & (0.01,0.98) & (0.18,0.59) & \\
%rP20$^{ell}$(Ly$\alpha$) & (0.15,0.68) & & & & &&& \\
Age & (0.52,0.08) &(0.01,0.98) & (-0.41,0.18) & (0.11,0.75) & (0.20,0.54) & (0.68,0.01) & (0.30,0.34)  & \\
sSFRuv\_corr& (-0.57,0.05) &  (0.07,0.84) &  (0.25,0.43)  & (0.00,0.99)  & (-0.30,0.34)  & (-0.74,0.01)  & (-0.16,0.61) & \\
M$_{*}$ & (0.74,0.01) & (0.06,0.86) &  (-0.33,0.30)  & (-0.02,0.96) &(0.09,0.78)  &(0.63,0.03)& (0.16,0.62) & \\
E(B-V)neb & (0.01,0.97)&  (-0.44,0.15)  & (-0.20,0.53) &  (-0.23,0.48) &  (-0.31,0.32) & (-0.13,0.69) & (-0.42,0.17) & \\
\hline
\hline
rest-optical & & & & & &&& \\
\hline
EW(Ly$\alpha$) & {\bf{(-0.60,0.04)}} & (0.24,0.46) &  (-0.03,0.93) & (-0.57,0.05) & (-0.13,0.70)  & (-0.36,0.25) &(-0.18,0.57) & \\
fesc(Ly$\alpha$) & (-0.38,0.22) &  (0.11,0.73)  & (0.24,0.44)  & (-0.31,0.33) &  (0.01,0.98) & (0.03,0.93)  &(0.24,0.46) & \\
%rP20$^{ell}$(Ly$\alpha$) & (-0.01,0.99)  & & & & &&& \\
Age & (0.58,0.05) & {\bf{(-0.62,0.03)}}& (0.05,0.88)  & (-0.17,0.60)  &(-0.39,0.21)  & (0.26,0.42) & (0.01,0.98)  & \\
sSFRuv\_corr& (-0.69,0.01) & {\bf{(0.62,0.03)}} &  (-0.16,0.62)  & (0.08,0.81)  & (0.50,0.10)  &  (-0.27,0.40) & (-0.01,0.98) & \\
M$_{*}$ & (0.66,0.02) &{\bf{(-0.55,0.07)}}  & (0.24,0.46) &(0.21,0.51)  & (-0.13,0.70)& (0.36,0.25) & (0.14,0.66) & \\
E(B-V)neb & (0.08,0.81) &  (-0.32,0.31) & (-0.42,0.17)  & (0.20,0.54) & (-0.10,0.75)  &  (-0.18,0.58) & (-0.41,0.18) & \\
\hline
\hline
 Ly$\alpha$ & & & & & &&& \\
\hline
EW(Ly$\alpha$) & (-0.22,0.53) & (0.51,0.09)  &  (0.03,0.93) & (0.59,0.04) &  {\bf{(-0.70,0.01)}}  &(-0.61,0.04) & (-0.42,0.17)& (0.25,0.49) \\
fesc(Ly$\alpha$) & (-0.52,0.13) &(0.20,0.53) & (0.42,0.17)  & (0.18,0.57) & (-0.67,0.02)  &  (-0.76,0.00) & (-0.19,0.56) & (-0.22,0.53)\\
%rP20$^{ell}$(Ly$\alpha$) & & & & & &&& \\
Age & (-0.01,0.99) & (0.60,0.04) &(0.22,0.50) &(-0.32,0.31) &{\bf{(0.61,0.04)}} & {\bf{(0.84,0.00)}} & {\bf{(0.57,0.05)}} & (-0.24,0.51)\\
sSFRuv\_corr& (0.02,0.96)&  (0.65,0.02) & (-0.38,0.23)  & (0.27,0.40)  & {\bf{(-0.56,0.06)}}   &{\bf{(-0.90,0.00)}} & {\bf{(-0.54,0.07)}}  &(0.49,0.15) \\
M$_{*}$ & (-0.03,0.93) & (-0.45,0.14) &  (0.31,0.33) & (-0.50,0.10) &  {\bf{(0.76,0.00)}} &  {\bf{(0.80,0.00)}}  & {\bf{(0.56,0.06)}} & (-0.47,0.17)\\
E(B-V)neb &(0.53,0.12) &  (-0.19,0.55) & (-0.34,0.29) &(0.02,0.95) &  (0.52,0.08) & (0.57,0.05) &(-0.06,0.85) & (0.31,0.38) \\
\hline
\end{tabular}
}
\tablefoot{The number pairs in the table correspond to the Spearman coefficient, r, and probability, p, for the correlations between physical (column, y) and morphological (row, x) parameters, measured in the rest-frame UV, optical, and Ly$\alpha$ images. For instance, for x=G$^{SB-rP20S}$ and y=sSFRuv\_corr, the Spearman(x,y) calculation tells us that the probability that x and y are uncorrelated datasets is only p=3\% and there is a positive (r $=0.62$) correlation. 
%y=physical
%x=morphol
Bold face indicates the most significant correlations.
}
\end{table*}

\subsection{LARS galaxies simulated to be at $z\sim2$}
\label{sec:highzLARS}

To investigate the detectability of LARS galaxies and haloes at high redshift, we performed simulations. The LARS galaxies were resampled at $z \sim 2$ as described in Sec 2.3. We find 
that the effects of pixel resampling and simulated survey noise are dependent on the irregular structures of each individual galaxy.
 
First of all we defined the detection limits in the continuum and line images (Table \ref{tab:limits}). 
In a survey shallower than the detection limit, the background noise dominates the continuum and line surface brightness profiles. Interestingly all LARS galaxies would be detected in the rest-frame UV in surveys like HUDF09 and GOODS, and at the rest-frame optical in a survey like CANDELS (deep), if located at $z\sim2$. 
However, even more interesting is that Ly$\alpha$ emission extended up to 5 kpc would be visible in 70\% of the sample at a 10$\sigma$ depth of 3E$-18$ erg sec$^{-1}$ cm$^{-2}$ (Table \ref{tab:REGsize} and Appendix C). 
The LARS galaxies characterized by the faintest integrated Ly$\alpha$ flux show an even more extended emission. This should be taken into account when designing Ly$\alpha$ spectroscopic observations (see also the discussion in Paper I), which aim to detect as much of the Ly$\alpha$ flux as possible. In the current Ly$\alpha$ surveys (typically characterized by F(Ly$\alpha)>2$E$-17$ erg sec$^{-1}$ cm$^{-2}$) only L14, the brightest LARS-LAE, would be detected.
Even if the Ly$\alpha$ size can be as large as 5 times the UV, the extremely detailed Ly$\alpha$ structures (Figs. \ref{RGB} and \ref{L01imagePSF}) could only be identified in the original HST-resolution images.

Second of all we quantified the morphological parameters of high-$z$ LARS galaxies. 
The z2LARS-LAEs tend to be smaller in the rest-frame UV, optical, and H$\alpha$ than the other LARS galaxies. This is consistent with the observed high-$z$ findings \citep{Gronwall2011,Bond2012}. Also, the z2LARS-LAEs present a large range of asymmetry values in UV and optical, but do show symmetric morphologies of Ly$\alpha$ emission. 

The low surface brightness structures tend to disappear within the background in shallow surveys, making a clumpier and sometimes more elliptical galaxy in the continuum. As a consequence the ellipticity values we measured in the rest-frame UV and optical of z2LARS-LAEs increase as the survey depth decreased. Also, the continuum clumpiness tends to significantly increase. Some of the high-$z$ observations of large ellipticity and clumpy systems could be explained in terms of depth \citep{Gronwall2011,Bond2012}, but we cannot explain the decrease in the merger fraction and ellipticity for EW(Ly$\alpha)>100$ {\AA} LAEs found by \citet{Shibuya2013}. 
However, the clumpiness measured in Ly$\alpha$ also increases for decreasing survey depth (Figs. \ref{stampsDEEP1}, \ref{stampsDEEP2}, and \ref{stampsDEEP3}), making the integrated EW more difficult to estimate. This happens for the LARS-LAEs as well, which eventually become close to impossible to detect as Ly$\alpha$ emitters. In our sample L08 is the LARS galaxy with the lowest rest-frame optical ellipticity in the original, high-$z$ simulated, and shallow-survey images. It is a massive face-on irregular galaxy in the rest-frame optical, with multiple star-formation clumps seen in the rest-frame UV (Fig. \ref{stampsDEEP2}), but it is characterized by the lowest EW(Ly$\alpha$) among the LARS-LAEs. 

Continuum G and M20 values are preserved after pixel resampling and adding noise (see also Appendix A). Therefore, we adopted these two parameters for characterizing LARS galaxies at high redshift (Fig. \ref{LARSreg_location_coHa}). LARS galaxies have a morphology consistent with merging systems even when simulated at high $z$. Some z2LARS-LAEs have both rest-frame optical and UV morphologies consistent with being mergers (Fig. \ref{LARSreg_location_coLa}). 

The asymmetry we estimated decreases after adding noise. 
A merger information calculated just adopting a large-asymmetry criterion could lead to a mis-interpretation of our sample in a shallow survey. This was also noticed in \citet{Shibuya2013,Gronwall2011} for low signal-to-noise sources.

In general the Ly$\alpha$ morphology tends to be significantly affected in shallow surveys, because Ly$\alpha$ detailed structures are very sensitive to depth and resolution. 

%%%%%%%%%%%%%%%%%%%%%%%%%%%
\subsection{Ly$\alpha$ haloes of LARS galaxies and their implications}
\label{z2LARShaloes}
%%%%%%%%%%%%%%%%%%%%%

There is still open debate about the conditions necessary for the formation of
Ly$\alpha$ haloes in high-$z$ galaxies, but it seems that HI scattering is the
main factor at the scale of LARS galaxies (Paper II).  At high redshift,
Ly$\alpha$ halo studies have been performed in stacked data obtained from
various samples of galaxies.  To be able to compare with high-$z$ results, we
simulated how the Ly$\alpha$ haloes of LARS galaxies would appear at $z\sim2$ and
$z\simeq5.7$. This was performed by stacking subsamples of our galaxies and
assuming a range of survey depths. In this test we have the advantage of knowing
the original morphology and halo profile. 

By simply examining the RGB mosaics in Fig. 1, we see that stacking images with
very different irregular structures can be very complicated, but it is commonly
done at high redshift to increase the signal-to-noise.  
Background noise, survey depth, and ground-based Point Spread Function are the primary limits of the detection of Ly$\alpha$ haloes. 
We find that a depth comparable to the M14
survey is ideal to recover these haloes at both $z\sim2$ and $z\simeq5.7$ (see
Sec.  \ref{sec:literature}, Fig.  \ref{liLa_LAE_av_med}).  
The stacked Ly$\alpha$ surface
brightness profile of the LARS-LAEs is peaked in the centre, whereas the
LARSaverage stack contains contributions from more diverse (asymmetric, patchy,
and eventually more extended) Ly$\alpha$ morphologies; the result is a more
irregular and extended profile.  It seems reasonable that the conditions under
which a galaxy may emit Ly$\alpha$ (mainly Ly$\alpha$ flux concentrated around
the UV-bright knots of star formation), would produce a consistent surface
brightness profile. However, variations in the dust and HI contents and distributions,
which may cause Ly$\alpha$ photons escape along the line of sight
(EW(Ly$\alpha)>0$ {\AA}, but not necessarily $>20$ {\AA}), would produce a more
complex, patchy, and extended emission. This is seen in the LARSaverage stack
and also in the stack observed in high-$z$ LBGs.  

Ground-based observations of high-$z$ galaxies allow us to construct Ly$\alpha$
profiles which extend much farther from the centre (see S11 and M14) than low-$z$ galaxies observed with the Hubble
Space Telescope. The continuum subtraction procedure \citep{Hayes2009}, described in Paper II to isolate Ly$\alpha$ emission, takes advantage of the HST resolution to provide a detailed Ly$\alpha$ mapping within and just outside the interstellar medium. Within the first 10 kpc Ly$\alpha$ scattering already begins to produce halo-like structures in LARS galaxies, while it is at radii above 10-20 kpc that high-$z$ studies are performed because no spatial information is available inside the PSF.  A fair comparison between low and
high-$z$ observations could only be made by applying the same procedure to extract Ly$\alpha$ from the rest-frame UV emission at high redshift and by investigating the medium on the same scales and the same PSF conditions.  

We convolved our HST images with typical ground-based PSFs and compared our
smoothed Ly$\alpha$ profile with ground-based observations at comparable scales.
Even in this case the $z\sim2$ LARS Ly$\alpha$ stacked profile is steeper than the one
derived for the complete sample by M14 and for the LAE-only subsample of S11.
In Feldmeier's survey, the LARS \lya\ profile would be indistinguishable from
the ground-based PSF on scales larger than 6 kpc. At $z\sim5.7$, LARS Ly$\alpha$ profile is as steep as the one from M14, which is brighter than the profile they obtained at $z\sim2$. 

With LARS we cannot probe as large scale as the current
high-$z$ observations do, due to smaller field-of-view of HST.
However, we could still expect some differences between LARS and Feldmeier's sample because of observational depth, and between LARS and the S11 sample because LARS are not
located in overdense regions.  Also, we may expect differences between LARS and M14 sample at large scale due to the difference in physical properties of the two samples.

The $z\simeq2.2$ LAE sample studied by Momose et al. is characterized by dust reddening (E(B-V) $<0.1$), metallicity (Z $\sim0.2$ Z$_{\odot}$), and stellar mass (M$_{*} <10^{10}$ M$_{\odot}$) lower than LARS. Since Ly$\alpha$ photons are sensitive to the
presence of dust grains and to the scattering on neutral hydrogen (e.g., Paper
II), Ly$\alpha$ morphology (extension and features) is expected to depend on the
larger dust and HI contents \citep[Paper III and][]{Rivera2014}.  The dust
grains, able to absorb Ly$\alpha$ photons close to the knots of star-formation,
also prevent their escape at large scales, where HI scattering plays the role of
making haloes.  There are other phenomena, such as gas kinematics, we are
investigating within the LARS survey, which could favour the escape of
Ly$\alpha$ photons from the HII regions and ultimately allow the formation of
Ly$\alpha$ haloes. \citet{Rivera2014}, Duval et al., and Orlitov\'a et al. (in prep) are dedicated to studying the gas kinematics in LARS galaxies from HST spectroscopy.

%%%%%%%%%%%%%%%%%%%%%%%%%%%%%%%%%%%%%%%%%%%%%%%%%%%%%%%%%%%%%%%%%%%%%%%%%%%%
\section{Conclusions}
\label{sec:conclusions}
%%%%%%%%%%%%%%%%%%%%%%%%%%%%%%%%%%%%%%%%%%%%%%%%%%%%%%%%%%%%%%%%%%%%%%%%%%%%

This paper is number IV of a series. In this work we have characterized and quantitatively studied the morphology of a sample of starburst galaxies at $z<0.2$: the Lyman alpha reference sample, LARS.

\begin{itemize}
\item LARS galaxies have continuum sizes and stellar masses similar to those of local Lyman break analogues and $2 < z < 3$ star-forming galaxies. The stellar mass and luminosities also match the two samples, respectively. Therefore, LARS galaxies can be studied as a reference of 10$^9$-10$^{11}$ M$_{\odot}$ high-$z$ star-forming galaxies; they also share some properties with the most massive Ly$\alpha$ emitters at high redshift.

\item The rest-frame optical morphology of LARS galaxies is the typical of merging systems. This is also valid for the LAEs within LARS. 

\item For the first time we were able to quantify the morphology of Ly$\alpha$ emission. LARS-LAEs are on average characterized by more concentrated and symmetrical, while LARS non-LAEs can present patchier and irregular Ly$\alpha$ emissions. LARS-LAEs are more compact in Ly$\alpha$, even when regridded to high redshift.

\item We have simulated LARS galaxies at high redshift and explored their detection: all LARS galaxies would be detected in the continuum in current deep surveys, but they would not be easily detected in the current Ly$\alpha$ surveys at $z\sim2$. %yes L14

\item In a shallow survey, it is the morphology of Ly$\alpha$ that is most affected by background noise, because the detailed Ly$\alpha$ structures strongly depend on depth and resolution. This may affect high-$z$ Ly$\alpha$ observations.

\item The measured ellipticity and clumpiness tend to increase in shallow surveys for most of the LARS galaxies. This could explain some of the high-$z$ observations of large ellipticity and clumpy systems in LAE samples.

\item We stacked the Ly$\alpha$ images of LARS galaxies simulated at high redshift. The LARS-LAE stack is peaked in the centre, whereas the LARSaverage stack contains contributions from more diverse Ly$\alpha$ morphologies resulting in a more irregular and extended profile. Variations in the dust and HI contents and distributions may produce more complex, patchy, and extended emission like the one seen in the LARSaverage stack and also in the stack observed in high-$z$ LBGs.  

\item The Ly$\alpha$ haloes we study in LARS galaxies probe much-smaller-scale media than high-$z$ observations. We find that LARS-halo-profile slope is steeper than $z\sim2$ and as steep as with $z\simeq5.7$ observations at $\sim$10-kpc scales, after applying ground-based PSF.   
\end{itemize}

A sample like LARS, at slightly larger redshift, could allow studying circumgalactic medium-scale haloes and relating them to other galaxy properties. Physical properties (as presented in Paper II), HI-mass estimations (from Paper III), and kinematics \citep[analysed in][Duval et al., and Orlitov\'a et al.]{Rivera2014}
%(analysed in \citet{Rivera2014}, Duval et al., and Orlitov\'a et al.) 
already helped in clarifying the Ly$\alpha$-photon propagation within the interstellar medium. With the large-scale halo information we would be able to also investigate the mechanisms that transport Ly$\alpha$ photons from the interstellar to the circumgalactic medium.

%%%%%%%%%%%%%%%%%%%%%%%%%%%%%%%%%%%%%%%%%%%%%%%%%%%%%%%%%%%%%%%%%%%%%%%%%%%%
\begin{acknowledgements}
We thank the Referee, Andrew Humphrey, for giving us useful comments which have improved our manuscript.
We would also like to thank Nick Bond, Claudia Scarlata, Masami Ouchi and his research group, Suraphong Yuma, Rieko Momose, Takatoshi Shibuya, Yoshiaki Ono, and Caryl Gronwall for useful discussions; Shiyin Shen and Christopher J. Conselice for key inputs on size and morphological parameter measurements. L.G. sincerely thanks Eric Gawiser and Nelson Padilla for their $always$ useful comments and support.
G.{\"O}. acknowledges financial support from the Swedish Research Council (VR) and the
Swedish National Space Board (SNSB).
J.M.M.H. has been funded by MINECO grant AYA2012-39362-C02-01.
H.O.F. was funded by a post-doctoral UNAM grant and is currently granted by a C\'atedra CONACyT para J\'ovenes Investigadores. 
I.O. has been supported by GACR grant 14-20666P of Czech Science Foundation, and long-term institutional grant RVO:67985815. 
P.L. acknowledges support from the ERC-StG grant EGGS-278202. Dark Cosmology Centre is funded by DNRF. 
D.K. has been financially supported by the CNES (Centre National d’Etudes Spatiales)/ CNRS 131425 grant. 
H.A. was supported by the European Research Council (ERC) advanced grant ``Light on the Dark'' (LIDA) and the Centre National d'Etudes Spatiales (CNES). 
\end{acknowledgements}

%%%%%%%%%%%%%%%%%%%%%%%%%%%%%%%%%%%%%%%%%%%%%%%%%%%%%%%%%%%%%%%%%%%%%%%%%%%%
%% references
\bibliographystyle{aa}    %% bibliography style file aa.bst from A&A
\bibliography{biblio}        %% your Bibtex entries copied from ADS

%%%%%%%%%%%%%%%%%%%%%%%%%%%%%%%%%%%%%%%%%%%%%%%%%%%%%%%%%%%%%%%%%%%%%%%% 

\clearpage
\onecolumn
\longtab{5}{
\begin{table*}
\caption{Size of high-$z$ simulated LARS galaxies}
\label{tab:REGsize}
\centering
\scalebox{0.9}{
\begin{tabular}{|c|c|c|c|c|c|c|}
\hline\hline 
%(1) & (2) & (3) & (4) & (5) &(6) & (7)\\ %& (9) &(10) & (11) & (12) & (13) & (14) & (15) &(16) & (17)& (18)\\
%\hline
%LARS &rP20$^{ell}$ & rP20$^{circ}$ & rP20$^{minA}$ & r$_{20}^{ell}$ & r$_{50}^{ell}$ & r$_{80}^{ell}$ \\
\hline
continuum & rP20$^{ell}$ & $\sigma$rP20$^{ell}$ & rP20$^{minA}$ & $\sigma$rP20$^{minA}$ & r$_{50}$ & $\sigma$r$_{50}$  \\
 & kpc & kpc & kpc & kpc & kpc & kpc  \\
\hline
line  & $\xi$(rP20$^{ell})$ & $\sigma \xi$(rP20$^{ell})$ & $\xi$(rP20$^{minA})$ & $\sigma \xi$(rP20$^{minA}$) & $\xi$(r$_{50}$) & $\sigma \xi$(r$_{50}$)  \\
\hline
\hline
Bond2010($z\sim3.1$) &  &  &  &  &  & \\
&  &  &  &  &  & \\
rest-UV$^{F606W}$ &  &  &  &  & 1.3 & \\
&  &  &  &  &  & \\
Ly$\alpha^{F502N}$ &  &  &  &  & $\sim$1 & \\
\hline
Finkelstein2011($z\sim4.5$) &  &   &  &  & & \\
&  &  &  &  &  & \\
rest-UV$^{F775W}$\_1 &  &   &  &  & 0.67$^{0.07}_{0.07}$ &\\
&  &  &  &  &  & \\
Ly$\alpha^{F658N}$\_1 &  &   &  &  & 1.1$^{3.5}_{0.4}$  & \\
\hline
&  &  &  &  &  & \\
rest-UV$^{F775W}$\_2 &  &   &  &  & 1.07 $^{0.27}_{0.20}$ &\\
&  &  &  &  &  & \\
Ly$\alpha^{F658N}$\_2 &  &   &  &  & 1.3 $^{1.1}_{0.6}$ &\\
\hline
&  &  &  &  &  & \\
rest-UV$^{F775W}$\_3 &  &   &  &  & 0.67 $^{0.07}_{0.00}$ & \\
&  &  &  &  &  & \\
Ly$\alpha^{F658N}$\_3 &  &   &  &  & 2.0 $^{1.3}_{0.7}$ &\\
\hline
\hline
L01 &  &  &  &  &  & \\
rest-UV &  &  &  &  &  & \\
30 & 2.58 & 0.20 & 1.33 & 0.33 & 1.17 & 0.17 \\
29 & 2.16 & 0.41 & 1.32 & 0.34 & 1.08 & 0.19 \\
28 & 1.83 & 0.47 & 1.32 & 0.50 & 0.94 & 0.26 \\
\hline
rest-optical &  &  &  &  &  & \\
29 & 4.38 & 0.20 & 3.18 & 0.37 & 2.00 & 0.17 \\
28 & 4.51 & 0.29 & 3.19 & 0.46 & 2.01 & 0.17 \\
27 & 4.37 & 0.49 & 3.17 & 0.80 & 1.98 & 0.20 \\
26 & 4.25 & 0.87 & 3.35 & 1.42 & 1.88 & 0.28 \\
\hline
Ly$\alpha$ &  &  &  &  &  & \\
5E-19 & 2.20 & 0.22 & 2.57 & 0.37 & 1.82 & 0.28 \\
3E-18 & 2.08 & 0.28 & 2.05 & 0.59 & 1.73 & 0.21 \\
8E-18 & 1.78 & 0.52 & 1.74 & 0.89 & 1.59 & 0.39 \\
\hline
H$\alpha$ &  &  &  &  &  & \\
2E-19 & 0.47 & 0.04 & 0.37 & 0.12 & 0.53 & 0.09 \\
6E-19 & 0.49 & 0.05 & 0.34 & 0.11 & 0.57 & 0.09 \\
1E-18 & 0.48 & 0.05 & 0.35 & 0.12 & 0.54 & 0.09 \\
3E-18 & 0.45 & 0.06 & 0.35 & 0.12 & 0.51 & 0.10 \\
%1E-17 & 290.273 & 290.273 & 384.464 & 384.464 & 535.327 & 535.327 \
\hline
\hline
L02 &  &  &  &  &  & \\
rest-UV &  &  &  &  &  & \\
30 & 3.57 & 1.42 & 1.86 & 0.45 & 2.10 & 0.91 \\
29 & 3.53 & 1.39 & 2.54 & 1.43 & 2.09 & 0.92 \\
\hline
rest-optical &  &  &  &  &  & \\
29 & 10.62 & 0.89 & 8.82 & 1.45 & 4.16 & 0.25 \\
28 & 9.76 & 1.48 & 6.89 & 2.02 & 4.08 & 0.40 \\
27 & 7.07 & 2.30 & 5.34 & 2.38 & 3.48 & 0.71 \\
\hline
Ly$\alpha$ &  &  &  &  &  & \\
5E-19 & 3.12 & 0.31 & 2.43 & 0.63 & 2.53 & 0.16 \\
3E-18 & 2.63 & 0.39 & 1.72 & 0.68 & 2.40 & 0.41 \\
8E-18 & 1.77 & 0.80 & 1.24 & 0.57 & 1.71 & 0.71 \\
\hline
H$\alpha$ &  &  &  &  &  & \\
2E-19 & 0.16 & 0.09 & 0.13 & 0.05 & 0.20 & 0.14 \\
6E-19 & 0.23 & 0.03 & 0.23 & 0.04 & 0.33 & 0.04 \\
1E-18 & 0.24 & 0.04 & 0.23 & 0.05 & 0.34 & 0.05 \\
\hline
\hline
L03 &  &  &  &  &  & \\
rest-UV &  &  &  &  &  & \\
30 &1.81 & 0.87 & 1.38 & 0.40 & 0.77 & 0.31 \\
29 & 2.15 & 4.58 & 1.79 & 3.65 & 0.89 & 1.38 \\
\hline
rest-optical &  &  &  &  &  & \\
29 & 9.05 & 2.23 & 5.82 & 0.43 & 3.63 & 2.00 \\
28 & 7.95 & 1.08 & 4.92 & 0.83 & 3.09 & 0.98 \\
27 & 7.23 & 1.71 & 4.35 & 0.83 & 2.98 & 1.40 \\
26 & 6.32 & 1.17 & 4.06 & 0.89 & 2.47 & 0.30 \\
25 & 5.47 & 1.57 & 4.06 & 0.89 & 2.50 & 0.54 \\
\hline
Ly$\alpha$ &  &  &  &  &  & \\
5E-19 & 7.90 & 4.45 & 8.97 & 11.29 & 9.19 & 9.00 \\
3E-18 & 5.21 & 11.83 & 4.00 & 9.81 & 8.00 & 16.92 \\
\hline
H$\alpha$ &  &  &  &  &  & \\
2E-19 & 0.20 & 0.02 & 0.29 & 0.06 & 0.18 & 0.05 \\
6E-19 & 0.20 & 0.02 & 0.29 & 0.06 & 0.18 & 0.05 \\
1E-18 & 0.21 & 0.02 & 0.29 & 0.06 & 0.18 & 0.05 \\
3E-18 & 0.22 & 0.02 & 0.29 & 0.08 & 0.21 & 0.05 \\
\hline
\end{tabular}
}
\tablefoot{Sizes in the continuum are reported as rP20$^{ell}$. 
Owing pixel re-sampling, the Petrosian radius can have a variation of 20\% with respect to the value estimated in the original image. 
The Ly$\alpha$ effective radii of two existent individual-source measurements are also shown \citep{Bond2010,Fin2011}. 
}
\end{table*}

\begin{table*}
\caption{Size of high-$z$ simulated LARS galaxies}
\label{tab:REGsize2}
\centering
\scalebox{0.9}{
\begin{tabular}{|c|c|c|c|c|c|c|}
\hline\hline 
continuum & rP20$^{ell}$ & $\sigma$rP20$^{ell}$ & rP20$^{minA}$ & $\sigma$rP20$^{minA}$ & r$_{50}$ & $\sigma$r$_{50}$  \\
 & kpc & kpc & kpc & kpc & kpc & kpc  \\
\hline
line  & $\xi$(rP20$^{ell})$ & $\sigma \xi$(rP20$^{ell})$ & $\xi$(rP20$^{minA})$ & $\sigma \xi$(rP20$^{minA}$) & $\xi$(r$_{50}$) & $\sigma \xi$(r$_{50}$)  \\
\hline
\hline
L05&  &  &  &  &  & \\
rest-UV &  &  &  &  &  & \\
30 & 2.50 & 0.17 & 2.33 & 0.34 & 1.32 & 0.17 \\
29 & 2.54 & 0.18 & 2.26 & 0.40 & 1.26 & 0.19 \\
28 & 2.66 & 0.22 & 2.26 & 0.69 & 1.33 & 0.19 \\
27 & 2.81 & 0.67 & 2.46 & 1.89 & 1.38 & 0.22 \\
\hline
rest-optical  &  &  &  &  &  & \\
29 & 2.73 & 0.20 & 2.33 & 0.33 & 1.27 & 0.19 \\
28 & 2.77 & 0.22 & 2.35 & 0.35 & 1.36 & 0.18 \\
27 & 2.87 & 0.27 & 2.57 & 0.61 & 1.46 & 0.22 \\
26 & 3.02 & 0.41 & 2.43 & 1.26 & 1.52 & 0.24 \\
\hline
Ly$\alpha$ &  &  &  &  &  & \\
5E-19 & 1.96 & 0.10 & 1.76 & 0.16 & 1.75 & 0.13 \\
3E-18 & 1.93 & 0.21 & 1.71 & 0.33 & 1.74 & 0.15 \\
8E-18 & 1.96 & 0.37 & 1.76 & 0.47 & 1.86 & 0.23 \\
\hline
H$\alpha$ &  &  &  &  &  & \\
2E-19 & 0.92 & 0.06 & 1.00 & 0.20 & 1.05 & 0.13 \\
6E-19 & 0.93 & 0.07 & 0.93 & 0.18 & 1.13 & 0.15 \\
1E-18 & 0.96 & 0.07 & 0.98 & 0.31 & 1.15 & 0.14 \\
3E-18 & 1.03 & 0.11 & 1.03 & 0.26 & 1.24 & 0.15 \\
\hline
\hline
L07 &  &  &  &  &  & \\
rest-UV &  &  &  &  &  & \\
30 & 1.52 & 0.18 & 1.00 & 0.34 & 0.67 & 0.17 \\
29 & 1.57 & 0.20 & 1.05 & 0.36 & 0.67 & 0.17 \\
28 & 1.57 & 0.24 & 1.12 & 0.39 & 0.72 & 0.18 \\
\hline
rest-optical  &  &  &  &  &  & \\
29 & 4.14 & 0.25 & 3.34 & 0.34 & 1.72 & 0.19 \\
28 & 4.02 & 0.21 & 3.39 & 0.38 & 1.67 & 0.17 \\
27 & 4.21 & 0.36 & 3.31 & 0.78 & 1.73 & 0.20 \\
26 & 3.99 & 0.65 & 3.00 & 1.12 & 1.62 & 0.27 \\
\hline
Ly$\alpha$ &  &  &  &  &  & \\
5E-19 & 4.54 & 1.30 & 3.97 & 0.39 & 4.15 & 2.19 \\
3E-18 & 3.59 & 0.59 & 3.85 & 0.84 & 3.22 & 0.39 \\
8E-18 & 3.12 & 0.72 & 3.45 & 1.28 & 3.08 & 0.58 \\
\hline
H$\alpha$ &  &  &  &  &  & \\
6E-19 & 0.32 & 0.04 & 0.30 & 0.10 & 0.48 & 0.10 \\
wE-19 & 0.32 & 0.04 & 0.30 & 0.10 & 0.48 & 0.10 \\
1E-18 & 0.32 & 0.04 & 0.31 & 0.10 & 0.47 & 0.10 \\
3E-18 & 0.34 & 0.08 & 0.33 & 0.12 & 0.48 & 0.10 \\
\hline
\hline
L08 &  &  &  &  &  & \\
rest-UV  &  &  &  &  &  & \\
30 & 7.94 & 2.62 & 6.55 & 1.13 & 4.40 & 1.69 \\
29 & 7.57 & 2.13 & 5.92 & 2.35 & 3.45 & 1.08 \\
28 & 6.67 & 2.05 & 5.85 & 2.75 & 3.71 & 0.99 \\
\hline
rest-optical  &  &  &  &  &  & \\
29 & 10.32 & 1.55 & 8.67 & 0.38 & 4.48 & 1.40 \\
28 & 9.86 & 1.89 & 8.53 & 0.54 & 4.33 & 1.64 \\
27 & 9.51 & 1.97 & 8.29 & 0.69 & 4.06 & 1.40 \\
26 & 9.17 & 1.82 & 7.67 & 0.96 & 3.94 & 1.28 \\
25 & 8.17 & 1.28 & 7.67 & 0.96 & 3.77 & 0.27 \\
\hline
Ly$\alpha$ &  &  &  &  &  & \\
5E-19 & 2.23 & 0.16 & 1.31 & 0.37 & 1.58 & 0.12 \\
3E-18 & 1.35 & 0.44 & 0.73 & 0.28 & 1.34 & 0.39 \\
8E-18 & 0.76 & 1.37 & 0.53 & 1.59 & 0.82 & 1.48 \\
\hline
H$\alpha$
2E-19 & 0.67 & 0.03 & 0.64 & 0.08 & 0.86 & 0.07 \\
6E-19 & 0.63 & 0.03 & 0.57 & 0.05 & 0.80 & 0.04 \\
1E-18 & 0.63 & 0.04 & 0.56 & 0.05 & 0.79 & 0.04 \\
3E-18 & 0.95 & 0.27 & 0.60 & 0.11 & 1.13 & 0.37 \\
\hline
\end{tabular}
}
\tablefoot{continuation of Tab. \ref{tab:REGsize}. Sizes in Ly$\alpha$ and H$\alpha$ images are given as $\xi$(Ly$\alpha$) = $\frac{size(Ly\alpha)}{size(rest-UV)}$ and $\xi$ (H$\alpha$) = $\frac{size(H\alpha)}{size(rest-optical)}$, where the sizes are rP20$^{ell}$, rP20$^{minA}$, and r50. 
%$\xi=$ rP20$^{ell}_{Ly\alpha}$/rP20$^{ell}_{rest-UV}$ and $\xi=$rP20$^{ell}_{H\alpha}$/rP20$^{ell}_{rest-optical}$. 
$\xi$(Ly$\alpha$) is the most meaningful due to the resonant nature of Ly$\alpha$.}
\end{table*}

\begin{table*}
\caption{Size of high-$z$ simulated LARS galaxies}
\label{tab:REGsize3}
\centering
\scalebox{0.9}{
\begin{tabular}{|c|c|c|c|c|c|c|}
\hline\hline 
continuum & rP20$^{ell}$ & $\sigma$rP20$^{ell}$ & rP20$^{minA}$ & $\sigma$rP20$^{minA}$ & r$_{50}$ & $\sigma_{r50}$  \\
 & kpc & kpc & kpc & kpc & kpc & kpc  \\
\hline
line  & $\xi$(rP20$^{ell})$ & $\sigma \xi$(rP20$^{ell})$ & $\xi$(rP20$^{minA})$ & $\sigma \xi$(rP20$^{minA}$) & $\xi$(r$_{50}$) & $\sigma \xi$(r$_{50}$)  \\
\hline
\hline
L09 &&&&&&\\
rest-UV &  &  &  &  &  & \\
30& 20.34 & 3.32 & 2.18 & 1.31 & 10.34 & 1.50 \\
29 & 5.60 & 2.65 & 5.67 & 2.57 & 3.19 & 1.84 \\
28 & 5.09 & 3.56 & 1.41 & 1.03 & 2.34 & 2.19 \\
\hline
rest-optical  &  &  &  &  &  & \\
29 & 21.71 & 0.41 & 18.01 & 0.33 & 12.17 & 0.17 \\
28 & 21.81 & 0.80 & 18.08 & 0.49 & 12.44 & 0.23 \\
27 & 15.72 & 8.35 & 10.87 & 5.30 & 9.00 & 4.68 \\
\hline
Ly$\alpha$ &  &  &  &  &  & \\
5E-19 & 1.27 & 0.04 & 8.24 & 0.77 & 1.61 & 0.04 \\
\hline
H$\alpha$ &  &  &  &  &  & \\
2E-19 & 0.66 & 0.19 & 0.19 & 0.06 & 0.50 & 0.14 \\
6E-19 & 0.80 & 0.12 & 0.17 & 0.06 & 0.66 & 0.09 \\
1E-18 & 0.73 & 0.25 & 0.19 & 0.08 & 0.67 & 0.31 \\
3E-18 & 0.43 & 0.19 & 0.13 & 0.05 & 0.37 & 0.22 \\
1E-17 & 0.19 & 0.08 & 0.10 & 0.04 & 0.14 & 0.06 \\
\hline
\hline
L10 &  &  &  &  &  & \\
rest-UV &  &  &  &  &  & \\
30 & 4.68 & 0.43 & 3.23 & 0.41 & 2.04 & 0.18 \\
29 & 4.64 & 0.66 & 3.34 & 0.80 & 2.10 & 0.23 \\
28 & 4.39 & 1.19 & 3.45 & 1.56 & 2.10 & 0.37 \\
\hline
rest-optical  &  &  &  &  &  & \\
29 & 9.65 & 0.36 & 6.37 & 0.37 & 4.08 & 0.20 \\
28 & 9.22 & 0.49 & 6.16 & 0.50 & 3.85 & 0.20 \\
27 & 9.09 & 0.75 & 5.72 & 0.86 & 3.85 & 0.27 \\
26 & 8.28 & 1.23 & 5.34 & 1.59 & 3.76 & 0.45 \\
\hline
Ly$\alpha$ &  &  &  &  &  & \\
5E-19 & 2.16 & 1.13 & 2.59 & 0.47 & 3.08 & 1.55\\
\hline
H$\alpha$ &  &  &  &  &  & \\
2E-19 & 0.52 & 0.07 & 0.48 & 0.09 & 0.54 & 0.05 \\
6E-19 & 0.52 & 0.07 & 0.48 & 0.09 & 0.54 & 0.05 \\
1E-18 & 0.50 & 0.08 & 0.43 & 0.12 & 0.53 & 0.06 \\
\hline
\hline
L11 &  &  &  &  &  & \\
rest-UV &  &  &  &  &  & \\
30 & 28.66 & 1.20 & 13.18 & 0.64 & 15.17 & 0.31 \\
29 & 25.51 & 2.83 & 9.64 & 3.72 & 13.50 & 0.72 \\
28 & 14.59 & 4.83 & 6.46 & 2.49 & 9.15 & 2.73 \\
\hline
rest-optical  &  &  &  &  &  & \\
29 & 25.29 & 6.73 & 17.64 & 3.71 & 12.49 & 2.77 \\
28 & 28.06 & 1.24 & 17.12 & 1.17 & 13.93 & 0.23 \\
27 & 24.95 & 1.90 & 13.64 & 2.29 & 13.67 & 0.66 \\
26 & 18.75 & 3.33 & 8.51 & 2.63 & 11.32 & 1.53 \\
\hline
Ly$\alpha$ &  &  &  &  &  & \\
5E-19 & 0.25 & 0.77 & 0.81 & 2.04 & 0.31 & 1.17 \\
3E-18 & 0.15 & 0.08 & 0.24 & 0.33 & 0.14 & 0.12 \\
\hline
H$\alpha$ &  &  &  &  &  & \\
2E-19 & 1.06 & 0.11 & 0.70 & 0.07 & 0.92 & 0.05 \\
6E-19 & 0.81 & 0.08 & 0.56 & 0.09 & 0.85 & 0.10 \\
1E-18 & 0.90 & 0.73 & 0.61 & 0.20 & 1.12 & 0.80 \\
3E-18 & 0.58 & 0.10 & 0.38 & 0.12 & 0.66 & 0.07 \\
\hline
\end{tabular}
}
\tablefoot{continuation of Tab. \ref{tab:REGsize}}
\end{table*}

\begin{table*}
\caption{Size of high-$z$ simulated LARS galaxies}
\label{tab:REGsize4}
%\tablewidth{0pt}
\centering
\scalebox{0.9}{
\begin{tabular}{|c|c|c|c|c|c|c|}
\hline\hline 
continuum & rP20$^{ell}$ & $\sigma$rP20$^{ell}$ & rP20$^{minA}$ & $\sigma$rP20$^{minA}$ & r$_{50}$ & $\sigma$r$_{50}$  \\
 & kpc & kpc & kpc & kpc & kpc & kpc  \\
\hline
line  & $\xi$(rP20$^{ell})$ & $\sigma \xi$(rP20$^{ell})$ & $\xi$(rP20$^{minA})$ & $\sigma \xi$(rP20$^{minA}$) & $\xi$(r$_{50}$) & $\sigma \xi$(r$_{50}$)  \\
\hline
\hline
L12&  &  &  &  &  & \\
rest-UV &  &  &  &  &  & \\
30 & 3.38 & 0.18 & 1.67 & 0.33 & 1.33 & 0.17 \\
29 & 3.44 & 0.20 & 1.67 & 0.34 & 1.33 & 0.17 \\
28 & 3.55 & 0.25 & 1.63 & 0.41 & 1.32 & 0.18 \\
27 & 3.12 & 0.94 & 1.70 & 0.68 & 1.10 & 0.35 \\
\hline
rest-optical &  &  &  &  &  & \\
29 & 4.79 & 0.21 & 1.68 & 1.02 & 2.05 & 0.18 \\
28 & 4.83 & 0.24 & 2.25 & 1.47 & 2.05 & 0.19 \\
27 & 5.09 & 0.30 & 2.82 & 1.59 & 2.19 & 0.22 \\
26 & 5.66 & 1.53 & 2.33 & 1.58 & 2.66 & 1.24\\
\hline
Ly$\alpha$ &  &  &  &  &  & \\
5E-19 & 4.14 & 0.55 & 6.87 & 0.31 & 5.65 & 0.87 \\
3E-18 & 3.81 & 1.70 & 6.55 & 1.07 & 5.67 & 2.43 \\
\hline
H$\alpha$ &  &  &  &  &  & \\
2E-19 & 0.74 & 0.17 & 2.04 & 0.38 & 0.86 & 0.41 \\
6E-19 & 0.71 & 0.04 & 1.99 & 0.20 & 0.81 & 0.08 \\
1E-18 & 0.73 & 0.04 & 1.99 & 0.20 & 0.82 & 0.09 \\
3E-18 & 0.73 & 0.07 & 1.84 & 0.34 & 0.84 & 0.09 \\
\hline
\hline
L13 &  &  &  &  &  & \\
rest-UV &  &  &  &  &  & \\
30 & 5.29 & 2.55 & 1.66 & 0.34 & 2.40 & 2.26 \\
29 & 4.56 & 2.08 & 1.59 & 0.37 & 1.80 & 1.92 \\
28 & 2.27 & 0.72 & 1.54 & 0.38 & 0.93 & 0.22 \\
\hline
rest-optical &  &  &  &  &  & \\
29 & 12.57 & 2.48 & 6.47 & 0.37 & 5.94 & 0.91 \\
28 & 13.43 & 0.33 & 6.35 & 0.37 & 6.44 & 0.20 \\
27 & 9.31 & 3.13 & 6.02 & 0.43 & 5.70 & 1.71 \\
\hline
Ly$\alpha$ &  &  &  &  &  & \\
5E-19 & 0.62 & 0.14 & 2.48 & 3.87 & 1.11 & 0.34 \\
3E-18 & 0.84 & 0.60 & 1.74 & 0.86 & 1.16 & 1.03 \\
8E-18 & 0.96 & 0.58 & 1.63 & 0.75 & 1.36 & 0.99 \\
\hline
H$\alpha$ &  &  &  &  &  & \\
2E-19 & 0.61 & 0.12 & 0.31 & 0.05 & 0.55 & 0.14 \\
6E-19 & 0.63 & 0.12 & 0.31 & 0.05 & 0.57 & 0.14 \\
1E-18 & 0.58 & 0.13 & 0.31 & 0.05 & 0.48 & 0.18 \\
3E-18 & 0.47 & 0.13 & 0.28 & 0.06 & 0.39 & 0.13 \\
\hline
\hline
L14 &  &  &  &  &  & \\
rest-UV  &  &  &  &  &  & \\
30 & 1.67 & 0.17 & 1.33 & 0.33 & 0.83 & 0.17 \\
29 & 1.83 & 0.18 & 1.33 & 0.33 & 0.84 & 0.17 \\
28 & 1.99 & 1.18 & 1.19 & 0.38 & 1.06 & 0.99 \\
\hline
rest-optical  &  &  &  &  &  & \\
29 & 4.53 & 0.18 & 1.20 & 0.37 & 1.50 & 0.17 \\
28 & 4.55 & 0.46 & 1.10 & 0.37 & 1.56 & 0.41 \\
27 & 4.88 & 2.60 & 1.48 & 0.59 & 3.11 & 2.02 \\
\hline
Ly$\alpha$ &  &  &  &  &  & \\
5E-19 & 3.55 & 0.33 & 2.50 & 0.25 & 3.00 & 0.22 \\
3E-18 & 3.87 & 0.77 & 2.49 & 0.27 & 3.31 & 1.17 \\
8E-18 & 3.50 & 0.53 & 2.45 & 0.29 & 3.09 & 0.25 \\
2E-17 & 3.02 & 0.51 & 2.44 & 0.31 & 2.87 & 0.27 \\
5E-17 & 2.96 & 0.61 & 2.44 & 0.49 & 2.93 & 0.38 \\
\hline
H$\alpha$ &  &  &  &  &  & \\
2E-19 & 0.51 & 0.04 & 1.39 & 0.28 & 0.79 & 0.13 \\
6E-19 & 0.54 & 0.04 & 1.39 & 0.28 & 0.86 & 0.12 \\
1E-18 & 0.56 & 0.04 & 1.39 & 0.28 & 0.88 & 0.12 \\
3E-18 & 1.17 & 0.60 & 1.39 & 0.30 & 2.34 & 1.57 \\
1E-17 & 0.45 & 0.07 & 1.22 & 0.32 & 0.70 & 0.12 \\
\hline
\end{tabular}
}
\tablefoot{continuation of Tab. \ref{tab:REGsize}}
\end{table*}

}

\clearpage

\onecolumn
\longtab{9}{
\begin{table*}
\caption{Morphological parameters of high-$z$ simulated LARS galaxies}
\label{tab:REG}
\centering
\scalebox{0.9}{
\begin{tabular}{|c|c|c|c|c|c|c|c|c|c|c|c|c|}
\hline\hline 
%(1) & (2) & (3) & (4) & (5) &(6) & (7) & (8) & (9) &(10) & (11) & (12) & (13) \\ %& (14) & (15) &(16) & (17)& (18)\\
%\hline
depth& G$^{rP20}$ & $\sigma$G$^{rP20}$ & C$^{minA}$ & $\sigma$C$^{minA}$ & M20 & $\sigma$M20 & A & $\sigma$A & S & $\sigma$S & ell & $\sigma$ell \\
\hline
L01 &  &  &  &  &  & & & & & & & \\
rest-UV &  &  &  &  &  & & & & & & & \\
30 & 0.73 & 0.01 & 2.39 & 0.00 & -1.08 & 0.03 & 0.36 & 0.01 & 0.20 & 0.01 & 0.56 & 0.02 \\
29 & 0.70 & 0.03 & 2.41 & 0.12 & -1.16 & 0.34 & 0.28 & 0.02 & 0.18 & 0.05 & 0.50 & 0.06 \\
28 & 0.67 & 0.04 & 2.46 & 0.45 & -1.14 & 0.41 & 0.17 & 0.04 & 0.20 & 0.08 & 0.43 & 0.09 \\
\hline
rest-optical &  &  &  &  &  & & & & & & & \\
29 & 0.60 & 0.00 & 2.13 & 0.00 & -0.91 & 0.08 & 0.26 & 0.01 & 0.06 & 0.01 & 0.58 & 0.01 \\
28 & 0.61 & 0.01 & 2.16 & 0.31 & -0.98 & 0.08 & 0.18 & 0.02 & 0.11 & 0.01 & 0.59 & 0.02 \\
27 & 0.61 & 0.02 & 2.59 & 0.57 & -1.03 & 0.10 & 0.10 & 0.03 & 0.22 & 0.04 & 0.58 & 0.03 \\
26 & 0.60 & 0.03 & 2.53 & 0.67 & -1.17 & 0.19 & 0.04 & 0.02 & 0.47 & 0.12 & 0.54 & 0.06 \\
\hline
Ly$\alpha$ &  &  &  &  &  & & & & & & & \\
5E-19 & 0.62 & 0.02 & 2.88 & 0.18 & -1.66 & 0.20 & 0.28 & 0.04 & 0.23 & 0.06 & 0.30 & 0.03 \\
3E-18 & 0.57 & 0.01 & 2.54 & 0.40 & -1.60 & 0.15 & 0.15 & 0.02 & 0.49 & 0.09 & 0.33 & 0.09 \\
8E-18 & 0.49 & 0.02 & 2.11 & 0.58 & -1.61 & 0.28 & 0.09 & 0.04 & 0.70 & 0.15 & 0.20 & 0.08 \\
\hline
H$\alpha$ &  &  &  &  &  & & & & & & & \\
2E-19 & 0.67 & 0.01 & 2.39 & 0.00 & -1.10 & 0.26 & 0.36 & 0.00 & 0.20 & 0.07 & 0.43 & 0.02 \\
6E-19 & 0.68 & 0.01 & 2.39 & 0.00 & -0.87 & 0.16 & 0.35 & 0.01 & 0.30 & 0.06 & 0.46 & 0.02 \\
1E-18 & 0.67 & 0.01 & 2.39 & 0.00 & -0.94 & 0.21 & 0.34 & 0.01 & 0.26 & 0.08 & 0.44 & 0.04 \\
3E-18 & 0.64 & 0.02 & 2.40 & 0.28 & -0.97 & 0.21 & 0.26 & 0.02 & 0.26 & 0.07 & 0.42 & 0.05  \\
\hline
\hline
L02 &  &  &  &  &  & & & & & & & \\
rest-UV &  &  &  &  &  & & & & & & & \\
30 & 0.64 & 0.02 & 2.40 & 0.64 & -1.41 & 0.22 & 0.21 & 0.03 & 0.36 & 0.06 & 0.69 & 0.03 \\
29 & 0.62 & 0.03 & 2.29 & 0.65 & -1.29 & 0.27 & 0.10 & 0.04 & 0.39 & 0.06 & 0.71 & 0.04 \\
\hline
rest-optical &  &  &  &  &  & & & & & & & \\
29 & 0.62 & 0.01 & 3.68 & 0.19 & -1.81 & 0.08 & 0.02 & 0.01 & 0.36 & 0.04 & 0.63 & 0.02 \\
28 & 0.60 & 0.01 & 3.38 & 0.32 & -1.78 & 0.12 & 0.01 & 0.01 & 0.54 & 0.08 & 0.66 & 0.03 \\
27 & 0.55 & 0.03 & 2.67 & 0.59 & -1.59 & 0.19 & 0.01 & 0.02 & 0.60 & 0.13 & 0.68 & 0.05 \\
\hline
Ly$\alpha$ &  &  &  &  &  & & & & & & & \\
5E-19 & 0.61 & 0.01 & 3.06 & 0.35 & -1.25 & 0.07 & 0.27 & 0.01 & 0.35 & 0.04 & 0.36 & 0.03 \\
3E-18 & 0.51 & 0.01 & 2.84 & 0.43 & -1.22 & 0.21 & 0.09 & 0.03 & 0.64 & 0.07 & 0.36 & 0.12 \\
8E-18 & 0.45 & 0.02 & 2.05 & 0.52 & -1.20 & 0.25 & 0.05 & 0.04 & 0.68 & 0.19 & 0.39 & 0.12 \\
\hline
H$\alpha$ &  &  &  &  &  & & & & & & & \\
2E-19 & 0.66 & 0.02 & 2.51 & 0.34 & -1.29 & 0.12 & 0.34 & 0.02 & 0.15 & 0.07 & 0.28 & 0.14 \\
6E-19 & 0.60 & 0.02 & 3.50 & 0.00 & -0.93 & 0.12 & 0.26 & 0.03 & 0.25 & 0.07 & 0.61 & 0.04 \\
1E-18 & 0.58 & 0.02 & 3.10 & 0.79 & -0.96 & 0.19 & 0.21 & 0.04 & 0.26 & 0.07 & 0.62 & 0.05 \\
\hline
\hline
L03 &  &  &  &  &  & & & & & & & \\
rest-UV &  &  &  &  &  & & & & & & & \\
30 &0.67 & 0.03 & 2.54 & 0.27 & -1.42 & 0.29 & 0.08 & 0.03 & 0.19 & 0.12 & 0.14 & 0.08 \\
29 & 0.64 & 0.15 & 2.82 & 1.66 & -1.58 & 1.49 & 0.06 & 0.14 & 0.38 & 1.02 & 0.20 & 0.50 \\
\hline
rest-optical &  &  &  &  &  & & & & & & & \\
29 & 0.63 & 0.06 & 3.38 & 0.10 & -2.01 & 0.32 & 0.31 & 0.03 & 0.09 & 0.07 & 0.46 & 0.07 \\
28 & 0.61 & 0.02 & 3.07 & 0.28 & -2.00 & 0.19 & 0.10 & 0.03 & 0.13 & 0.03 & 0.46 & 0.04 \\
27 & 0.63 & 0.03 & 2.98 & 0.36 & -1.96 & 0.33 & 0.06 & 0.01 & 0.29 & 0.07 & 0.34 & 0.07 \\
26 & 0.61 & 0.03 & 3.06 & 0.34 & -2.04 & 0.13 & 0.03 & 0.01 & 0.53 & 0.11 & 0.27 & 0.07 \\
25 & 0.57 & 0.02 & 3.06 & 0.34 & -1.87 & 0.19 & 0.03 & 0.01 & 0.64 & 0.13 & 0.37 & 0.10 \\
\hline
Ly$\alpha$ &  &  &  &  &  & & & & & & & \\
5E-19 & 0.47 & 0.01 & 2.70 & 0.12 & -1.78 & 0.15 & 0.10 & 0.02 & 0.62 & 0.03 & 0.30 & 0.06 \\
3E-18 & 0.41 & 0.09 & 2.04 & 2.74 & -0.93 & 0.90 & 0.02 & 0.10 & 0.76 & 0.60 & 0.50 & 0.75 \\
\hline
H$\alpha$ &  &  &  &  &  & & & & & & & \\
2E-19 & 0.67 & 0.00 & 3.01 & 0.00 & -1.58 & 0.25 & 0.25 & 0.00 & 0.08 & 0.01 & 0.21 & 0.03 \\
6E-19 & 0.66 & 0.01 & 3.01 & 0.00 & -1.57 & 0.25 & 0.21 & 0.01 & 0.10 & 0.01 & 0.27 & 0.03 \\
1E-18 & 0.67 & 0.01 & 3.02 & 0.05 & -1.56 & 0.27 & 0.19 & 0.01 & 0.11 & 0.01 & 0.29 & 0.03 \\
3E-18 & 0.65 & 0.01 & 2.99 & 0.33 & -1.60 & 0.26 & 0.14 & 0.03 & 0.17 & 0.04 & 0.35 & 0.05 \\
\hline
\end{tabular}
}
\tablefoot{Non-parametric measurements are derived by following the equations in Appendix A. Only G$^{rP20}$ and C$^{minA}$ are reported here, because they are the ones used to compare z2LARS with high-$z$ observations. The uncertainty on the parameters is obtained from the standard deviation of 100 realizations of a simulated survey depth.}
\end{table*}

\begin{table*}
\caption{Morphological parameters of high-$z$ simulated LARS galaxies}
\label{tab:REG2}
\centering
\scalebox{0.9}{
\begin{tabular}{|c|c|c|c|c|c|c|c|c|c|c|c|c|}
\hline\hline 
depth& G$^{rP20}$ & $\sigma$G$^{rP20}$ & C$^{minA}$ & $\sigma$C$^{minA}$ & M20 & $\sigma$M20 & A & $\sigma$A & S & $\sigma$S & ell & $\sigma$ell \\
\hline
L05&  &  &  &  &  & & & & & & & \\
rest-UV &  &  &  &  &  & & & & & & & \\
30 & 0.70 & 0.01 & 3.88 & 0.07 & -1.22 & 0.07 & 0.27 & 0.01 & 0.23 & 0.04 & 0.67 & 0.01 \\
29 & 0.70 & 0.01 & 3.73 & 0.21 & -1.13 & 0.10 & 0.19 & 0.02 & 0.35 & 0.12 & 0.71 & 0.02 \\
28 & 0.70 & 0.01 & 3.68 & 0.48 & -1.13 & 0.12 & 0.11 & 0.03 & 0.32 & 0.14 & 0.73 & 0.03 \\
27 & 0.68 & 0.03 & 2.82 & 0.78 & -1.22 & 0.25 & 0.05 & 0.04 & 0.36 & 0.13 & 0.73 & 0.05 \\
\hline
rest-optical  &  &  &  &  &  & & & & & & & \\
29 & 0.74 & 0.01 & 3.89 & 0.00 & -1.42 & 0.17 & 0.27 & 0.01 & 0.29 & 0.05 & 0.73 & 0.01 \\
28 & 0.74 & 0.01 & 3.90 & 0.06 & -1.40 & 0.18 & 0.19 & 0.02 & 0.27 & 0.07 & 0.76 & 0.02 \\
27 & 0.75 & 0.02 & 3.93 & 0.24 & -1.28 & 0.24 & 0.12 & 0.03 & 0.25 & 0.11 & 0.79 & 0.02 \\
26 & 0.74 & 0.03 & 3.31 & 1.04 & -1.26 & 0.29 & 0.06 & 0.03 & 0.33 & 0.17 & 0.81 & 0.03 \\
\hline
Ly$\alpha$ &  &  &  &  &  & & & & & & & \\
5E-19 & 0.58 & 0.01 & 1.95 & 0.09 & -1.42 & 0.06 & 0.26 & 0.00 & 0.11 & 0.01 & 0.13 & 0.03 \\
3E-18 & 0.52 & 0.02 & 1.98 & 0.38 & -1.48 & 0.08 & 0.13 & 0.02 & 0.39 & 0.08 & 0.17 & 0.05 \\
8E-18 & 0.45 & 0.01 & 2.18 & 0.38 & -1.40 & 0.13 & 0.05 & 0.03 & 0.64 & 0.11 & 0.27 & 0.10 \\
\hline
H$\alpha$ &  &  &  &  &  & & & & & & & \\
2E-19 & 0.59 & 0.01 & 2.21 & 0.20 & -0.79 & 0.01 & 0.31 & 0.01 & 0.27 & 0.01 & 0.44 & 0.02 \\
6E-19 & 0.57 & 0.01 & 2.25 & 0.19 & -0.76 & 0.02 & 0.26 & 0.03 & 0.24 & 0.04 & 0.52 & 0.02 \\
1E-18 & 0.56 & 0.01 & 2.21 & 0.20 & -0.76 & 0.02 & 0.21 & 0.04 & 0.22 & 0.04 & 0.56 & 0.03 \\
3E-18 & 0.52 & 0.02 & 2.24 & 0.24 & -0.75 & 0.09 & 0.11 & 0.04 & 0.27 & 0.06 & 0.64 & 0.04 \\
\hline
\hline
L07 &  &  &  &  &  & & & & & & & \\
rest-UV &  &  &  &  &  & & & & & & & \\
30 & 0.72 & 0.01 & 2.39 & 0.00 & -1.20 & 0.11 & 0.31 & 0.01 & 0.02 & 0.05 & 0.47 & 0.02 \\
29 & 0.72 & 0.01 & 2.39 & 0.00 & -1.12 & 0.23 & 0.27 & 0.01 & 0.07 & 0.10 & 0.50 & 0.03 \\
28 & 0.72 & 0.02 & 2.43 & 0.24 & -1.02 & 0.29 & 0.19 & 0.04 & 0.14 & 0.12 & 0.51 & 0.06 \\
\hline
rest-optical  &  &  &  &  &  & & & & & & & \\
29 & 0.63 & 0.01 & 3.01 & 0.03 & -1.54 & 0.12 & 0.23 & 0.01 & 0.05 & 0.01 & 0.59 & 0.03 \\
28 & 0.62 & 0.01 & 3.05 & 0.13 & -1.51 & 0.12 & 0.15 & 0.01 & 0.08 & 0.01 & 0.59 & 0.02 \\
27 & 0.63 & 0.01 & 2.97 & 0.38 & -1.54 & 0.13 & 0.07 & 0.02 & 0.20 & 0.04 & 0.61 & 0.02 \\
26 & 0.63 & 0.02 & 2.70 & 0.65 & -1.51 & 0.18 & 0.04 & 0.03 & 0.46 & 0.12 & 0.55 & 0.09 \\
\hline
Ly$\alpha$ &  &  &  &  &  & & & & & & & \\
5E-19 & 0.64 & 0.01 & 3.44 & 0.10 & -2.17 & 0.21 & 0.24 & 0.01 & 0.18 & 0.05 & 0.36 & 0.04 \\
3E-18 & 0.57 & 0.01 & 3.18 & 0.27 & -2.18 & 0.12 & 0.11 & 0.02 & 0.49 & 0.09 & 0.13 & 0.07 \\
8E-18 & 0.51 & 0.02 & 2.85 & 0.48 & -2.08 & 0.16 & 0.06 & 0.03 & 0.66 & 0.12 & 0.17 & 0.08 \\
\hline
H$\alpha$ &  &  &  &  &  & & & & & & & \\
2E-19 & 0.65 & 0.01 & 0.88 & 0.00 & -0.72 & 0.06 & 0.34 & 0.01 & 0.23 & 0.05 & 0.24 & 0.07 \\
6E-19 & 0.65 & 0.01 & 0.88 & 0.00 & -0.72 & 0.06 & 0.34 & 0.01 & 0.23 & 0.05 & 0.24 & 0.07 \\
1E-18 & 0.66 & 0.01 & 0.88 & 0.00 & -0.69 & 0.05 & 0.33 & 0.01 & 0.25 & 0.04 & 0.29 & 0.06 \\
3E-18 & 0.64 & 0.02 & 1.16 & 0.58 & -0.67 & 0.10 & 0.25 & 0.04 & 0.27 & 0.09 & 0.37 & 0.0 \\
\hline
\hline
L08 &  &  &  &  &  & & & & & & & \\
rest-UV  &  &  &  &  &  & & & & & & & \\
30 & 0.61 & 0.04 & 2.81 & 0.51 & -0.84 & 0.06 & 0.10 & 0.04 & 0.39 & 0.09 & 0.37 & 0.04 \\
29 & 0.60 & 0.03 & 2.58 & 0.55 & -1.02 & 0.12 & 0.05 & 0.04 & 0.68 & 0.12 & 0.25 & 0.11 \\
28 & 0.55 & 0.03 & 2.12 & 0.61 & -1.07 & 0.22 & 0.01 & 0.02 & 0.72 & 0.13 & 0.54 & 0.10 \\
\hline
rest-optical  &  &  &  &  &  & & & & & & & \\
29 & 0.59 & 0.03 & 2.71 & 0.03 & -1.21 & 0.06 & 0.37 & 0.00 & 0.07 & 0.03 & 0.20 & 0.01 \\
28 & 0.59 & 0.03 & 2.69 & 0.07 & -1.24 & 0.09 & 0.16 & 0.02 & 0.13 & 0.03 & 0.16 & 0.02 \\
27 & 0.60 & 0.02 & 2.63 & 0.11 & -1.25 & 0.07 & 0.08 & 0.01 & 0.25 & 0.04 & 0.16 & 0.05 \\
26 & 0.59 & 0.02 & 2.50 & 0.20 & -1.26 & 0.09 & 0.04 & 0.01 & 0.47 & 0.07 & 0.17 & 0.07 \\
25 & 0.54 & 0.02 & 2.50 & 0.20 & -1.17 & 0.09 & 0.04 & 0.01 & 0.62 & 0.10 & 0.24 & 0.05 \\
\hline
Ly$\alpha$ &  &  &  &  &  & & & & & & & \\
5E-19 & 0.55 & 0.00 & 2.77 & 0.35 & -1.75 & 0.12 & 0.29 & 0.03 & 0.62 & 0.03 & 0.15 & 0.07 \\
3E-18 & 0.45 & 0.01 & 2.06 & 0.43 & -1.33 & 0.25 & 0.10 & 0.04 & 0.74 & 0.11 & 0.37 & 0.09 \\
8E-18 & 0.41 & 0.11 & 1.81 & 2.61 & -0.75 & 0.81 & 0.04 & 0.27 & 0.74 & 0.82 & 0.54 & 0.70 \\
\hline
H$\alpha$
2E-19 & 0.60 & 0.01 & 1.67 & 0.04 & -0.91 & 0.04 & 0.36 & 0.01 & 0.23 & 0.02 & 0.16 & 0.06 \\
6E-19 & 0.57 & 0.01 & 1.57 & 0.12 & -0.87 & 0.02 & 0.33 & 0.01 & 0.24 & 0.01 & 0.13 & 0.02 \\
1E-18 & 0.56 & 0.01 & 1.63 & 0.17 & -0.88 & 0.03 & 0.28 & 0.01 & 0.29 & 0.03 & 0.11 & 0.02 \\
3E-18 & 0.50 & 0.01 & 1.67 & 0.22 & -1.00 & 0.13 & 0.11 & 0.02 & 0.60 & 0.13 & 0.47 & 0.11 \\
\hline
\end{tabular}
}
\tablefoot{continuation of Tab. \ref{tab:REG}}
\end{table*}

\begin{table*}
\caption{Morphological parameters of high-$z$ simulated LARS galaxies}
\label{tab:REG3}
\centering
\scalebox{0.9}{
\begin{tabular}{|c|c|c|c|c|c|c|c|c|c|c|c|c|}
\hline\hline 
depth& G$^{rP20}$ & $\sigma$G$^{rP20}$ & C$^{minA}$ & $\sigma$C$^{minA}$ & M20 & $\sigma$M20 & A & $\sigma$A & S & $\sigma$S & ell & $\sigma$ell \\
\hline
L09 &  &  &  &  &  & & & & & & & \\
rest-UV &  &  &  &  &  & & & & & & & \\
30& 0.70 & 0.02 & 3.59 & 0.63 & -1.23 & 0.10 & 0.18 & 0.05 & 0.46 & 0.06 & 0.79 & 0.08 \\
29 & 0.61 & 0.03 & 1.54 & 1.09 & -0.90 & 0.22 & 0.19 & 0.08 & 0.46 & 0.10 & 0.57 & 0.17 \\
28 & 0.69 & 0.05 & 2.46 & 0.56 & -1.74 & 0.65 & 0.04 & 0.03 & 0.51 & 0.25 & 0.38 & 0.17 \\
\hline
rest-optical  &  &  &  &  &  & & & & & & & \\
29 & 0.67 & 0.01 & 2.85 & 0.05 & -0.98 & 0.01 & 0.36 & 0.00 & 0.33 & 0.00 & 0.81 & 0.00 \\
28& 0.65 & 0.01 & 2.84 & 0.12 & -0.98 & 0.01 & 0.21 & 0.01 & 0.36 & 0.01 & 0.83 & 0.00 \\
27 & 0.62 & 0.06 & 2.23 & 0.71 & -1.01 & 0.26 & 0.15 & 0.05 & 0.50 & 0.13 & 0.57 & 0.22 \\
\hline
Ly$\alpha$ &  &  &  &  &  & & & & & & & \\
5E-19 & 0.47 & 0.00 & 0.82 & 0.10 & -0.84 & 0.01 & 0.28 & 0.07 & 0.57 & 0.02 & 0.48 & 0.02 \\
\hline
H$\alpha$ &  &  &  &  &  & & & & & & & \\
2E-19 & 0.70 & 0.02 & 4.38 & 0.27 & -1.46 & 0.14 & 0.23 & 0.03 & 0.33 & 0.12 & 0.67 & 0.03 \\
6E-19 & 0.67 & 0.01 & 4.26 & 0.29 & -1.49 & 0.23 & 0.19 & 0.01 & 0.40 & 0.04 & 0.79 & 0.00 \\
1E-18 & 0.63 & 0.03 & 4.13 & 0.29 & -1.49 & 0.50 & 0.16 & 0.02 & 0.46 & 0.09 & 0.71 & 0.15 \\
3E-18 & 0.58 & 0.03 & 3.69 & 0.61 & -2.04 & 0.84 & 0.11 & 0.03 & 0.49 & 0.16 & 0.59 & 0.12 \\
1E-17 & 0.56 & 0.04 & 2.96 & 0.60 & -1.81 & 0.44 & 0.06 & 0.04 & 0.58 & 0.20 & 0.36 & 0.14 \\
\hline
\hline
L10 &  &  &  &  &  & & & & & & & \\
rest-UV &  &  &  &  &  & & & & & & &  \\
30 & 0.59 & 0.02 & 2.11 & 0.25 & -1.27 & 0.14 & 0.14 & 0.02 & 0.21 & 0.04 & 0.54 & 0.02 \\
29 & 0.57 & 0.03 & 2.33 & 0.41 & -1.30 & 0.13 & 0.07 & 0.02 & 0.37 & 0.09 & 0.60 & 0.05 \\
28 & 0.53 & 0.04 & 2.27 & 0.60 & -1.31 & 0.23 & 0.02 & 0.02 & 0.57 & 0.15 & 0.62 & 0.09 \\
\hline
rest-optical  &  &  &  &  &  & & & & & & & \\
29 & 0.57 & 0.01 & 3.34 & 0.26 & -1.97 & 0.04 & 0.12 & 0.00 & 0.09 & 0.00 & 0.55 & 0.02 \\
28 & 0.59 & 0.01 & 3.35 & 0.16 & -2.04 & 0.04 & 0.08 & 0.01 & 0.20 & 0.02 & 0.50 & 0.02 \\
27 & 0.59 & 0.01 & 3.18 & 0.21 & -1.98 & 0.09 & 0.04 & 0.01 & 0.41 & 0.05 & 0.52 & 0.03 \\
26 & 0.57 & 0.01 & 2.95 & 0.34 & -1.92 & 0.10 & 0.02 & 0.01 & 0.62 & 0.08 & 0.53 & 0.07 \\
\hline
Ly$\alpha$ &  &  &  &  &  & & & & & & & \\
5E-19 & 0.45 & 0.06 & 1.08 & 0.40 & -0.82 & 0.66 & 0.17 & 0.13 & 0.74 & 0.18 & 0.41 & 0.11\\
\hline
H$\alpha$ &  &  &  &  &  & & & & & & & \\
2E-19 & 0.58 & 0.01 & 2.38 & 0.70 & -1.06 & 0.14 & 0.24 & 0.02 & 0.30 & 0.06 & 0.59 & 0.03 \\
6E-19 & 0.58 & 0.01 & 2.38 & 0.70 & -1.06 & 0.14 & 0.24 & 0.02 & 0.30 & 0.06 & 0.59 & 0.03 \\
1E-18 & 0.55 & 0.02 & 2.56 & 0.68 & -1.05 & 0.15 & 0.18 & 0.03 & 0.35 & 0.09 & 0.65 & 0.03 \\
\hline
\hline
L11 &  &  &  &  &  & & & & & & & \\
rest-UV &  &  &  &  &  & & & & & & & \\
30 & 0.57 & 0.02 & 3.20 & 0.10 & -1.21 & 0.08 & 0.30 & 0.04 & 0.39 & 0.03 & 0.88 & 0.08 \\
29 & 0.56 & 0.04 & 2.58 & 0.53 & -1.11 & 0.18 & 0.06 & 0.05 & 0.47 & 0.07 & 0.88 & 0.09 \\
28 & 0.57 & 0.04 & 2.53 & 0.74 & -0.70 & 0.18 & 0.04 & 0.03 & 0.69 & 0.10 & 0.62 & 0.15 \\
\hline
rest-optical  &  &  &  &  &  & & & & & & & \\
29 & 0.51 & 0.01 & 2.95 & 0.46 & -1.23 & 0.10 & 0.28 & 0.07 & 0.26 & 0.13 & 0.82 & 0.13 \\
28 & 0.51 & 0.02 & 3.03 & 0.08 & -1.33 & 0.03 & 0.18 & 0.06 & 0.28 & 0.01 & 0.87 & 0.00 \\
27 & 0.49 & 0.02 & 2.94 & 0.17 & -1.31 & 0.07 & 0.05 & 0.05 & 0.43 & 0.05 & 0.87 & 0.05 \\
26 & 0.51 & 0.02 & 2.60 & 0.37 & -0.89 & 0.12 & 0.01 & 0.01 & 0.64 & 0.07 & 0.75 & 0.12 \\
\hline
Ly$\alpha$ &  &  &  &  &  & & & & & & & \\
5E-19 & 0.44 & 0.34 & 1.62 & 1.54 & -0.73 & 1.69 & 0.43 & 0.32 & 0.72 & 0.87 & 0.63 & 0.98 \\
3E-18 & 0.39 & 0.05 & 1.81 & 0.54 & -0.65 & 0.16 & 0.29 & 0.07 & 0.63 & 0.25 & 0.64 & 0.07 \\
\hline
H$\alpha$ &  &  &  &  &  & & & & & & & \\
2E-19 & 0.58 & 0.02 & 2.61 & 0.08 & -0.89 & 0.10 & 0.23 & 0.01 & 0.37 & 0.03 & 0.88 & 0.01 \\
6E-19 & 0.52 & 0.01 & 2.39 & 0.16 & -1.11 & 0.14 & 0.13 & 0.02 & 0.37 & 0.11 & 0.81 & 0.18 \\
1E-18 & 0.48 & 0.10 & 2.58 & 0.46 & -0.67 & 0.36 & 0.10 & 0.12 & 0.73 & 0.29 & 0.44 & 0.64 \\
3E-18 & 0.44 & 0.01 & 2.28 & 0.40 & -1.01 & 0.11 & 0.03 & 0.02 & 0.63 & 0.10 & 0.73 & 0.13 \\
\hline
\end{tabular}
}
\tablefoot{continuation of Tab. \ref{tab:REG}}
\end{table*}

\begin{table*}
\caption{Morphological parameters of high-$z$ simulated LARS galaxies}
\label{tab:REG4}
%\tablewidth{0pt}
\centering
\scalebox{0.9}{
\begin{tabular}{|c|c|c|c|c|c|c|c|c|c|c|c|c|}
\hline\hline 
depth& G$^{rP20}$ & $\sigma$G$^{rP20}$ & C$^{minA}$ & $\sigma$C$^{minA}$ & M20 & $\sigma$M20 & A & $\sigma$A & S & $\sigma$S & ell & $\sigma$ell \\
\hline
L12&  &  &  &  &  & & & & & & & \\
rest-UV &  &  &  &  &  & & & & & & & \\
30 & 0.69 & 0.00 & 3.33 & 0.23 & -1.43 & 0.03 & 0.37 & 0.01 & 0.10 & 0.02 & 0.38 & 0.01 \\
29 & 0.69 & 0.01 & 3.32 & 0.24 & -1.44 & 0.04 & 0.25 & 0.03 & 0.11 & 0.01 & 0.41 & 0.01 \\
28 & 0.70 & 0.01 & 3.29 & 0.30 & -1.47 & 0.22 & 0.16 & 0.04 & 0.18 & 0.03 & 0.45 & 0.02 \\
27 & 0.69 & 0.03 & 3.11 & 0.53 & -1.84 & 0.41 & 0.11 & 0.04 & 0.39 & 0.19 & 0.27 & 0.17 \\
\hline
rest-optical &  &  &  &  &  & & & & & & & \\
29 & 0.65 & 0.01 & 3.09 & 0.22 & -1.71 & 0.15 & 0.37 & 0.07 & 0.15 & 0.01 & 0.49 & 0.01 \\
28 & 0.65 & 0.01 & 3.20 & 0.30 & -1.65 & 0.10 & 0.25 & 0.01 & 0.15 & 0.01 & 0.52 & 0.01 \\
27 & 0.66 & 0.01 & 3.35 & 0.39 & -1.80 & 0.18 & 0.14 & 0.01 & 0.20 & 0.01 & 0.54 & 0.01 \\
26 & 0.68 & 0.05 & 3.21 & 0.60 & -1.54 & 0.48 & 0.07 & 0.04 & 0.41 & 0.12 & 0.54 & 0.09\\
\hline
Ly$\alpha$ &  &  &  &  &  & & & & & & & \\
5E-19 & 0.62 & 0.01 & 1.27 & 0.11 & -2.26 & 0.25 & 0.37 & 0.03 & 0.44 & 0.10 & 0.27 & 0.03 \\
3E-18 & 0.54 & 0.06 & 1.38 & 0.30 & -1.81 & 0.87 & 0.13 & 0.07 & 0.75 & 0.15 & 0.47 & 0.13 \\
\hline
H$\alpha$ &  &  &  &  &  & & & & & & & \\
2E-19 & 0.61 & 0.05 & 2.99 & 0.11 & -1.54 & 0.22 & 0.34 & 0.05 & 0.19 & 0.14 & 0.39 & 0.06 \\
6E-19 & 0.62 & 0.01 & 3.01 & 0.00 & -1.66 & 0.16 & 0.28 & 0.00 & 0.19 & 0.06 & 0.40 & 0.02 \\
1E-18 & 0.62 & 0.01 & 3.01 & 0.00 & -1.83 & 0.20 & 0.26 & 0.01 & 0.17 & 0.05 & 0.41 & 0.02 \\
3E-18 & 0.60 & 0.01 & 2.93 & 0.29 & -1.60 & 0.30 & 0.17 & 0.01 & 0.23 & 0.03 & 0.44 & 0.03 \\
\hline
\hline
L13 &  &  &  &  &  & & & & & & & \\
rest-UV &  &  &  &  &  & & & & & & & \\
30 & 0.72 & 0.03 & 1.60 & 0.36 & -1.24 & 0.34 & 0.27 & 0.01 & 0.17 & 0.12 & 0.37 & 0.11 \\
29 & 0.71 & 0.03 & 2.30 & 0.73 & -1.35 & 0.24 & 0.23 & 0.03 & 0.23 & 0.13 & 0.38 & 0.08 \\
28 & 0.65 & 0.03 & 2.36 & 0.65 & -1.37 & 0.19 & 0.18 & 0.03 & 0.20 & 0.12 & 0.15 & 0.08 \\
\hline
rest-optical &  &  &  &  &  & & & & & & & \\
29 & 0.64 & 0.02 & 3.64 & 0.03 & -1.32 & 0.22 & 0.31 & 0.00 & 0.24 & 0.15 & 0.47 & 0.06 \\
28 & 0.62 & 0.01 & 3.52 & 0.08 & -1.45 & 0.06 & 0.21 & 0.01 & 0.23 & 0.01 & 0.53 & 0.02 \\
27 & 0.49 & 0.05 & 3.31 & 0.11 & -0.98 & 0.16 & 0.16 & 0.01 & 0.49 & 0.15 & 0.53 & 0.11 \\
\hline
Ly$\alpha$ &  &  &  &  &  & & & & & & & \\
5E-19 & 0.53 & 0.03 & 2.10 & 1.28 & -0.52 & 0.12 & 0.40 & 0.07 & 0.61 & 0.09 & 0.40 & 0.12 \\
3E-18 & 0.48 & 0.02 & 1.74 & 0.50 & -1.18 & 0.32 & 0.28 & 0.07 & 0.55 & 0.09 & 0.24 & 0.10 \\
8E-18 & 0.44 & 0.02 & 1.63 & 0.56 & -1.22 & 0.25 & 0.12 & 0.06 & 0.72 & 0.13 & 0.40 & 0.11 \\
\hline
H$\alpha$ &  &  &  &  &  & & & & & & & \\
2E-19 & 0.67 & 0.01 & 1.99 & 0.00 & -1.90 & 0.09 & 0.16 & 0.00 & 0.17 & 0.03 & 0.23 & 0.07 \\
6E-19 & 0.66 & 0.01 & 1.99 & 0.00 & -1.87 & 0.10 & 0.16 & 0.01 & 0.24 & 0.03 & 0.25 & 0.07 \\
1E-18 & 0.65 & 0.02 & 1.99 & 0.00 & -1.93 & 0.43 & 0.14 & 0.01 & 0.30 & 0.11 & 0.28 & 0.17 \\
3E-18 & 0.60 & 0.02 & 1.80 & 0.30 & -1.91 & 0.32 & 0.09 & 0.02 & 0.44 & 0.15 & 0.34 & 0.16 \\
\hline
\hline
L14 &  &  &  &  &  & & & & & & & \\
rest-UV  &  &  &  &  &  & & & & & & & \\
30 & 0.64 & 0.00 & 2.39 & 0.00 & -0.99 & 0.04 & 0.36 & 0.00 & 0.04 & 0.02 & 0.41 & 0.02 \\
29 & 0.64 & 0.01 & 2.39 & 0.00 & -0.77 & 0.11 & 0.32 & 0.01 & 0.11 & 0.07 & 0.52 & 0.03 \\
28 & 0.65 & 0.05 & 2.38 & 0.09 & -0.94 & 0.20 & 0.29 & 0.05 & 0.08 & 0.10 & 0.47 & 0.15 \\
\hline
rest-optical  &  &  &  &  &  & & & & & & & \\
29 & 0.78 & 0.00 & 2.39 & 0.00 & -1.06 & 0.09 & 0.39 & 0.00 & 0.19 & 0.01 & 0.09 & 0.02 \\
28 & 0.78 & 0.01 & 2.39 & 0.00 & -1.17 & 0.12 & 0.29 & 0.02 & 0.27 & 0.04 & 0.22 & 0.05 \\
27 & 0.67 & 0.13 & 1.97 & 0.51 & -0.93 & 0.34 & 0.28 & 0.11 & 0.36 & 0.18 & 0.50 & 0.12 \\
\hline
Ly$\alpha$ &  &  &  &  &  & & & & & & & \\
5E-19 & 0.61 & 0.03 & 2.13 & 0.00 & -1.47 & 0.07 & 0.29 & 0.00 & 0.16 & 0.01 & 0.17 & 0.01 \\
3E-18 & 0.61 & 0.02 & 2.13 & 0.01 & -1.51 & 0.09 & 0.24 & 0.01 & 0.20 & 0.03 & 0.24 & 0.04 \\
8E-18 & 0.56 & 0.03 & 2.12 & 0.04 & -1.48 & 0.09 & 0.19 & 0.01 & 0.28 & 0.07 & 0.21 & 0.04 \\
2E-17 & 0.50 & 0.02 & 2.04 & 0.16 & -1.44 & 0.09 & 0.12 & 0.02 & 0.40 & 0.11 & 0.15 & 0.05 \\
5E-17 & 0.43 & 0.02 & 1.99 & 0.23 & -1.33 & 0.12 & 0.05 & 0.03 & 0.60 & 0.11 & 0.23 & 0.09 \\
\hline
H$\alpha$ &  &  &  &  &  & & & & & & & \\
2E-19 & 0.55 & 0.01 & 1.50 & 0.00 & -1.07 & 0.03 & 0.20 & 0.00 & 0.03 & 0.01 & 0.39 & 0.01 \\
6E-19 & 0.57 & 0.01 & 1.50 & 0.00 & -1.08 & 0.01 & 0.21 & 0.00 & 0.03 & 0.00 & 0.42 & 0.01 \\
1E-18 & 0.58 & 0.01 & 1.50 & 0.00 & -1.09 & 0.01 & 0.18 & 0.00 & 0.04 & 0.00 & 0.45 & 0.02 \\
3E-18 & 0.65 & 0.07 & 1.63 & 0.21 & -0.91 & 0.30 & 0.19 & 0.09 & 0.23 & 0.18 & 0.43 & 0.13 \\
1E-17 & 0.52 & 0.03 & 1.48 & 0.12 & -1.35 & 0.12 & 0.10 & 0.02 & 0.14 & 0.07 & 0.19 & 0.08 \\
\hline
\end{tabular}
}
\tablefoot{continuation of Tab. \ref{tab:REG}}
\end{table*}

}

\appendix

\section{Non-parametric measurements}
\label{sec:appendix}

In this section we explain the way we measured galaxy sizes and non-parametric quantities in detail, and show the results for galaxies with known-profiles.

In Fig. \ref{eqMORPH} we summarize the equations adopted in this analysis and first introduced by \citet{Conselice2000} and \citet{Lotz2004}.

The code we developed makes a basic use of the ELLIPSE task in $iraf.stsdas.isophote$ and the PHOT task in $iraf.digiphot.apphot$. 
We first ran the Source Extractor (SExtractor) software \citep{bertin1996} on one galaxy image. This provided the centroid and the elliptical aperture containing the entire galaxy. We adopted configuration parameters like in \citet{Bond2009} (DETECT\_THRESH =1.65 and DEBLEND\_MINCONT=1). 
We followed the choice of DETECT\_MINAREA=30 for the high-$z$ simulations. Those parameters were optimized to provide significant morphological measurements in deep HST-band observations. A larger value of contiguous pixels was adopted to prevent SExtractor from breaking up the clumpy, resolved, original $z\sim0$ LARS galaxies into smaller fragments.

We adopted SExtractor centroid, orientation angle, and ellipticity as the fixed ELLIPSE parameters and the SEx AUTO photometry semi-major axis as the reference semi-major axis length (sma0). We then measured flux within ellipses by varying the semi-major axis (sma). 
The task was able to fit elliptical isophotes at a pre-defined, fixed sma, and works better for well-defined galaxy profiles. As LARS galaxies are irregular, a better convergence of the task was performed by fixing the ellipse orientation and ellipticity.

The ELLIPSE task outputs surface brightness (I(r)) and integrated flux (F) within every sma (r$_{i}$, r$_{i+1}$, ...). We used the given surface brightness to derive the Petrosian ratio \citep[$\eta=$I(r)/$<$I(r)$>$]{Bershady2000} as a function of sma and the integrated flux to estimate r$_{20}$, r$_{50}$, r$_{80}$, the radii containing 20\%, 50\%, and 80\% of the total source flux. The Petrosian semi-major axis (rP20$^{ell}$) corresponds to the sma where $\eta=0.2$. We defined an elliptical concentration (C$^{ell}$), proportional to r$_{80}$/r$_{20}$. Applying a smoothing kernel with width equal to rP20/5, we also estimated a smoothed-image Petrosian radius \citep[rP20S,][]{Lotz2004}. 

The PHOT task outputs fluxes integrated within circular apertures. We derived the corresponding I(r) and estimated the circular Petrosian radius (rP20$^{circ}$), the circular r$_{20}^{circ}$, and r$_{80}^{circ}$. The circular concentration (C$^{circ}$) was then proportional to r$_{80}^{circ}$/r$_{20}^{circ}$.

We also defined the signal-to-noise (SN) per pixel as \\

SN$_{pixel} = \frac{1}{n} \Sigma_{i=1}^{n} \frac{S_i}{\sqrt{\sigma_{sky}^2+|S_i|}}$,

where $\sigma_{sky}$ is the standard deviation of means measured in more than three boxes around the galaxy, S$_i$ the signal, and $n$ the number of pixels belonging to a galaxy. The total SN of the galaxy was then obtained by multiplying the SN$_{pixel}$ by $\sqrt{n}$. If $\sigma_{sky}^2>|S_i|$,

SN$_{pixel} = \frac{1}{n} \Sigma_{i=1}^{n} \frac{S_i}{\sigma_{sky}}$.

The asymmetry (A) was calculated as the minimum value of the normalized difference between the galaxy image (I$_{0}$) and the same one rotated by 180$^{o}$ (I$_{180}$). We adopted the background (B) correction advised by \citet{Conselice2000} for low SN galaxies. 
Based on this definition, $0<$ A $<1$ \citep[e.g.,][]{Scarlata2007,Aguirre2013,Law2012}. After calculating the position of minimum asymmetry, we ran PHOT on that position and calculated the Petrosian radius (rP20$^{minA}$), r$_{20}^{minA}$, and r$_{80}^{minA}$, corresponding to the minimum of asymmetry. 
The point of the galaxy of minimum asymmetry is generally close to its brightest pixel, but not necessarily to the SExtractor centroid. The concentration (C$^{minA}$) at minimum asymmetry was then calculated from the r$_{80}^{minA}$/r$_{20}^{minA}$ ratio \citep{Conselice2000,Lotz2006,Jiang2013,Holwerda2014}.

\citet{Bershady2000} defined C by measuring photometry inside circular apertures. They estimated that C could be underestimated up to 30\%, in the case of an ellipticity of $\sim0.75$ and circular aperture, but that it was within 10-15\% for early-type galaxies. C$^{minA}$ was the quantity we mainly used throughout the paper.
 
A galaxy was assumed to be composed of a certain number of pixels, constituting a segmentation map. The non-parametric measurements and SN estimation were performed counting the flux of pixels belonging to that segmentation map. We defined the segmentation map in two ways, one (the fixed-size segmentation map) is an ellipse with sma $=$ rP20 and orientation given by SEx \citep[see][]{Scarlata2007}; one described in \citet{Lotz2004}, where the pixels belonging to the segmentation map have surface brightness larger than the value at rP20S. The fixed-size map was mainly concentrated in the central part of the galaxy, the second one could contain bright pixels in the galaxy outskirts. 

The Gini coefficient (G) and M20 were also calculated by following the equations in Fig. \ref{eqMORPH}. X$_{i}$ (i $=1$ to $n$) correspond to the pixel values, sorted in increasing order, and $\bar{X}$ the average pixel value, within the chosen segmentation map. f$_{i}$ (i $=1$ to $n$) correspond to the pixel values within the chosen segmentation map, but sorted in decreasing order; x$_{c}$ and y$_{c}$ are the pixels corresponding to SEx centroid. 
A first value of G (G$^{rP20}$) was estimated within the first segmentation map \citep{Lotz2004, Lotz2006, Jiang2013, Holwerda2014}. 
A second value (G$^{SB-rP20S}$) was estimated within the second segmentation map. The latter is sensitive to multiple knots in a galaxy full of structures. 
Within the fixed-sized segmentation map we measured M20 \citep{Scarlata2007,Jiang2013,Aguirre2013,Holwerda2014} and S.

The clumpiness (S) was defined by \citet{Conselice2003} as the normalized difference between the galaxy image (I) and the smoothed image (I$_{0.3xrP20}$, where the smoothing kernel sigma was $0.3\times$ rP20). The pixels belonging to the galaxy image are the ones within 0.3 and 1.5 times the rP20, i.e. we excluded the very central pixels, which are often unresolved. 

\begin{figure}
\resizebox{\hsize}{!}{\includegraphics{FigA1.png}}
\caption{Basic equations adopted to calculate galaxy size and non-parametric measurements of the LARS original and high-$z$ simulated galaxies.}
\label{eqMORPH}
\end{figure}
\begin{figure}
\resizebox{\hsize}{!}{\includegraphics[width=9cm]{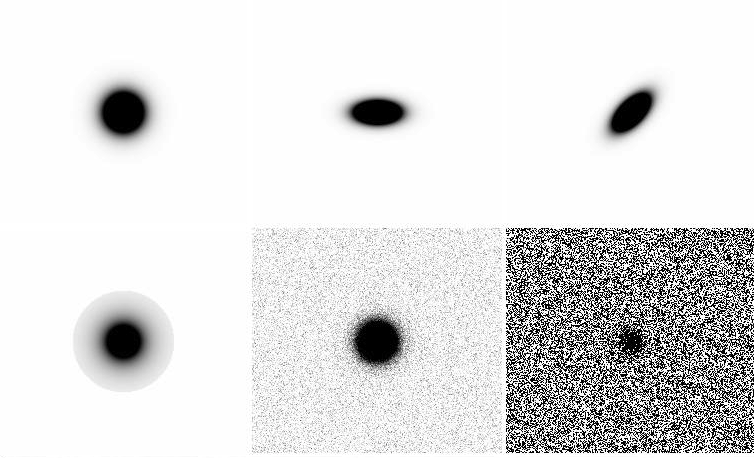}}
\caption{Simulated profiles used as a test of our code performance. From the $upper~left$ to the $lower~ right$ panels: symmetric exponential profile, exponential profile with ellipticity equal to 0.5 and position angle 0$^{o}$, exponential profile with ellipticity equal to 0.5 and position angle 45$^{o}$, deVacouleur profile, asymmetric exponential profile after adding a noise corresponding to a 10$\sigma$ detection limit 7 times fainter (noise in Tables \ref{tab:testsize} and \ref{tab:test}) and 3 times fainter than the fake galaxy flux. They were generated by running the MKOBJECT task in $iraf.artdata$ and noise was added by running the MKNOISE task in $iraf.artdata$.}
\label{summarytestIMAGE}
\end{figure}
\begin{figure*}
\centering
\includegraphics[width=18cm]{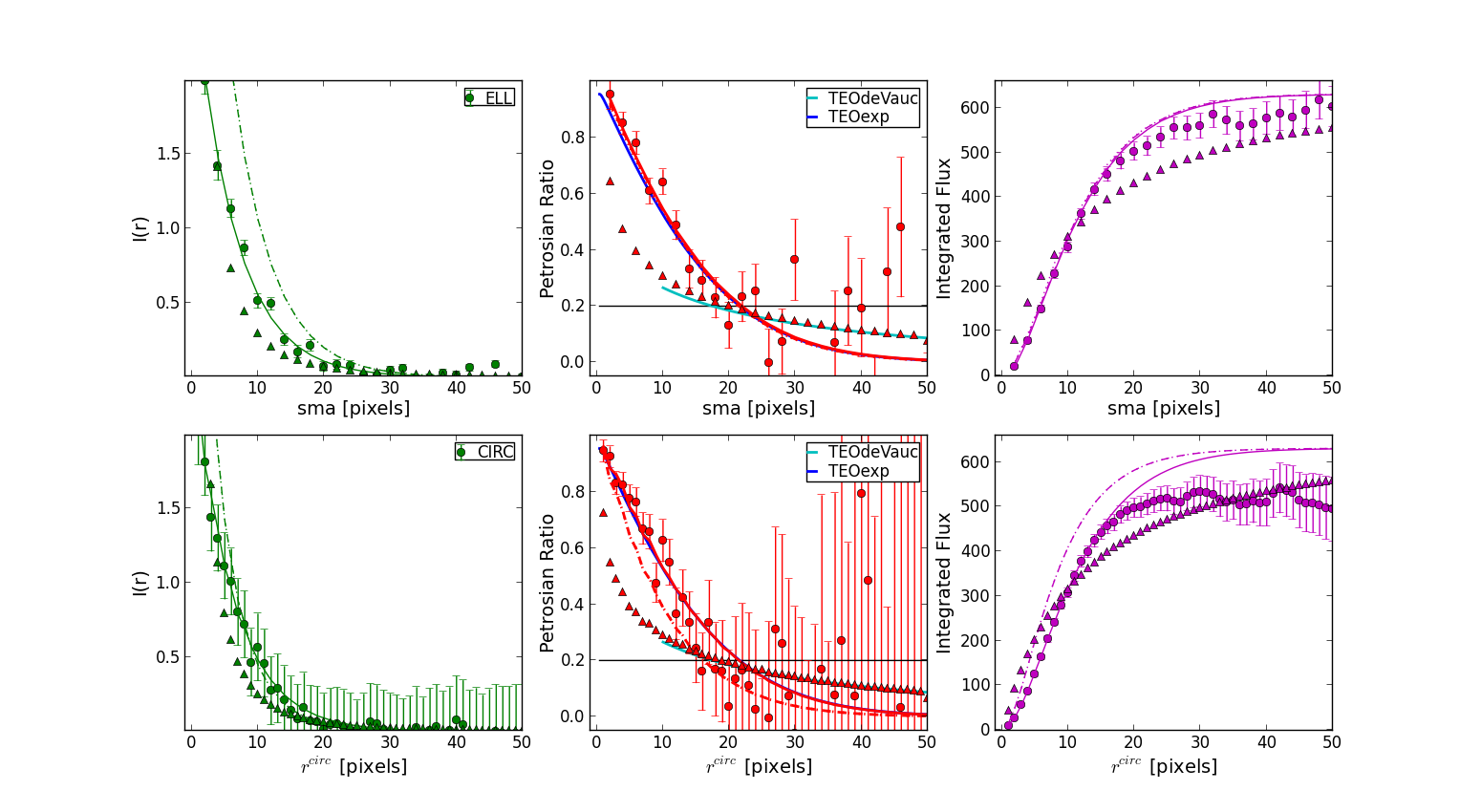}
\caption{From the $left$ to the $right$ panels surface brightness, Petrosian ratio, and integrated flux. The $upper(lower)$ row is obtained by using elliptical(circular) aperture and the $iraf$ ELLIPSE(PHOT) task. For an effective radius, Re $=10$ pixels, we estimated photometry in the case of a symmetric exponential profile (solid line), exponential profile with ellipticity equal to 0.5 and either position angle 0 (dashed line) or 45 (dotted line) degrees, and of a deVacouleur profile (triangles). The case of adding noise is represented by filled circles with and without errorbars. The analytical Petrosian ratios for exponential (THEOexp) and deVacouleur (THEOdeVauc) profiles are presented in the second-column panels as blue and cyan solid curves respectively.}
\label{summarytestPLOT}
\end{figure*}

We tested our code on the frames showed in Fig. \ref{summarytestIMAGE}. The code calculations are shown in Fig. \ref{summarytestPLOT}. First of all we analytically calculated the theoretical (THEO in the figure) Petrosian ratio and radius of an exponential and deVacouleur profiles. Then, we ran SExtractor and calculated the non-parametric measurements described above.

The results are listed in Tables \ref{tab:testsize} and \ref{tab:test}. 
The rP20$^{ell}$ size better recovers the analytical value of an exponential profile. It also well recovers the value in the case of added noise and of a deVacouleur profile. For this reason, we adopted rP20$^{ell}$ as the main size estimator throughout the paper.
By comparing with the analytical solution and the estimations by \citet{Bershady2000} and \citet{Lotz2006}, we noticed that we could underestimate C$^{circ}$ of a deVacouleur profile up to 30\%, C$^{minA}$ of an exponential(deVacouleur) profile up to 10(20)\%, and overestimate G$^{SB-rP20S}$ up to 10(5)\% for an exponential(deVacouleur) profile. M20 is well-recovered for all the profiles within 3\%.

Also, we tested our code on the public galaxy stamps from COSMOS and compared the results with the ones obtained with the ZEST software \citep{Scarlata2007}. When assuming the brightest pixel as the centre of a galaxy, we recovered G$^{rP20}$ values in more than 80\% of the cases. As the fixed-size segmentation map is a better choice for redshift comparisons \citep{Scarlata2007}, we tended to prefer G$^{rP20}$ rather than G$^{SB-rP20S}$ throughout the paper.  

\begin{table*}

%\tabletypesize{\scriptsize}
%\rotate
\centering
\caption{Sizes from known profile galaxies of Re=10 pixels}
%\tablewidth{0pt}
\label{tab:testsize}

\scalebox{1.0}{
\begin{tabular}{|c|c|c|c|c|c|c|}
\hline

profile & rP20$^{ell}$ & rP20$^{circ}$ & rP20$^{minA}$ & r$_{20}$ & r$_{50}$ & r$_{80}$  \\
\hline
kpc  &kpc & kpc & kpc& kpc& kpc& kpc \\ 
\hline
exp  & &  & & & &  \\ 
ell0 & 22.0 & 21.0 & 19.0 & 5.5 & 10.0 & 17.5 \\
ell0.5,pa0 & 21.5 & 16.0 & 15.0 & 5.0 & 10.0 & 17.5 \\
ell0.5,pa45 & 21.5 & 16.0 & 15.0 & 5.0 & 10.0 & 17.0 \\
noise & $19.0_{-1}^{+5}$ & $16.0_{-2}^{+7}$ & 23.0 & 5.0 & 10.0 & 16.0 \\
\hline
deVauc  & &  & & & &  \\   
& 20.5 & 17.0 & 11.0 & 2.5 & 7.5 & 17.0 \\
\hline

\end{tabular}
}
\tablefoot{The analytical values for the ratio of rP20 and effective radius is 2.16 for an exponential and 1.82 for a deVacouleur profile \citep{Bershady2000}.}

\end{table*}

\begin{table*}

%\tabletypesize{\scriptsize}
%\rotate
\centering
\caption{Morphology measurements from known profile galaxies of Re $=10$ pixels}
%\tablewidth{0pt}
\label{tab:test}

\scalebox{1.0}{
\begin{tabular}{|c|c|c|c|c|c|c|c|c|c|}
\hline
profile & G$^{rP20}$ & G$^{SB-rP20S}$ & M20 & ell & C$^{circ}$ & C$^{ell}$ & C$^{minA}$ & A & S \\
\hline
exp  & &  & & & & & & & \\
ell0 & 0.4888 & 0.5229 & -1.752 & 0.05 & 2.657& 2.513 & 2.386 & 0.0 & 0.016 \\
ell0.5,pa0 & 0.4923 & 0.5326 & -1.836 & 0.5 & 2.559& 2.720 & 2.559 & 0.0 & 0.036 \\
ell0.5,pa45 & 0.4874 & 0.5333 & -1.796& 0.5 & 2.559 & 2.657 & 2.559 & 0.0 & 0.034 \\
noise & 0.4914 & 0.5164 & -1.164 & 0.114 & 2.236 & 2.526 & 2.236 & -0.0135 & 0.393\\
\hline
deVauc  & &  & & & &  & & & \\  
& 0.6183 & 0.6271 & -2.398 & 0.05 & 3.495 & 4.162 & 3.495 & 0.0 & 0.025 \\
\hline

\end{tabular}
}
\tablefoot{The analytical values of C$^{circ}$ is 2.7 and 5.2 for an exponential and a deVacouleur profile respectively \citep{Bershady2000}. \citet{Lotz2006} estimated G$^{SB-rP20S}$=0.473, M20=-1.8, C$^{minA}$=2.71 for a noise-free image of an exponential profile and G$^{SB-rP20S}$=0.6, M20=-2.47, C$^{minA}$=4.34 for a noise-free image of a deVacouleur profile (half-light-radius equal to 600 resolution elements).}

\end{table*}

\subsection{Comparison between original and simulated galaxy morphological parameters}
\label{ORsimPARAM}

Following the method described in the previous section, we estimated sizes and morphological parameters in the continuum and line images. We adopted the same method in the case a galaxy was simulated to be at $z\sim2$ and observed in different depth surveys. To quantify the morphological parameter variations owing pixel resampling and added noise, we defined $\Delta$(param),\\

\begin{equation}
\Delta (param) = \frac{param(z\sim2,noise) - param(original)}{param(original)} \\
\label{eq1}
\end{equation}

which represented the difference between the measurement of a parameter in the simulated and in the original image, normalized to the value in the original image. 
By definition, $\Delta$(param) tended to be zero when the two measurements were very similar, tended to be equal to $-1$ when the measurement in the simulated image was much smaller than in the original one, and equal to 1 when the measurement in the simulated image was twice that in the original one.
We calculated $\Delta$ (param) for every galaxy and estimated the mean and the standard deviation. 

\begin{figure*}
\centering
\includegraphics[width=12cm]{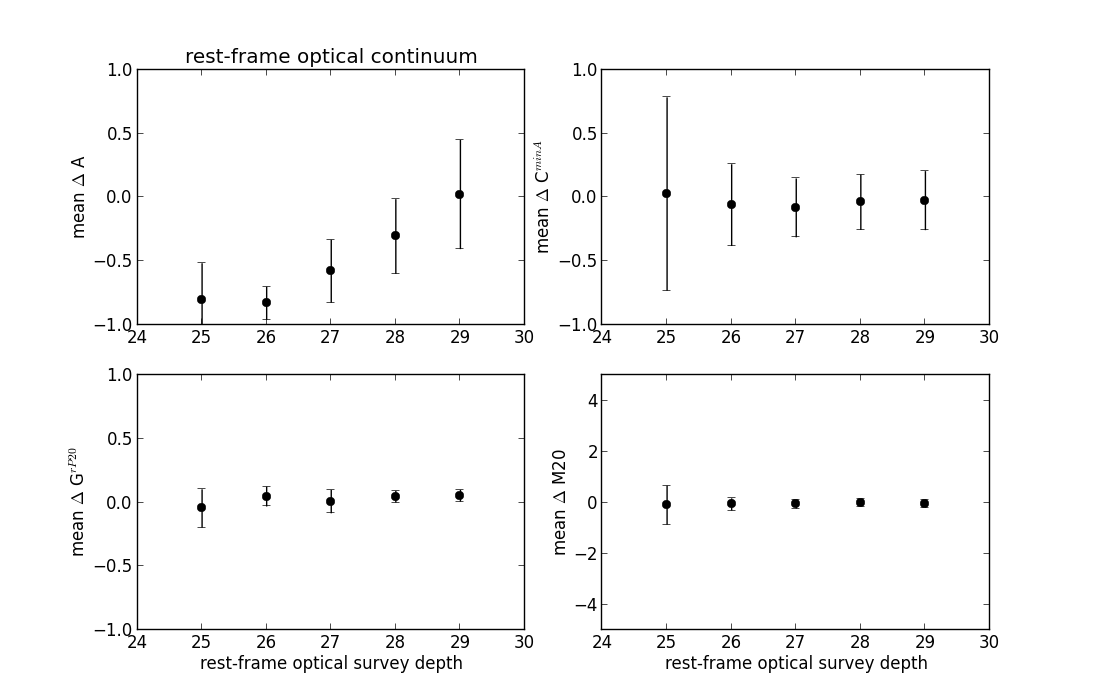}
\includegraphics[width=12cm]{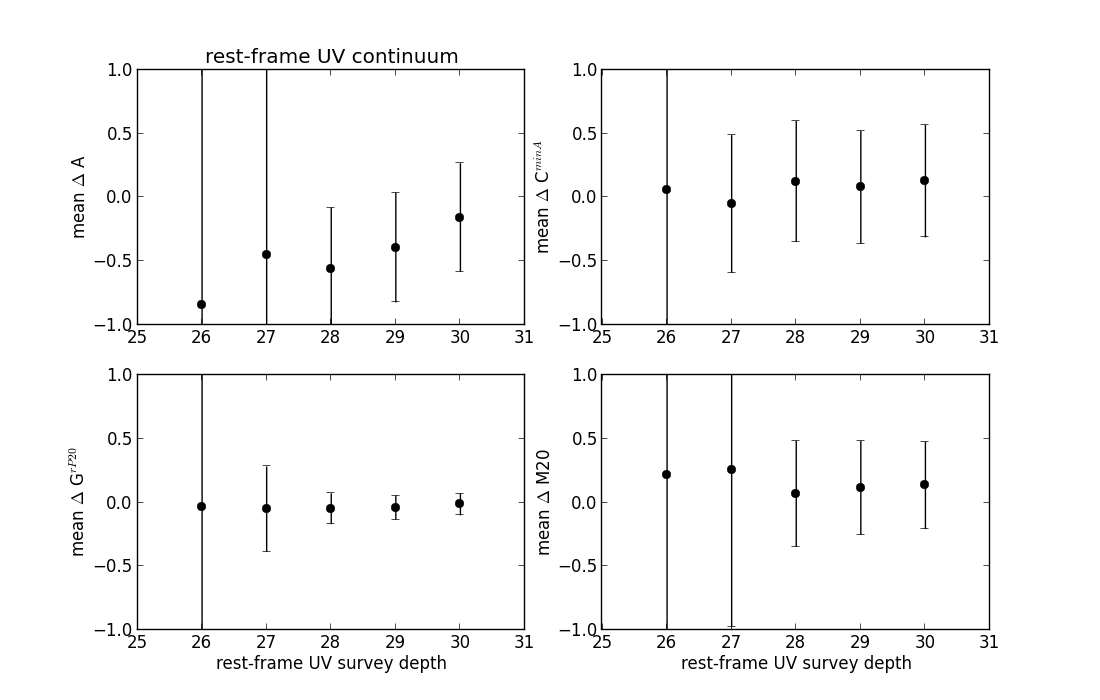}
\includegraphics[width=12cm]{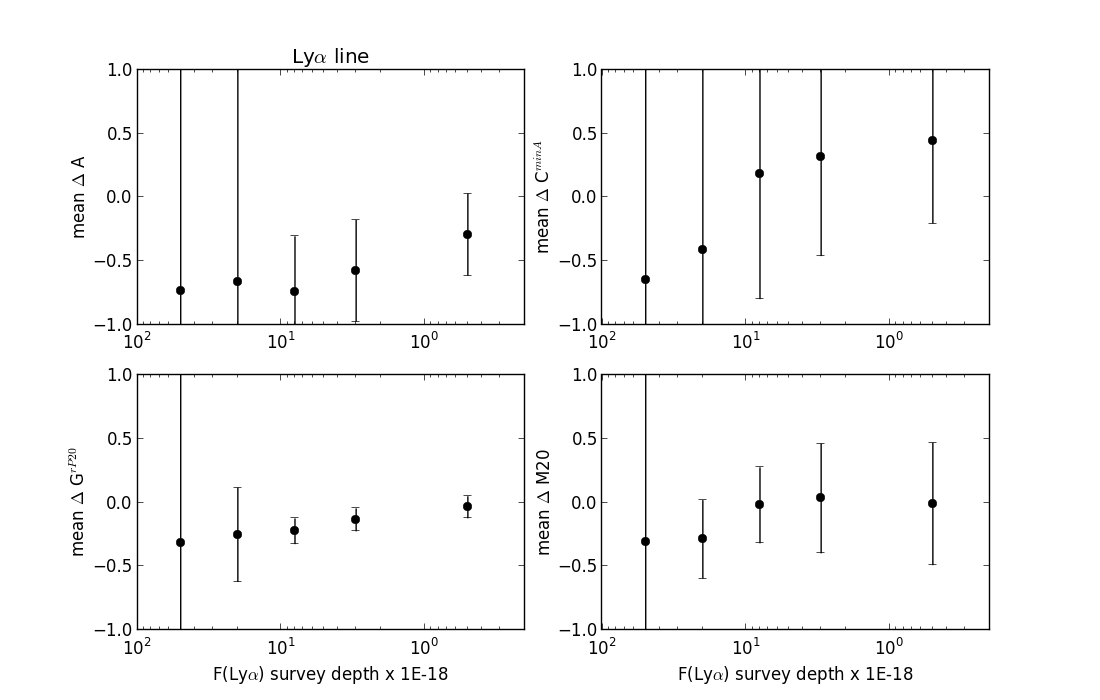}
  \caption{Mean $\Delta (param)$ as defined in eq. \ref{eq1} vs simulated survey depth. $param$ is A ($upper~left$), C$^{minA}$ ($upper~right$), G$^{rP20}$ ($lower~left$), and M20 ($lower~right$) for the rest-frame optical, UV, and Ly$\alpha$ simulated images. The deepest simulated survey is represented by the most right data point in each panel. The error bars represent the standard deviation among all the high-$z$ simulated LARS galaxies. In the case fewer-than-twelve galaxies are detected at a specific depth, the error bars is increased proportionally to the number of undetected galaxies. The numbers on the top right corner of each panel represent the Spearman test coefficient, r, and probability, p, of uncorrelated datasets, assuming that the depth vector indicates the deepest survey on the right of the x-axis.}
  \label{DeltacoLa}
\end{figure*}

In Fig. \ref{DeltacoLa} we show the results. A, C$^{minA}$, G$^{rP20}$, and M20 were all preserved after pixel resampling in the continuum images. Owing survey depth, A decreased. 

\citet{HuertasCompany2014} studied the variation of A, G, and M20 due to resampling the rest-frame optical of local galaxies to $z>1$. They found that A can increase up to 50\%, G up to 10\%, while M20 can decrease up to 10\%. These trends were observed to be more pronounced for early-type galaxies; \citet{Lotz2006} resampled the rest-frame UV of a few sources from $z=1.5$ to $z=4$, finding that G and M20 were preserved, consistent with our results.
\citet{Overzier2010} investigated the change of A, C, G, and M20 when resampling $z=0.2$ Lyman break analogues to $z=2$. They gave an R-value scale, where $|R|\sim0$ when the difference between the median parameter measured in the resampled and in the original image was small and $|R|\sim1$ when was comparable to the sample scatter. They found $R^{A}=-1.1$, R$^{C}=0.02$, R$^{G}=-0.34$, and R$^{M20}=-0.43$ in the rest-frame UV and $R^{A}=-0.46$, R$^{C}=-0.52$, R$^{G}=-1.1$, and R$^{M20}=-0.26$ in the rest-frame optical. 
Recently, \citet{Petty2014} explored the variations of G and M20 of local LIRGS when simulated to be at $z=$ 0.5, 1.5, 2, 3 and in a survey with the HUDF depth. Some of the galaxies in their sample, characterized by clumps and filamentary structures (typical of merging systems) in high-resolution HST images, tended to appear as disk-like galaxies by $z=2$. Some others maintained merging systems' morphology. All these trends are in agreement with our findings, but they also tell us that the variations do not follow a specific trend for irregular galaxies. 

In the Ly$\alpha$ images, A tended to decrease and C$^{minA}$ could increase due to pixel resampling. Owing survey depth, all the four parameters tended to decrease and A significantly decreased. 

\begin{figure*}
\centering
\includegraphics[width=12cm]{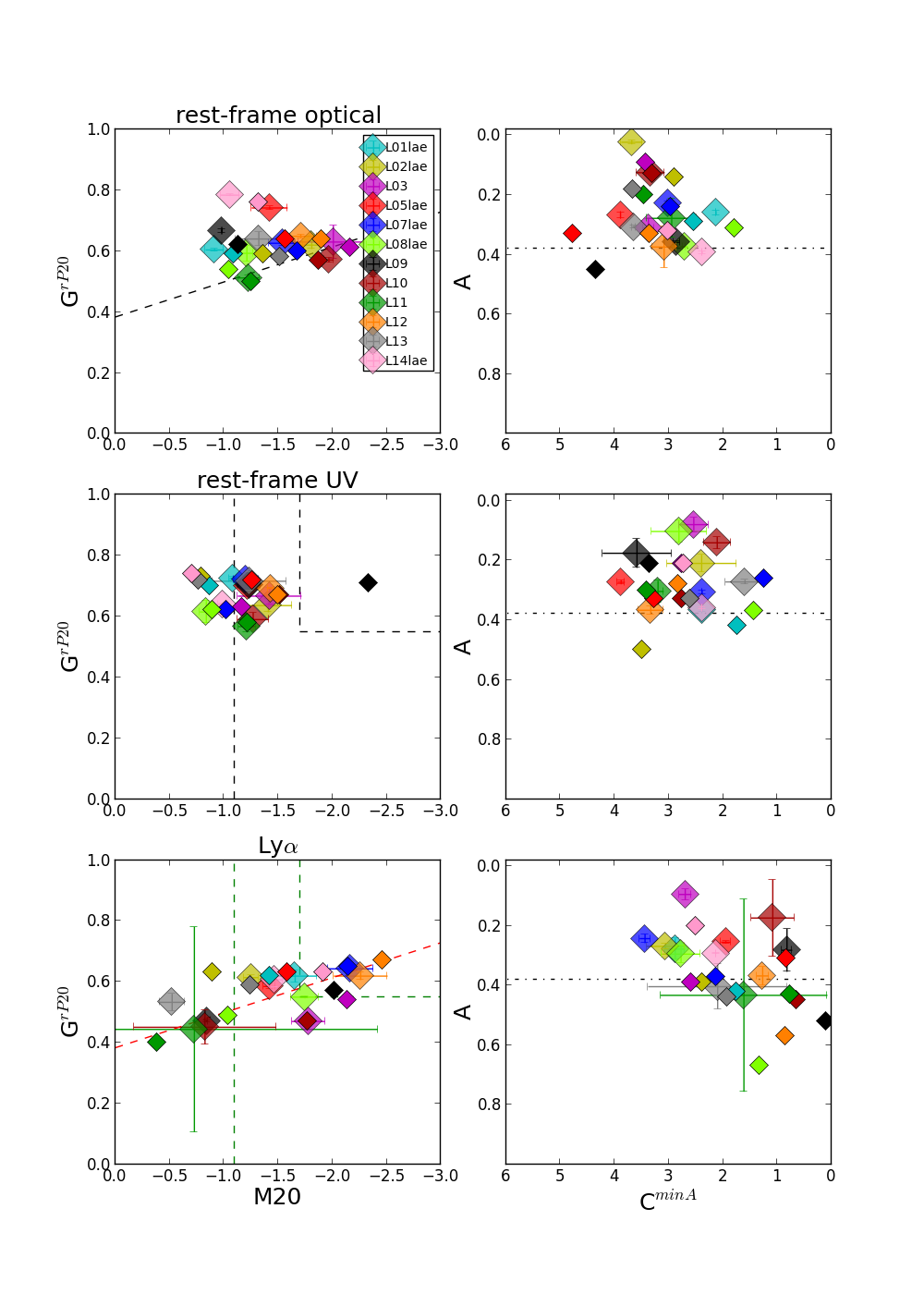}
  \caption{G$^{rP20}$ vs M20 (first column) and A vs C$^{minA}$ (second column) measured in the rest-frame optical ($upper$), UV ($middle$), and Ly$\alpha$ ($lower$) images. Small and big diamonds with error bars indicate the parameters of the original and (deepest survey) high-$z$ simulated LARS galaxies. 
The dashed and dotted-dashed lines indicate the separation between merging system and normal galaxies' parameter space. Owing pixel resampling the galaxies stay in the same region of the G$^{rP20}$ vs M20 diagram.}
  \label{GMAC_coHa_liLa_z2orig}
\end{figure*}

Another way to study the change of morphological parameters is to look at pairs of them, like G$^{rP20}$ vs M20 and A vs C$^{minA}$. In Fig. \ref{GMAC_coHa_liLa_z2orig} we show the diagrams of two-parameters, measured in the rest-frame optical, UV, and Ly$\alpha$ images. Our analysis showed that the G and M20 estimations were not affected by resolution and survey depth. Therefore, they were useful for comparisons at different redshifts and different survey depths throughout the paper. 

\clearpage

\section{LARS galaxies simulated at high redshift in a deep and shallow survey}
\label{sec:appendix2}

\begin{figure*} 
\centering
\includegraphics[width=20cm]{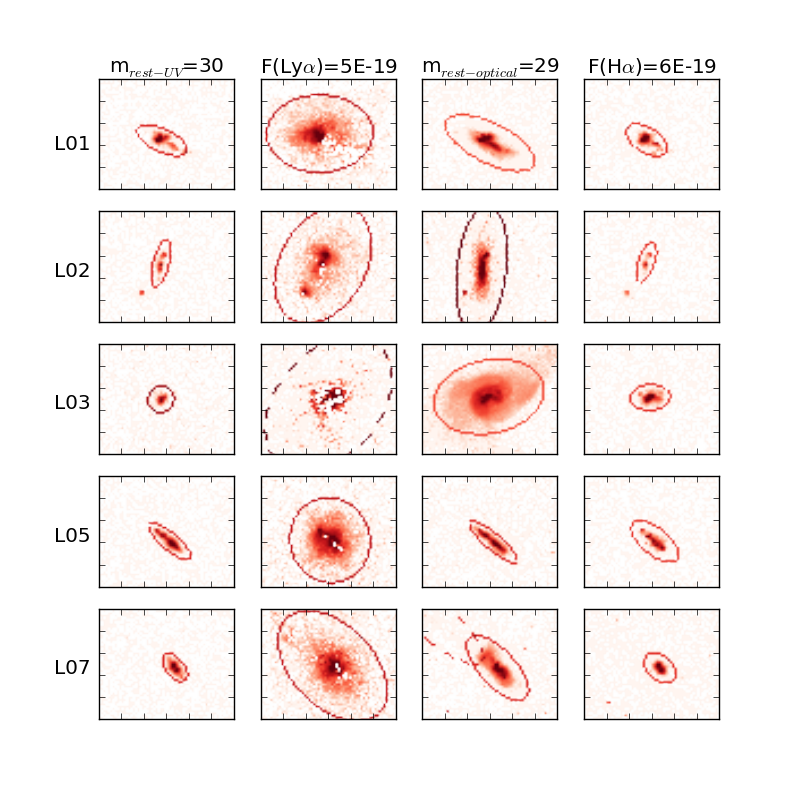}
\caption{Rest-frame UV (first column), Ly$\alpha$ (second column), rest-frame optical (third column), and H$\alpha$ (fourth column panels) images for z2L01, z2L02, z2L03, z2L05, and z2L07 in the deepest survey, probed here. Every panel is 20x17 kpc wide. The reddish ellipses indicate SEx apertures (dashed curves indicate flagged sources according to SEx convention), corresponding to the assumed detection parameters: DETECT THRESH =1.65, DETECT MINAREA=30, and DEBLEND MINCONT=1 from Bond et al. (2009). The colour scaling is logarithmic and chosen to show a visually consistent background noise.}
\label{stampsDEEP1}
\end{figure*}

\begin{figure*} 
\centering
\includegraphics[width=20cm]{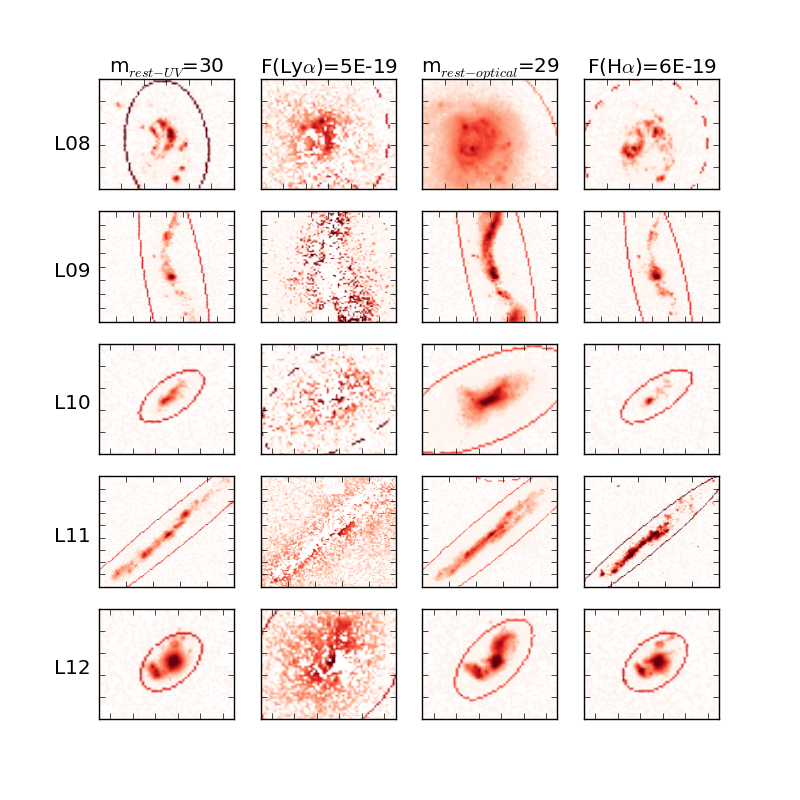}
\caption{As Fig. \ref{stampsDEEP1} for z2L08, z2L09, z2L10, z2L11, and z2L12. In the case of z2L09 and z2L11 the size is 27x27 kpc and 33x30 kpc respectively to fit their elongated shapes. For these galaxies SEx apertures in Ly$\alpha$ happen to be outside the shown region.}
\label{stampsDEEP2}
\end{figure*}

\begin{figure*} 
\centering
\includegraphics[width=20cm]{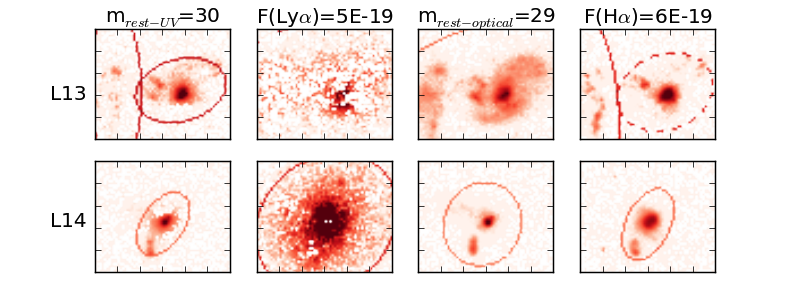}
\caption{As Fig. \ref{stampsDEEP1} for z2L13 and z2L14. For z2L13 SEx aperture in Ly$\alpha$ happens to be outside the shown region. Also, two main sources connected by undetectable (m$_{rest-UV}>$30 and F(H$\alpha)<6$E-19 erg sec$^{-1}$ cm$^{-2}$) surface brightness structures are seen in UV continuum and H$\alpha$. By using the chosen detection parameters, SEx found two sources as separated. As in Ly$\alpha$ and optical continuum the photometric measurements are done centring the aperture close to the right clump, we locate the aperture on that one for the photometry in UV and H$\alpha$ as well.}
\label{stampsDEEP3}
\end{figure*}

\begin{figure*} 
\centering
\includegraphics[width=20cm]{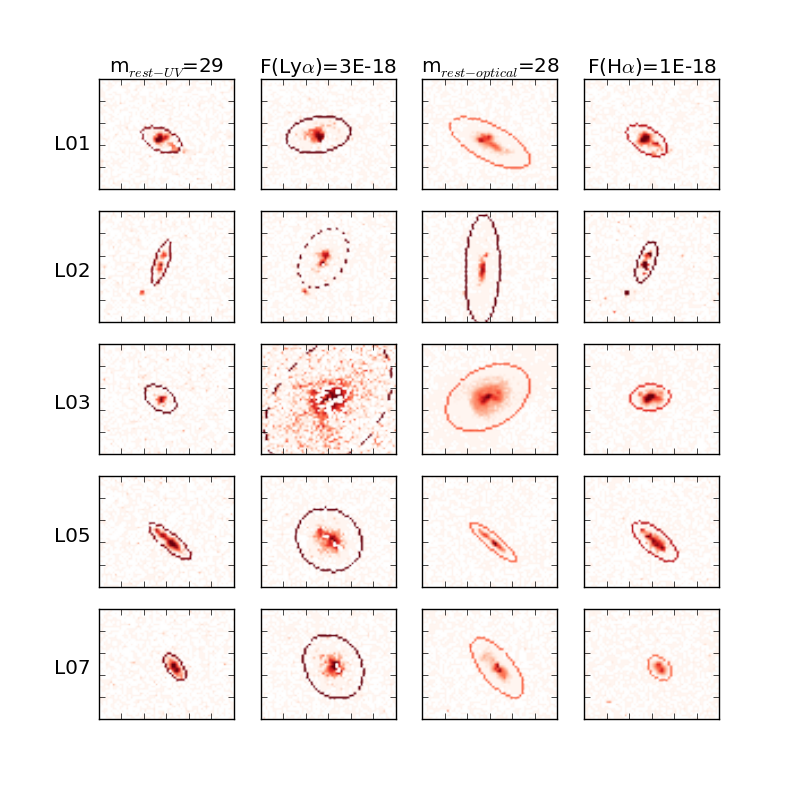}
\caption{As B.1, but for a shallow simulated survey. The dashed-line apertures indicate that some detected source is blended to another \citep{bertin1996}.}
\label{stampsSHALLOW1}
\end{figure*}

\begin{figure*} 
\centering
\includegraphics[width=20cm]{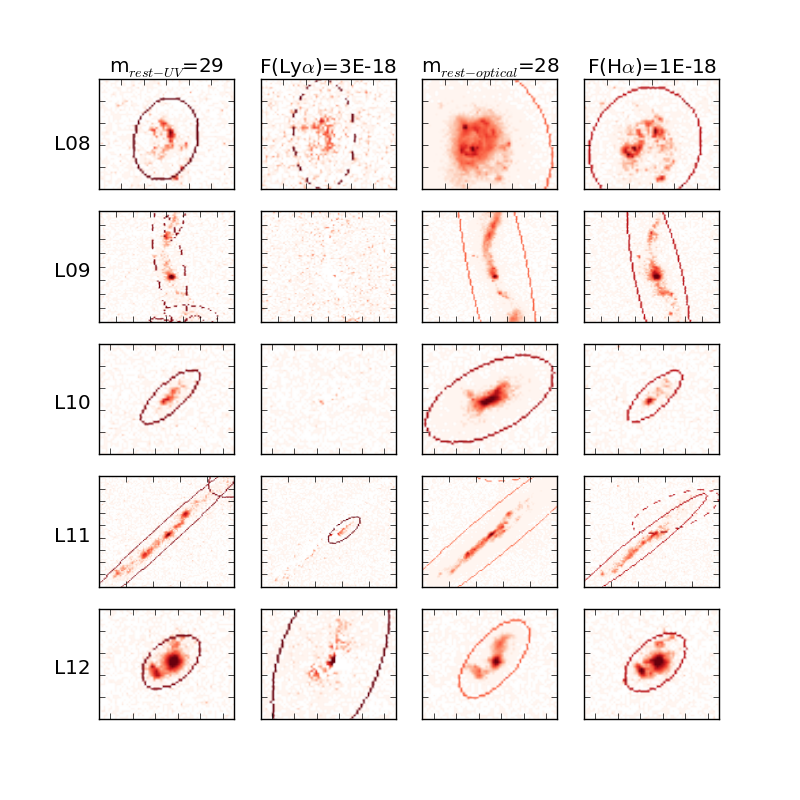}
\caption{As Fig. \ref{stampsDEEP2}, but for a shallow simulated survey. z2L09 and z2L10 are not detected in Ly$\alpha$ at a 10$\sigma$ detection limit of 3E-18 erg sec$^{-1}$ cm$^{-2}$.}
\label{stampsSHALLOW2}
\end{figure*}

\begin{figure*}
\centering 
\includegraphics[width=20cm]{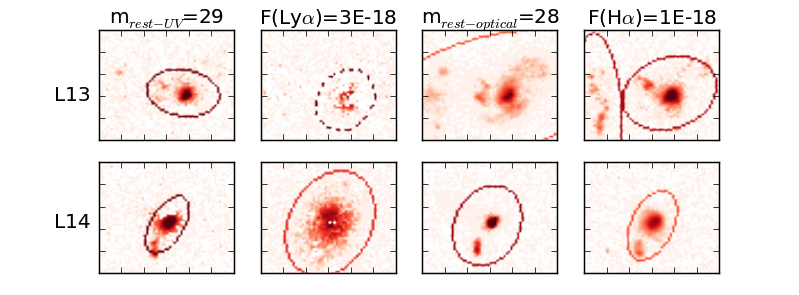}
\caption{As Fig. \ref{stampsDEEP3}, but for a shallow simulated survey. Only the bright right clump of z2L13 is detected in UV at a 10$\sigma$ detection limit of m$_{rest-UV}=29$. Its Ly$\alpha$ emission is localized around the lower side of the right clump. In H$\alpha$ the two overlapping sources become well separated.}
\label{stampsSHALLOW3}
\end{figure*}

In this appendix, we show the continuum and line maps of the high-$z$ simulated LARS galaxies. Figures \ref{stampsDEEP1}, \ref{stampsDEEP2}, and \ref{stampsDEEP3} present rest-frame UV, Ly$\alpha$,  rest-frame optical, and H$\alpha$ images in the deepest survey depth probed here, together with SEx detection apertures. In Figs. \ref{stampsSHALLOW1}, \ref{stampsSHALLOW2}, and \ref{stampsSHALLOW3}, we show the same stamps but for a shallower simulated survey.

\section{Surface brightness profiles of original and high-$z$ simulated LARS galaxies}
\label{sec:appendix3}

We present here the surface brightness profiles of eleven LARS galaxies studied in this paper. The profiles of L01 were shown in Fig. \ref{SBL01}. Every figure in this appendix shows four panels, rest-frame UV and optical continua, Ly$\alpha$ and H$\alpha$ lines. The profiles are normalized to 2 kpc to compare continuum and line profiles.
\begin{figure*} 
\centering
\includegraphics[width=16cm]{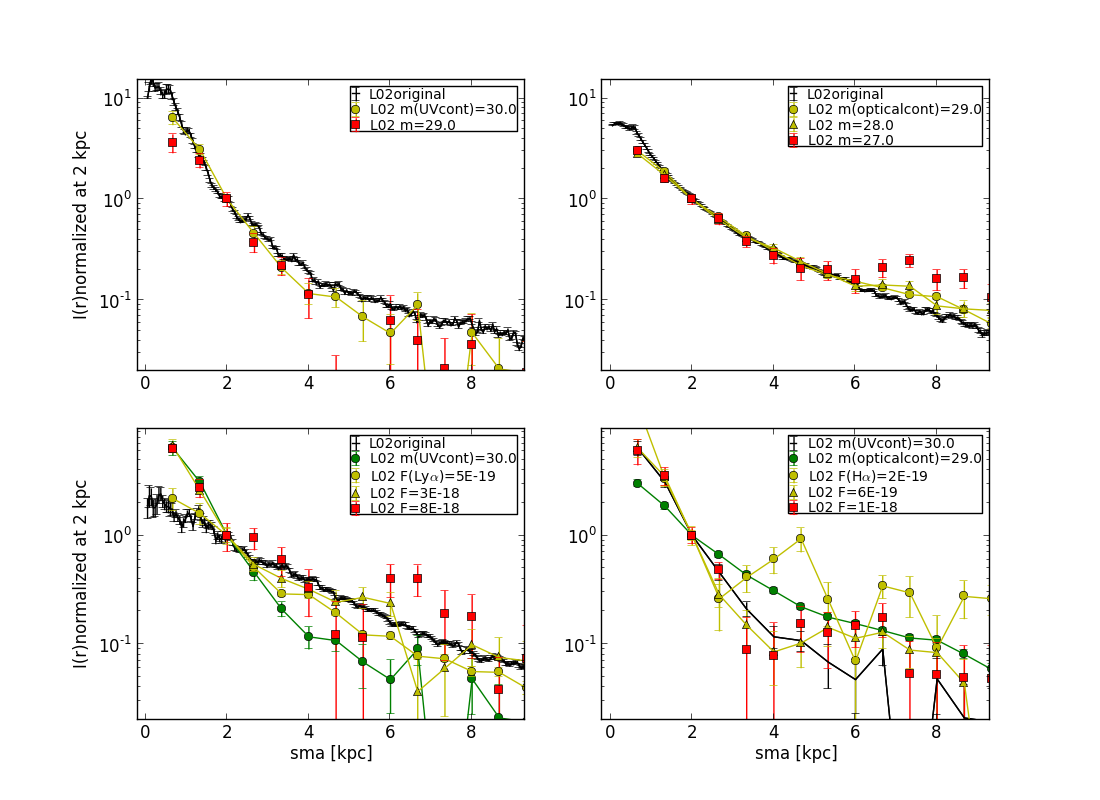}
\caption{Normalized surface brightness profiles of L02. Black points with error bars correspond to the surface brightness profile of the original LARS images in the rest-frame UV, optical, and Ly$\alpha$ as explained in the text (Fig. \ref{SBL01}). The red squares represent the profile affected by background noise, for a certain shallow survey.
}
\label{SBL02}
\end{figure*}

\clearpage

\begin{figure*}
\centering
\includegraphics[width=9cm]{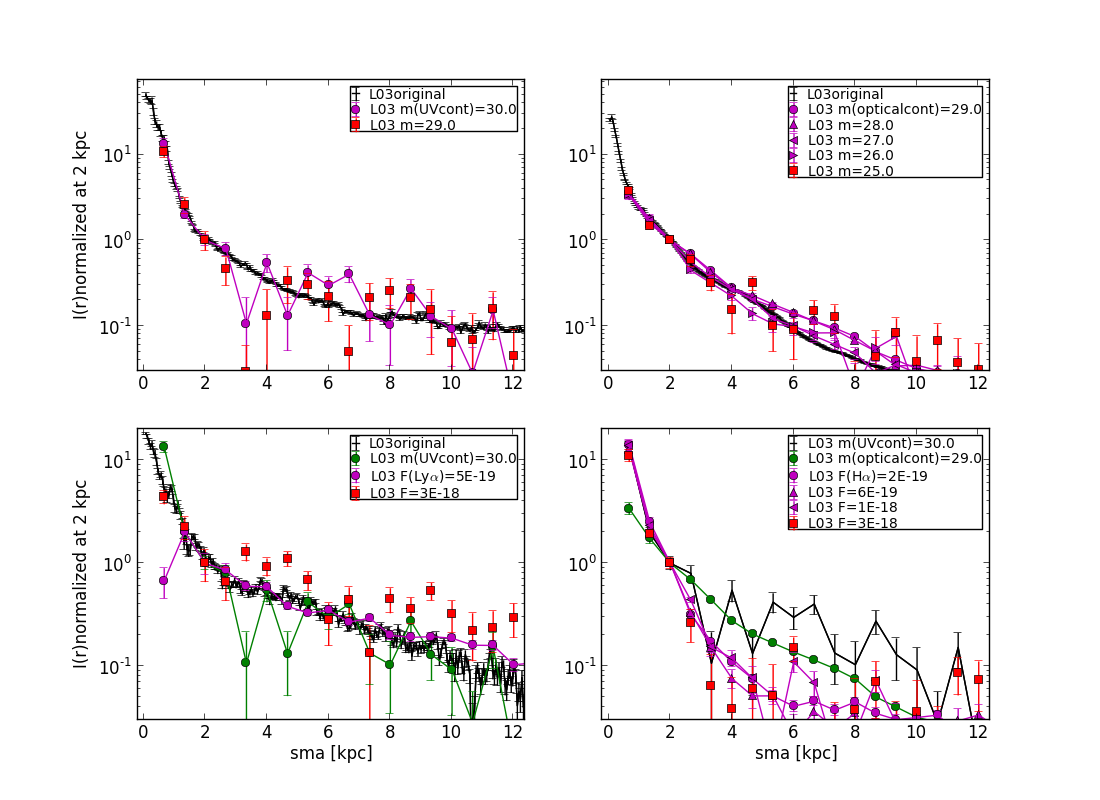}
\includegraphics[width=9cm]{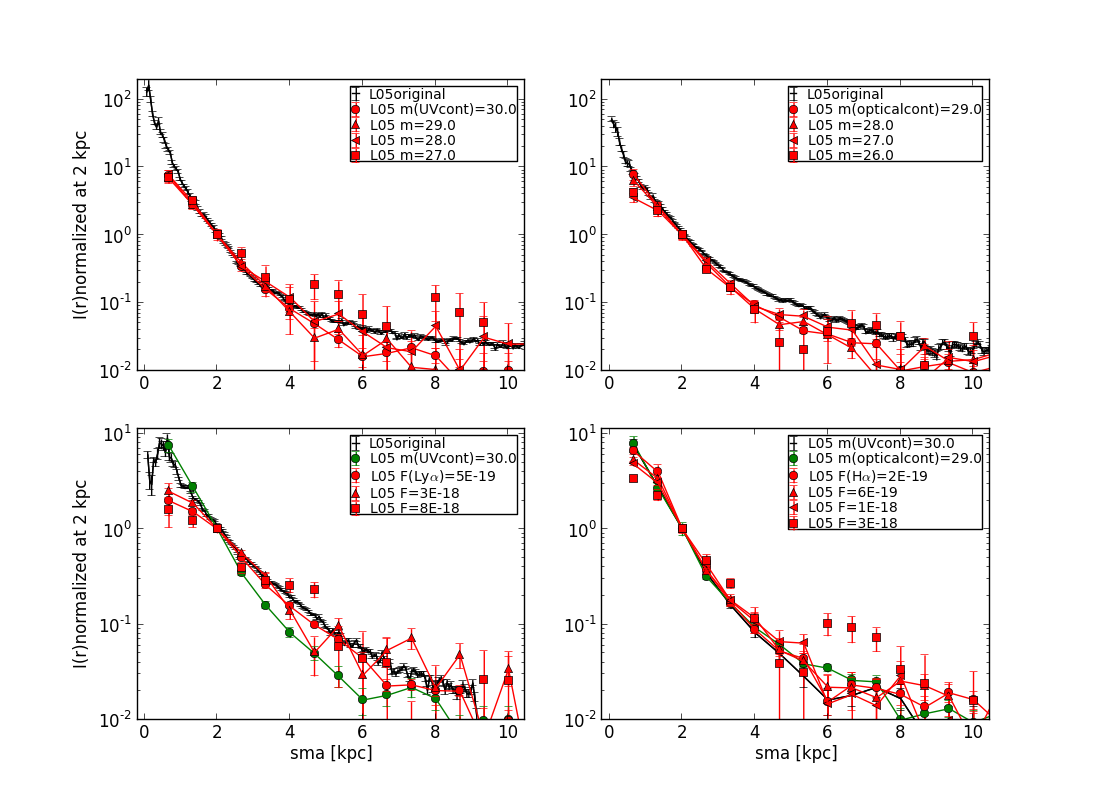}
  \caption{Same colour coding as in Fig. \ref{SBL02}, but for z2L03 and z2L05. The deepest and intermediate depth surveys are shown for z2L03(z2L05) in magenta(red) symbols with error bars, as throughout the paper.}
  \label{SBL03L05}
\end{figure*}
\begin{figure*}
\centering
\includegraphics[width=9cm]{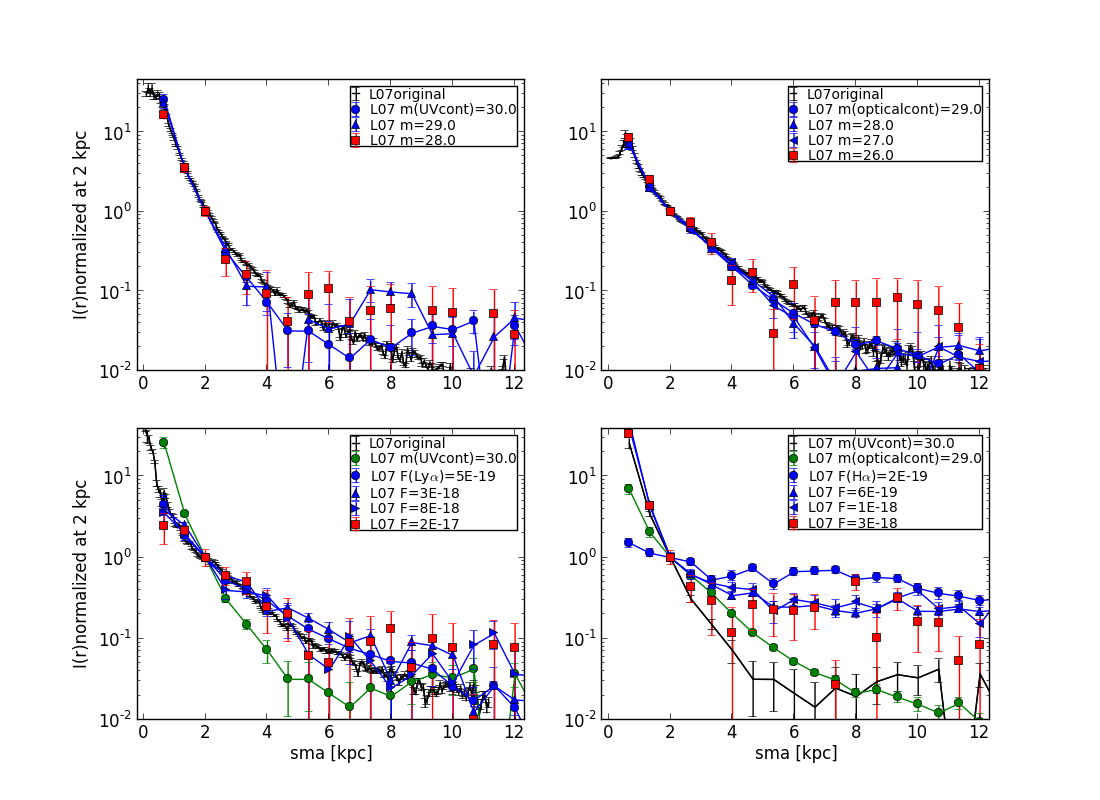}
\includegraphics[width=9cm]{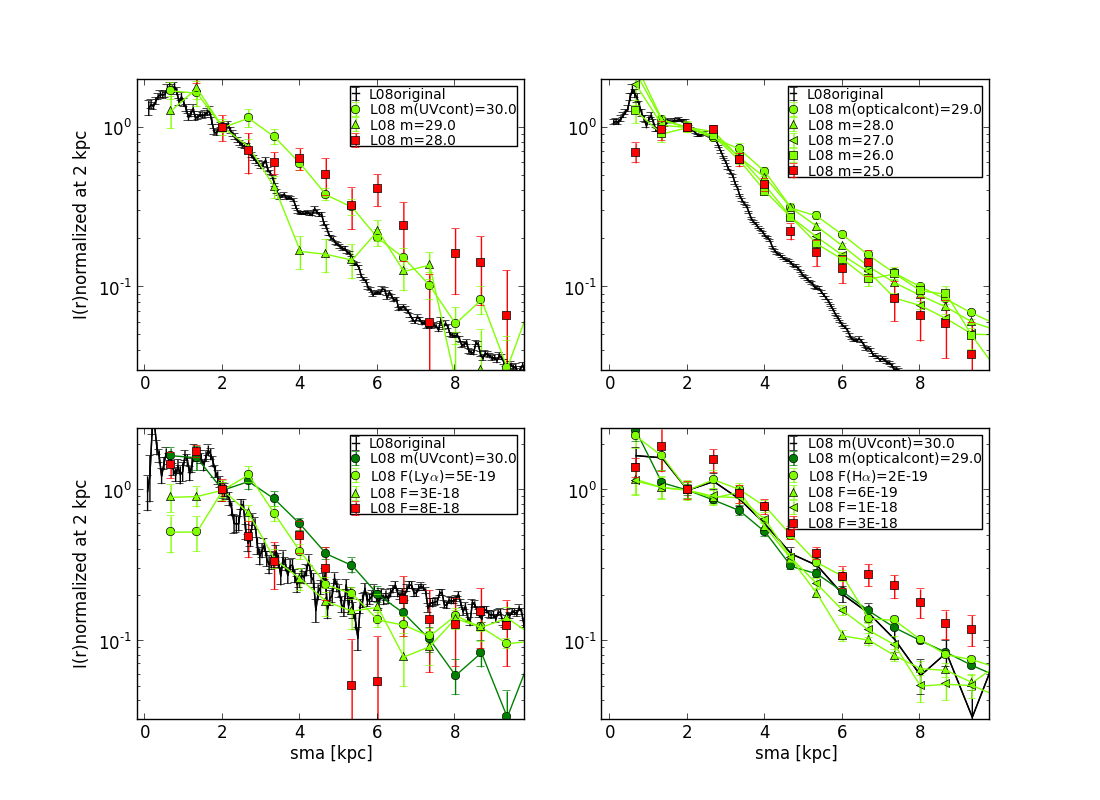}
  \caption{Same colour coding as in Fig. \ref{SBL02}, but for z2L07 and z2L08. The deepest and intermediate depth surveys are shown for z2L07(z2L08) in blue(light green) symbols with error bars.}
  \label{SBL07L08}
\end{figure*}
\begin{figure*}
\centering
\includegraphics[width=9cm]{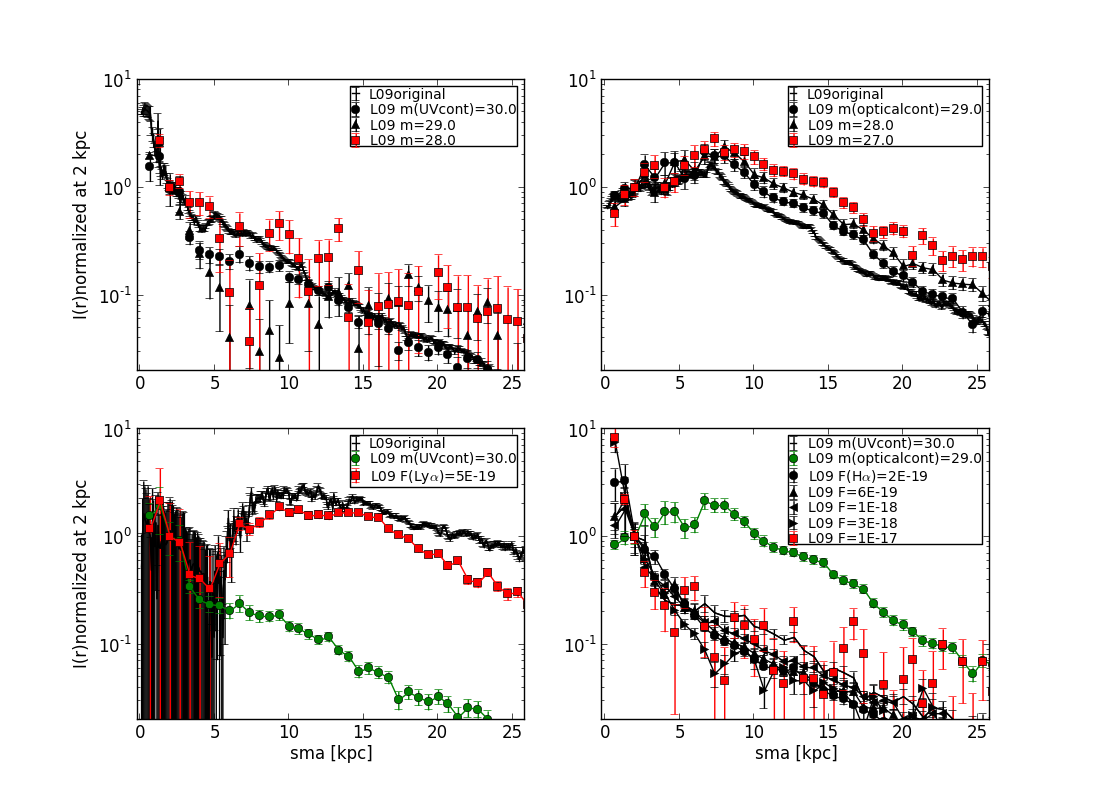}
\includegraphics[width=9cm]{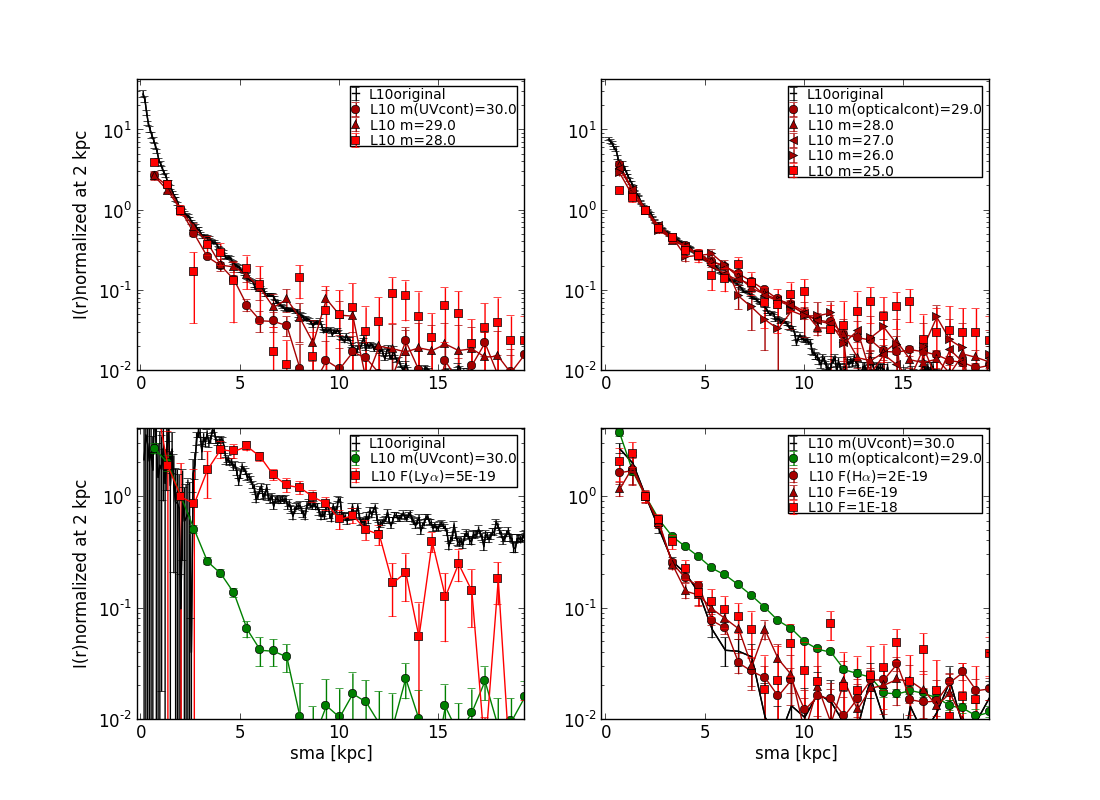}
  \caption{Same colour coding as in Fig. \ref{SBL02}, but for z2L09 and z2L10. The deepest and intermediate depth surveys are shown for z2L09(z2L10) in black(dark red) symbols with error bars.}
  \label{SBL09L10}
\end{figure*}
\begin{figure*}
\centering
\includegraphics[width=9cm]{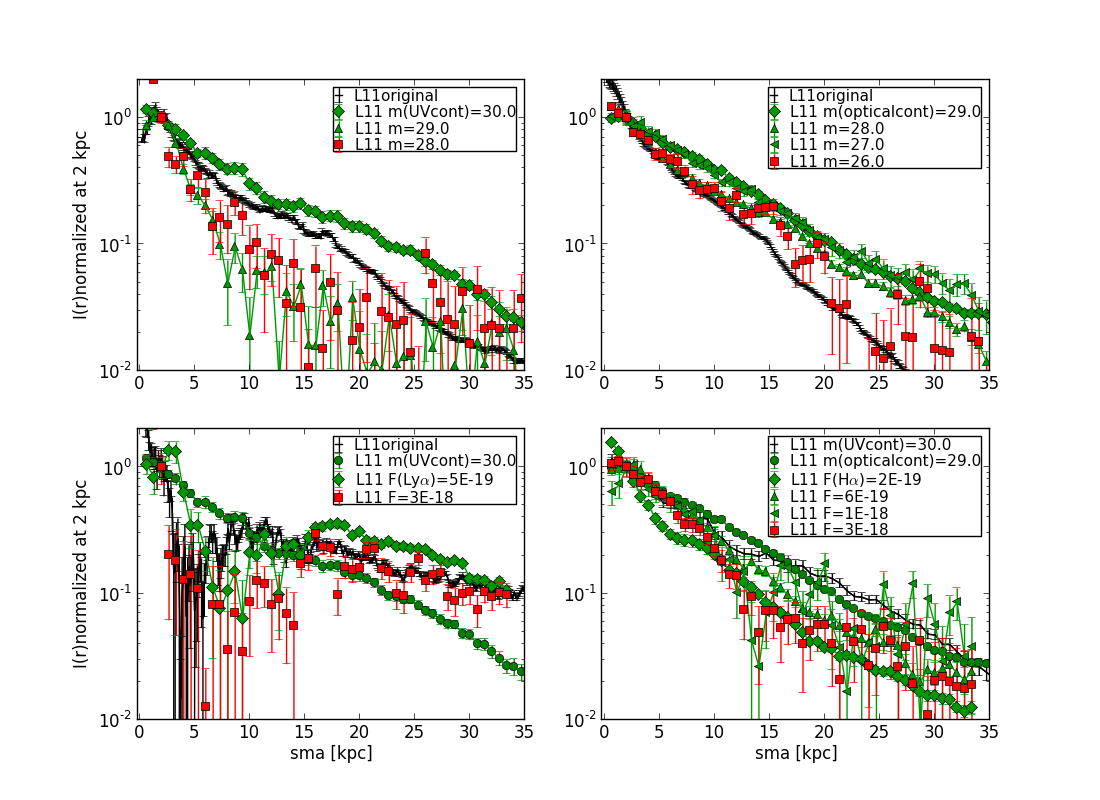}
\includegraphics[width=9cm]{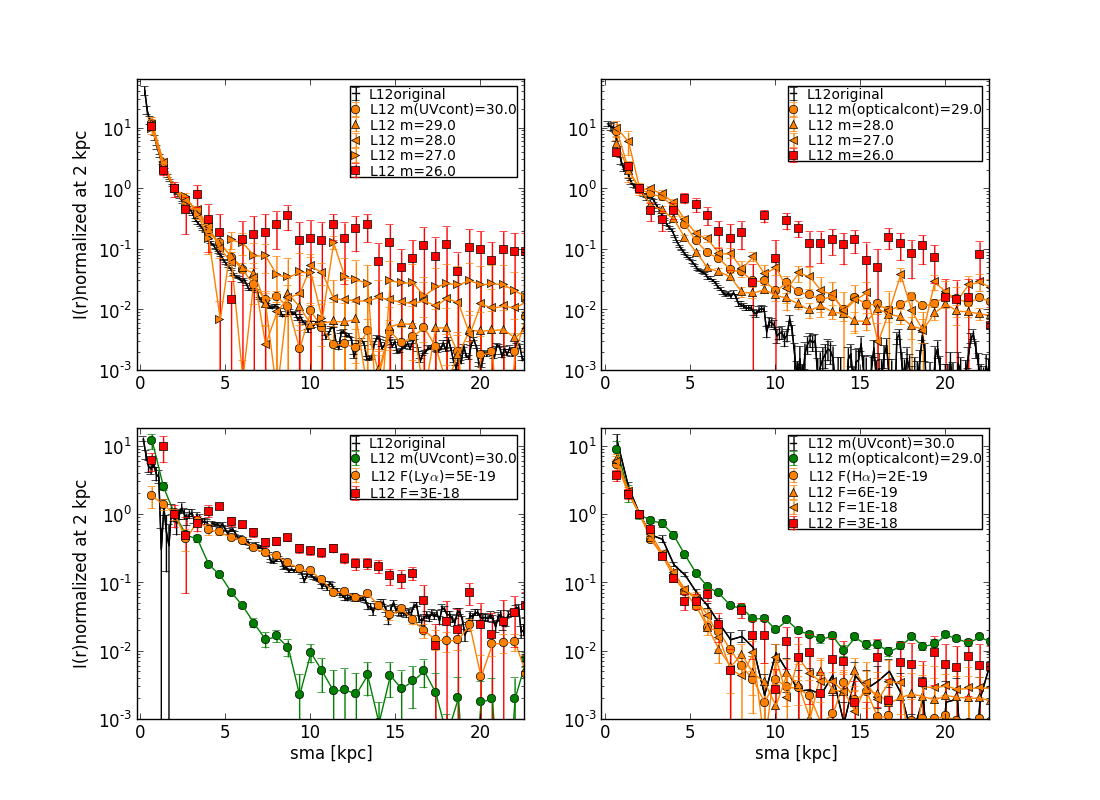}
  \caption{Same colour coding as in Fig. \ref{SBL02}, but for z2L11 and z2L12. The deepest and intermediate depth surveys are shown for z2L11(z2L12) in dark green(orange) symbols with error bars.}
  \label{SBL11L12}
\end{figure*}
\begin{figure*}
\centering
\includegraphics[width=9cm]{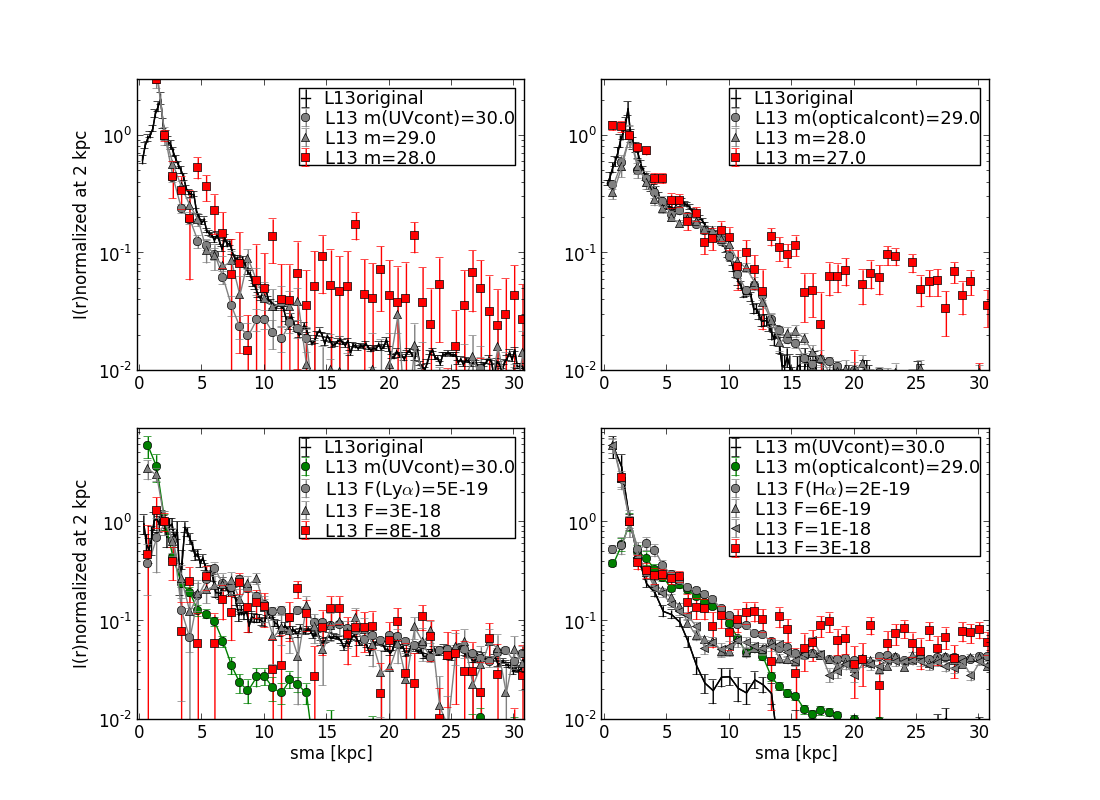}
\includegraphics[width=9cm]{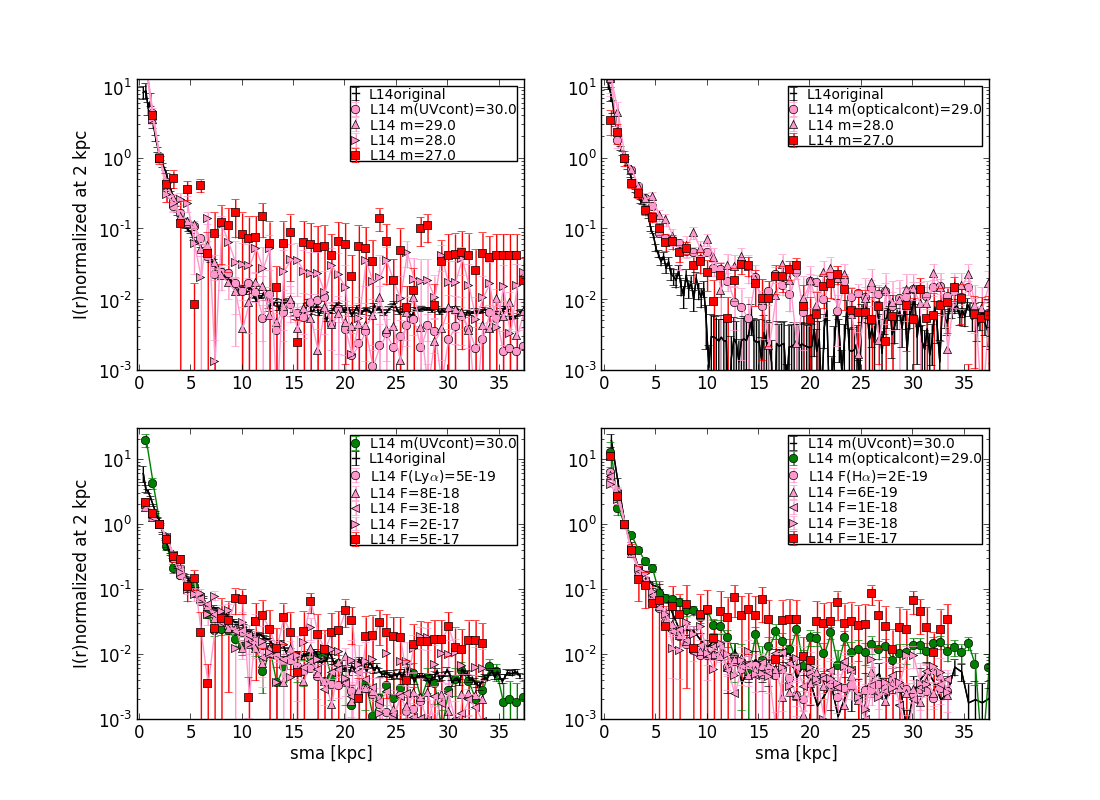}
  \caption{Same colour coding as in Fig. \ref{SBL02}, but for z2L13 and z2L14. The deepest and intermediate depth surveys are shown for z2L13(z2L14) in grey(pink) symbols with error bars.}
  \label{SBL13L14}
\end{figure*}

\end{document}